\documentclass[12pt]{JHEP3}
\usepackage{mathrsfs}
\usepackage{amsmath,amssymb}
\usepackage{epsfig}
\usepackage{relsize}
\usepackage{graphicx,here}

\def\ssp{\hspace{0.3mm}}

\newcommand{\E}{\mathcal E}
\def\tH{\widetilde{{\cal E}}}

\def\sl(2){\alg{sl}(2)}

\def\det{\hbox{det}}
\def\be{\begin{equation}}
\def\ee{\end{equation}}

\newcommand{\bea}{\begin{eqnarray}}
\newcommand{\eea}{\end{eqnarray}}
\newcommand{\beml}{\begin{multline}}
\newcommand{\eeml}{\end{multline}}

\def\a {\alpha}

\def\s {\sigma}
\def\pa {\partial}

\def\g {\gamma}
\def\om {\omega}

\def\la{\label}
\def\e{\epsilon}
\def\ov{\over}

\def\tp{{\widetilde p}}

\newcommand{\alg}[1]{\mathfrak{#1}}
\newcommand{\su}{\alg{su}}

\newcommand{\psu}{\alg{psu}}

\newcommand{\AdS}{{\rm  AdS}_5\times {\rm S}^5}

\newcommand{\bem}{\left (\begin{matrix}}
\newcommand{\eem}{\end{matrix} \right )}

\def\ka{\kappa}
\def\hstar{\,\hat{\star}\,}
\def\cstar{\,\check{\star}\,}

\def\({\left(}
\def\){\right)}

\def\ts{\tilde{s}}

\author{Gleb Arutyunov$^a$\footnote{Email: G.E.Arutyunov@uu.nl, frolovs@maths.tcd.ie, rsuzuki@maths.tcd.ie} {}\footnote{Correspondent fellow at Steklov
Mathematical Institute, Moscow.}\,, \  Sergey Frolov$^{b\,\dagger}$
\, and\,  Ryo Suzuki$^b$
 \\ $^{a}$ {\it Institute for Theoretical
Physics and Spinoza Institute,\\ ~~Utrecht University, 3508 TD
Utrecht, The Netherlands} \\ $^b$ {\it Hamilton Mathematics Institute and School of Mathematics, \\
~~Trinity College, Dublin 2, Ireland} }

\abstract{ We apply the contour deformation trick to the
Thermodynamic  Bethe Ansatz equations for  the $\AdS$ mirror
model, and obtain the integral equations determining the energy of
two-particle excited states dual to ${\cal N}=4$ SYM operators
from the $\sl(2)$ sector. We show that each state/operator is
described by its own set of TBA equations. Moreover, we provide
evidence that for each state there are infinitely-many critical
values of 't Hooft coupling constant $\lambda$, and the excited
states integral equations have to be modified each time one
crosses one of those. In particular, estimation   based on the
large $L$ asymptotic solution gives $\lambda\approx774$ for the
first critical value corresponding to the Konishi operator. Our
results indicate that the related calculations and conclusions of
Gromov, Kazakov and Vieira should be interpreted with caution. The
phenomenon we discuss might potentially explain the mismatch
between their recent computation of the scaling dimension of the
Konishi operator and the one done by Roiban and Tseytlin by using
the string theory sigma model.

}

\title{Exploring the mirror TBA}

\preprint{%\smaller{\smaller{\smaller{???}}}\\[-.5ex]
          \smaller{\smaller{\smaller{ITP-UU-09-54}}}\\[-.5ex]
          \smaller{\smaller{\smaller{SPIN-09-44}}}\\[-.5ex]
          \smaller{\smaller{\smaller{TCDMATH 09-24}}}\\[-.5ex]
          \smaller{\smaller{\smaller{HMI-09-10}}}}
\begin{document}c

\renewcommand{\thefootnote}{\arabic{footnote}}
\setcounter{footnote}{0}
%%%%%%%%%%%%%%%%%%%%%%%%%%%%%%%
%%%%%%%%%%%%%%%%%%%%%%%%%%%%%%%%%

%\vskip 0.5cm

\section{Introduction}
%%%%%%%%%%%%%%%%%%%%%%%%%%%%%%%%%

An important open problem of the AdS/CFT correspondence
\cite{Maldacena} is to understand the finite-size spectrum of the
$\AdS$ superstring. Recently, there has been further significant
progress in this direction. First, the four-loop anomalous
dimension of the Konishi operator was computed  \cite{BJ08} by
means of generalized L\"uscher's formulae
\cite{Luscher85,JL07,BJ08} (see also
\cite{HS08a}-\cite{BJ09} for other applications of
L\"uscher's approach), and the result exhibits a stunning
agreement with a direct field-theoretic computation
\cite{Sieg,Vel}. Second, the groundwork for constructing the
Thermodynamic Bethe Ansatz (TBA) \cite{Zamolodchikov90}, which
encodes the finite-size spectrum for all values of the 't Hooft
coupling, has been laid down, based on the mirror theory
approach\footnote{The TBA approach in the AdS/CFT spectral problem
was advocated  in \cite{AJK} where it was used to explain wrapping
effects in gauge theory.} \cite{AF07}. Most importantly, the
string hypothesis for the mirror model was formulated \cite{AF09a}
and used to derive
 TBA equations for the ground state
\cite{AF09b}-\cite{GKKV09}.  Also, a  corresponding Y-system \cite{ZamY}  was conjectured
\cite{GKV09}, and its general solution was obtained \cite{Heg}.
The AdS/CFT Y-system has unusual properties, and, in particular,  is
defined on an infinite-genus  Riemann surface \cite{AF09b,FS}.

\smallskip

The derivation of the TBA equations is not yet complete, because
the equations pertain only to the ground state energy (or Witten's
index in the case of periodic fermions) and do not capture the
energies of excited states. Therefore, one has to find a
generalization of the TBA equations that can account for the
complete spectrum of the string sigma model, including all excited
states.

\smallskip

Here we continue to explore the mirror TBA approach. In
particular, we will be interested in finding the TBA integral
equations which describe the spectrum of string states in the
$\sl(2)$ sector. An attempt in this direction has been already
undertaken in \cite{GKKV09} and the emerging integral equations
have been used for numerical computation of the anomalous
dimension of the Konishi operator \cite{GKV09b}. However, the
subleading term in the strong coupling expansion in this result
disagrees with the result by \cite{RT09k} obtained by string
theory means. There exists yet another prediction \cite{AF05} for
this subleading term, which differs from both \cite{GKV09b} and
\cite{RT09k}. All these results are based on certain assumptions
which require further justification. This makes urgent to
carefully analyze the issue of the TBA equations for excited states, and
to better understand what happens on the string theory side.

\smallskip

In this paper we analyze two-particle states in the $\sl(2)$
sector. First, we show that each state is governed by its
own set of the TBA equations.
 Second, we provide evidence that for each state there are
infinitely-many critical values of 't Hooft coupling constant
$\lambda$, and that the excited states integral equations have to
be modified each time one of these critical values is crossed.\footnote{Existence of such critical values was observed
in the excited-state TBA equations for perturbed minimal models
\cite{DT97}. We thank Patrick Dorey for this comment.}
Performing careful analysis of two-particle states in a region
between any two neighboring critical points, we propose the
corresponding integral equations.

\smallskip

The problem of finite-size spectrum of two-dimensional integrable
models has been studied in many works, see e.g.
\cite{Kuniba:1993cn}-\cite{GKV08}. To explain our findings, we
start with recalling that for some integrable models the inclusion of
excited states  in the framework of the TBA approach  has been achieved by applying a certain analytic
continuation procedure \cite{DT96, DT97}. This can be understood
from the fact that the convolution terms entering the integral TBA
equations exhibit a singular behavior in the complex rapidity
plane, the structure of these singularities does depend on the
value of the coupling constant. This
leads to a modification of the ground-state TBA equations, which,
indeed, describe the profile and energies of excited states.
Here we intend a similar strategy for the string sigma model.

\smallskip

To derive the TBA equations for excited states, we propose to use
a contour deformation trick. In other words, we assume that the
TBA equations for excited states have the same form as those for
the ground state with the only exception  that the integration
contour in the convolution terms is different. Returning the
contour back to the real rapidity line of the mirror theory, one
picks up singularities of the convolution terms which leads to
modification of the final equations. The original contour should
be drawn in such a way, that the arising TBA equations would
reproduce the large $L$ asymptotic solution (where $L$ is the size
of the system).

\smallskip

Recall that the TBA equations for the string mirror model
\cite{AF09b} are written in terms of the following Y-functions:
$Y_Q$-functions associated with $Q$-particle bound states,
auxiliary functions $Y_{Q|vw}$ for $Q|vw$-strings, $Y_{Q|w}$ for
$Q|w$-strings, and $Y_{\pm}$ for $y_\pm$-particles. The
Y-functions depend on the 't Hooft coupling $\lambda$ related to
the string tension $g$ as $\lambda=4\pi^2 g^2$. As we will see,
the analytic structure of these Y-functions depends on $g$ and
plays a crucial role in obtaining the TBA equations for excited
states.

\smallskip

Most conveniently, the large $L$ asymptotic solution for the
Y-functions is written in terms of certain transfer-matrices
associated with an underlying symmetry group of the model
\cite{Kuniba:1993cn,Tsuboi}. In the context of the string sigma
model the corresponding asymptotic solution was presented in
\cite{GKV09}. We will use this solution to check the validity of
our TBA equations.

\smallskip

Our analysis starts from describing physical two-particle states
in the $\sl(2)$ sector. It appears that for the $\sl(2)$
states the functions $Y_{Q|vw}(u)$ play the primary role in
formulating the excited state TBA equations. Analyzing the
asymptotic solution, we find that each $Y_{Q|vw}$ has four zeroes
in the complex $u$-plane. With $g$ changing, the zeroes change
their position as well, and at certain critical values $g=g_{cr}$
they give rise to new singularities in the TBA equations which
resolution results in the appearance of new driving terms.
 The critical values $g_{cr}$ are defined as values of $g$ at
which $Y_{Q|vw}(u)$ acquires two zeros at $u=\pm i/g_{cr}$:
$Y_{Q|vw}(\pm i/g_{cr})=0$.

We show that at weak coupling $g\sim 0$ all two-particle states
can be organized into an infinite tower of classes
$k={\rm I}\,, {\rm II}\,, \ldots\,, \infty$, see Table 1.
\begin{center}
{\small
 {
 \renewcommand{\arraystretch}{1.5}
\renewcommand{\tabcolsep}{0.2cm}
\begin{tabular}{|c|l|l|}
\hline Type of a state & Y-functions & Number of zeros\\
\hline I & $Y_{1|vw}$   & 2\\
\hline II & $Y_{1|vw}$, $Y_{2|vw}$ & 2+2 \\
\hline III & $Y_{1|vw}$, $Y_{2|vw}$, $Y_{3|vw}$ & 4+2+2 \\
\hline IV & $Y_{1|vw}$, $Y_{2|vw}$, $Y_{3|vw}$,
$Y_{4|vw}$  & 4+4+2+2 \\
\hline \vdots & \vdots & \vdots \\
\hline $k\to\infty$ & $Y_{1|vw}$, $Y_{2|vw}$,\quad\ldots &
4+4+\quad\ldots
\\ \hline
\end{tabular}
}
}

\vspace{0.5cm}
\parbox{13cm}
{\small Table 1. Classification of two-particle states in the $\sl(2)$-sector
at $g\sim 0$. The right column shows the number of zeros which the
corresponding asymptotic $Y_{M|vw}$-functions from the middle
column have in the physical strip $|{\rm Im}(u)|<1/g$. States of
type I are called ``Konishi-like".
States of type II, III, IV, $\ldots$ correspond to larger value of  $\kappa$, see section 3.}
\end{center}
\vspace{0.3cm}
Each
class is unambiguously determined by number of zeroes of
$Y_{M|vw}$-functions  in the strip $|{\rm
Im}(u)|<1/g$. In
particular, for states of type I only
$Y_{1|vw}$-function has two zeroes in the physical strip. We call
all these states ``Konishi-like" because they share this property
with the particular string state corresponding to the Konishi
operator.

\smallskip

Our results disagree with that by \cite{GKKV09,GKV09b}
in the following two aspects: First, the integral equations for
excited states in the $\sl(2)$ sector do not have a universal
form, even for two-particle states. Second, we face the issue of critical points. When a critical point is
crossed, the compatibility of the asymptotic solution with the
integral TBA equations requires modification of the latter. The
equations proposed in \cite{GKKV09} capture only type I states and
only below the first critical point.

\smallskip

To find approximate locations of the critical values, we first
solve the asymptotic Bethe-Yang equations \cite{BS05} (which
include the BES/BHL dressing phase \cite{BES,BHL06}) for some
states numerically from weak to strong coupling and obtain the
corresponding interpolating curve $u\equiv u_{1}(g)$, where
$u_{1}$ is the rapidity of an excited particle in string theory.
The second particle has rapidity $u_{2}=-u_{1}$ due to the level
matching condition. Further, we compute the Y-functions on the
large $L$ asymptotic solution corresponding to this two-particle
excited state and study their analytic properties considered as
functions of $g$, and, in particular, determine approximately the
critical values.

\smallskip

We also note that with the coupling increasing more and more
critical points get crossed which leads to accumulation of zeroes
of $Y_{M|vw}$'s in the physical strip. Apparently, as the
asymptotic solution indicates, when $g$ tends to infinity the
zeroes move towards the points $\pm 2$, so that the latter points
behave as an attractor for zeroes of all $Y_{M|vw}$-functions.

\smallskip

Estimation based on the large $L$ asymptotic solution gives
$\lambda\approx774$ for the first critical value corresponding to
the Konishi operator. In the weak-coupling region below the first
critical point, the integral equations for Konishi operator we
obtain seem to agree with that of \cite{GKKV09}. However, these
weak-coupling equations become inconsistent with the known large
$L$ asymptotic solution  once the first critical point is crossed,
and have to be modified. Consequently, the derivation of the
anomalous dimension for the Konishi operator at strong coupling
requires re-examination. Of course, the existence of critical
points is not expected to violate analyticity of the energy
$E(\lambda)$ of a string state considered as the function of
$\lambda$, but it poses a question about the precise analytic
behavior of $E(\lambda)$ in the vicinity of critical points.

\smallskip

We discuss both the canonical and simplified TBA equations. The
canonical equations \cite{AF09b,Bombardelli:2009ns,GKKV09} follow
from the string hypothesis for the mirror model \cite{AF09a} by
using the standard procedure, see e.g. \cite{Korepin}. The
simplified equations \cite{AF09b,AF09d} obtained from the
canonical ones have more close relation to the Y-system. It turns
out that the simplified equations are sensitive only to the
critical points defined above. In contrast, the canonical
equations have to be modified when crossing not only a critical
point but also what we call a subcritical point $\bar{g}_{cr}$. A
subcritical point $\bar{g}_{cr}$ is defined as the value of $g$ at
which the function $Y_{Q|vw}(u)$ acquires zero at $u=0$.
Hence, in comparison to the canonical equations, the simplified
equations exhibit a more transparent analytic structure. In
addition to locality, this is yet another reason why we attribute
to the simplified equations a primary importance and carry out
their analysis in the main text. To study the exact Bethe
equations which determine the exact, {\it i.e.} non-asymptotic,
location of the Bethe roots, we find it advantageous to use a
so-called {\it hybrid form} of the TBA equations for
$Y_Q$-functions. This form is obtained by exploiting both the
canonical and simplified TBA equations.

%The more complicated discussion
%of the canonical equations is postponed to the appendix.

\smallskip

Recently, the finite-gap solutions of semi-classical string theory
have been nicely derived \cite{Gromov} from the TBA equations
\cite{GKKV09}. This raises a question why modifications of the TBA
equations we find in this paper were not relevant for this
derivation. We have not studied this question thoroughly. However,
one can immediately see that there is a principle difference
between states with finite number of particles and semiclassical
states composed of infinitely many particles. Namely, at strong
coupling the rapidities of two-particle states fall inside the
interval $[-2,2]$, while those of semi-classical states are always
outside this interval. Thus, the modification of the TBA equations
discussed in this paper might not be necessary for semi-classical
states. It would be important to better understand this issue.

\smallskip

The paper is organized as follows. In the next section we explain
our criteria for a choice of the integration contour in the
excited states TBA equations. In section 3 we discuss two-particle
states in the $\sl(2)$ sector and the corresponding asymptotic
Y-functions. By analysing analytic properties of the Y-functions,
we determine the critical and subcritical values of the coupling
constant both for the Konishi and for some other states. In
section 4 we present the simplified TBA
equations for Konishi-like states and we explain why and how their
form depends on the value of the coupling constant. In section 5
we generalize this discussion to arbitrary two-particle states
from the $\sl(2)$ sector.  In section 6 we summarize the most
essential properties of the AdS/CFT Y-system implied by the TBA equations
under study. Finally, in Conclusions we mention some interesting
open problems. The definitions,  treatment of canonical equations, and further technical details are relegated
to eight appendices.

%%%%%%%%%%%%%%%%%%%%%%%%%%%%%%%%%%%%%%

%%%%%%%%%%%%%%%%%%%%%%%%%%%%%%%%%%
\section{Contour deformation trick}

The TBA equations for the $\AdS$ mirror model are written for Y-functions which depend on the real momentum of the mirror model. The energy of string excited states  obviously depends on real momenta of string theory particles, and to formulate the TBA equations for
excited states one also needs to  continue analytically the Y-functions to the
string theory kinematic region. To visualize the analytic continuation it is convenient to use the $z_Q$-tori because
the kinematic regions of the mirror and string theory Q-particle bound states (Q-particles for short) are subregions of the $z$-torus, see Figure 1.
In addition,  the Q-particle energy, and many of the kernels
appearing in the set of TBA equations are meromorphic functions  on
the corresponding torus. The mirror Q-particle region can be mapped onto a $u$-plane with the cuts running from the points $\pm2\pm{i\ov g}Q$ to $\pm\infty$, and
the string Q-particle region can be mapped onto a $u_*$-plane with the cuts
connecting the points $-2\pm{i\ov g}Q$ and $2\pm{i\ov g}Q$, see Figure 1. Since the cut structure on the planes is different for each Y-function, they cannot be considered as different sheets of a Riemann surface. The $z_Q$-torus can be glued  either from four  mirror $u$-planes or four string $u_*$-planes.

%\vspace{-6cm}

\begin{figure}[t]
\begin{center}
\includegraphics*[width=0.8\textwidth]{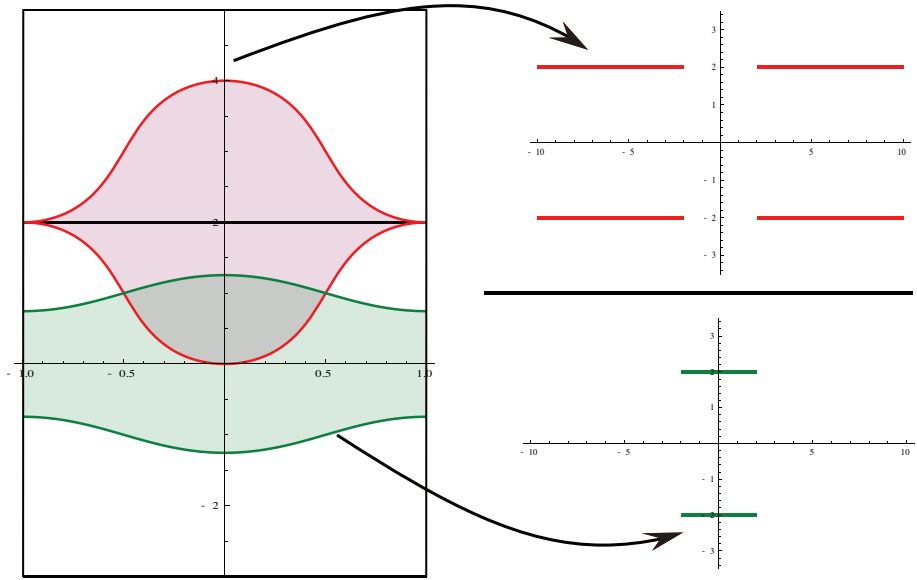}
\end{center}
\caption{These are the mirror and string regions on the $z$-torus. They are in one-to-one correspondence with the $u$-planes. The boundaries of the regions are mapped to the cuts.}
\end{figure}

As was shown in \cite{AF07}, $z$-torus variables corresponding to real momenta of the mirror and string theory are related to each other by the shift by a quarter of the imaginary period of the torus
\bea
z = z_*  + {\om_2\ov 2} \,,
\eea
where $z$ is the variable parametrizing the real momenta of the mirror theory, and $z_*$ is the variable parametrizing the real momenta of the string theory.
The line $(-\infty,\infty)$ in the string $u_*$-plane is mapped to the interval  Re$(z_*)\in (-{\om_1\ov 2},{\om_1\ov 2})$, Im$(z_*)=$const on the $z$-torus,  and we choose $z_*$ in the string region to be real. Then,
 the interval Re$(z)\in (-{\om_1\ov 2},{\om_1\ov 2})$, Im$(z)={\om_2\ov 2i}$ of the mirror region is mapped onto the real line of the mirror $u$-plane.

It is argued in  \cite{DT96, DT97} that the TBA equations for
excited states can be obtained from the ones for the ground state
by analytically continuing in the coupling constants and picking up the singularity of proper convolution terms. We prefer however to employ a slightly different procedure which we refer to as the contour deformation trick. We believe it is equivalent to \cite{DT96, DT97}.  It is based on the following assumptions
\begin{itemize}
\item  The form of TBA equations for any excited state and the expression for the energy  are universal.
TBA equations for excited states differ from each other only by a choice of
 integration contours  of convolution terms  and the length parameter $L$ which depend on a state.

\item The choice of the integration contours and $L$ is fixed by requiring that the large $L$ solution of the excited state TBA equations be given by the generalized L\"uscher formulae, that is all the Y-functions can be written in terms of the eigenvalues of the transfer matrices. The integration contour depends on the excited state under consideration, and  in general on the values of  't Hooft's coupling and $L$.

\item An excited state is completely characterized by the five charges it carries and a set of $N$ real numbers $p_k$  which are in one-to-one correspondence with momenta $p_k^o$ of $N$ Q-particles in the small coupling limit $g\to 0$.  The momenta $p_k^o$ are found by using the one-loop Bethe equations for fundamental particles and their bound states.
For finite values of $g$ the set of $p_k$ is determined by exact Bethe equations which state that  at any $p=p_k$ the corresponding $Y_Q$-functions are equal to $-1$.

\end{itemize}

We consider only the simplest case of two-particle excited states in the $\sl(2)$ sector of the string theory because there are no bound states  in this sector and the complete two-particle spectrum can be readily classified.
The physical states satisfy the level-matching condition which for
two-particle states takes a very simple form: $p_1 = -p_2$, or $u_1 = -u_2$ or $z_{*1}=-z_{*2}$\,, depending on the coordinates employed. The TBA equations we propose in next sections are valid only for physical states.

%%%%%%%%%%%%%%%%%%%%%%%%%%%%%%%%%
\section{States and Y-functions in the $\sl(2)$ sector}

To fix the integration contour one should choose a state and analyze the analytic structure of the large $L$ Y-functions which we refer to as the asymptotic Y-functions. We begin with a short discussion of two-particle states in  the $\sl(2)$ sector.

%%%%%%%%%%%%%%%%%%%%%%%%%%%%%%%%%
\subsection{Bethe-Yang equations for the $\sl(2)$ sector}

There is only a single Bethe-Yang (BY) equation in the $\sl(2)$-sector  for two-particle physical configurations satisfying the vanishing total momentum condition $p_1 +p_2=0$ that can be written in the form \cite{St04}
\begin{eqnarray}\la{BYeJ1}
1=e^{ip J}S_{\sl(2)}(p,-p)\  \Longrightarrow\
e^{ip (J+1)}=\frac{1+ {1\ov x_s^+{}^2}}{1+{1\ov x_s^-{}^2}}\sigma(p,-p)^2
\, ,
\end{eqnarray}
where $p\equiv p_1 >0$, $J$ is the charge carried by the state, $\s$ is the dressing factor, and $x_s^\pm$ are defined in appendix \ref{app:rapidity}.
Taking the logarithm  of the equation, one gets
\bea\la{BYe}
i p(J+1) - \log \frac{1+{1\ov x_s^+{}^2}}{1+{1\ov x_s^-{}^2}} - 2i \,\theta(p,-p) =2\pi i\, n\,,
\eea
where $\theta = {1\ov i}\log\s$ is the dressing phase, and $n$ is a positive integer because we have assumed $p$ to be positive.  As was shown in  \cite{AFS}, at large values of $g$ the integer $n$ is equal to the string level of the state.

As is well known, in the small $g$ limit the equation has the obvious solution
\bea\la{BYeJsol0}
p^o_{J,n} ={2\pi n\ov J+1}\,,\quad n=1\,,\ldots\,, \left[{J+1\ov 2}\right]\,,
\eea
where $[x]$ denotes the integer part of $x$, and the range of $n$ is bounded because
the momentum $p$ can only take values from $0$ to $\pi$. The corresponding  rapidity variable $u_{J,n}$ in the small $g$ limit takes the following form
\bea\la{uJmo}
u_{J,n} \to {1\ov g} u^o_{J,n}\,,\quad u^o_{J,n}=\cot{\pi n\ov J+1}\,.
\eea
Thus, any two-particle state in the $\sl(2)$ sector is completely characterized by the two integers $J$ and $n$.
In particular, in the simplest $J=2$ case corresponding to a descendent of the Konishi state $n$ can take only one value $n=1$, and the small $g$ solution is
\bea\la{BYeJsol02}
p^o_{2,1} ={2\pi\ov 3}\,,\quad u^o_{2,1}={1 \ov \sqrt3}\,.
\eea
The BY equation \eqref{BYe} can be easily solved perturbatively up to any desired order in $g$, and numerically up to very large values of $g$.
We have used the BES series representation \cite{BES} for the dressing phase for perturbative computations, and the DHM integral representation \cite{DHM} for the numerical ones\footnote{The DHM representation can be also readily used for perturbative computations.}.
For the Konishi state, the perturbative solution up to  $g^{16}$-th can be found in appendix \ref{appBY}.

 We have solved numerically the BY equation for the Konishi state for ${1\ov 10}\le g\le 1000$ with the step ${1\ov 10}$ for ${1\ov 10}\le g\le 10$, the step $1$ for $10\le g\le 100$, and the step $10$ for $100\le g\le 1000$. In Figure 2 we show the results up to $g=100$. For greater values of $g$ nothing interesting happens, and  the solution can be approximated by asymptotic formulae from \cite{AFS,AF05,RS09}, see appendix \ref{appBY} for more details.
\begin{figure}[t]
\begin{center}
\includegraphics[width=.48\textwidth]{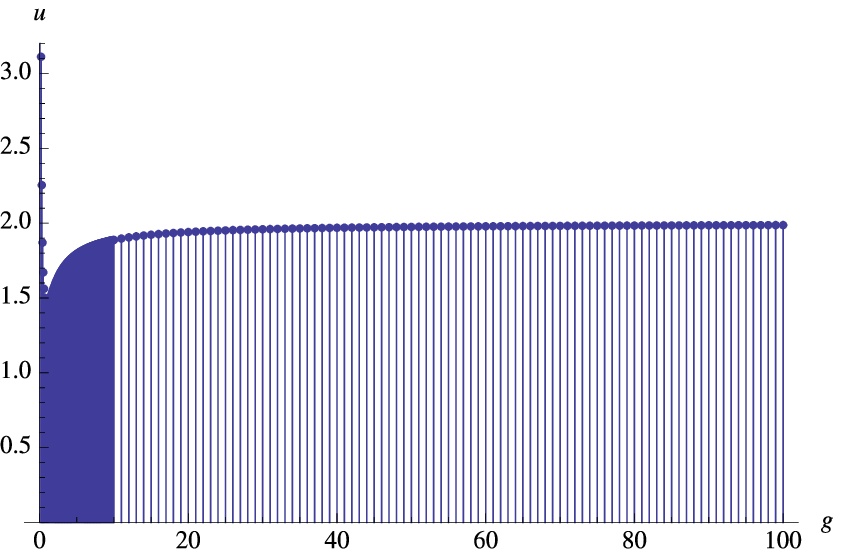}\quad \includegraphics[width=.48\textwidth]{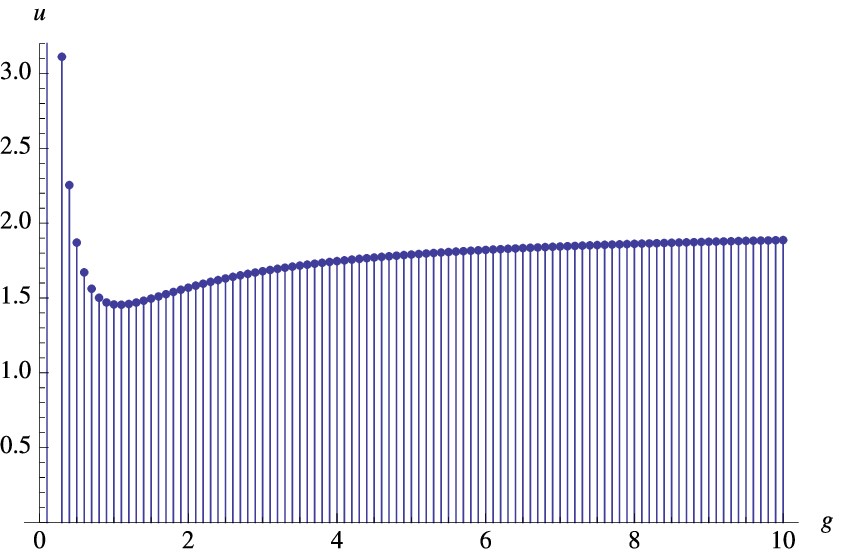}
\end{center}
\caption{These are the plots of $u$ which solves the BY equation for the Konishi state.  }
\end{figure}
We then applied the Interpolation function in Mathematica to have $u_{2,1}$ as  a smooth function of $g$. Using the function, one can find that $u_{2,1}(g)$ decreases up to $g\sim 0.971623$, and then begins to increase and at large $g$ it asymptotes to $u=2$.

The functions $u_{J,n}(g)$ for other values of $J$ and $n$ have similar $g$ dependence. The only exception is the case $J=2n-1$ where the exact solution of the BY equation  is $p_{2n-1,n}=\pi$ and, therefore, $u_{2n-1,n}=0$. The TBA equations we propose below are not in fact valid for these $(2n-1,n)$ states.

The perturbative and numerical solutions for $u_{J,n}(g)$ can be easily used to analyze the behavior of Y-functions
considered as functions of $g$. In particular it is easy to determine if some of them become negative for large enough values of $g$.

\subsection{Y-functions in the $\sl(2)$ sector}

Let us recall that the TBA equations for the $\AdS$ mirror model involve
$Y_Q$-functions for momentum carrying $Q$-particle bound states, and auxiliary functions $Y_{Q|vw}^{(\a)}$ for $Q|vw$-strings, $Y_{Q|w}^{(\a)}$ for $Q|w$-strings, and $Y_{\pm}^{(\a)}$ for $y_\pm$-particles. The index $\a=1,2$ reflects two $\su(2|2)$ algebras in the symmetry algebra of the light-cone string sigma model.
The TBA equations \cite{AF09b} depend also on the parameters $h_\a$ which take care of the periodicity condition of  the fermions of  the model \cite{AFrev}.
For the $\sl(2)$ sector the fermions are periodic, and from the very beginning
 one can set the parameters $h_\a$ to 0 because there is no singularity at $h_\a=0$ in the excited states TBA equations.

For the $\sl(2)$ states there is a symmetry between the left and right $\su(2|2)$ auxiliary roots, and, therefore, all Y-functions satisfy the condition
\bea
Y^{(1)}_{\forall} = Y^{(2)}_{\forall}   = Y_{\forall} \,,
\eea
where $\forall$ denotes a Y-function of any kind.

The string theory spectrum in the $\sl(2)$ sector is characterized by a set of $N$ real numbers $z_{*k}$ or $u_k$  corresponding to momenta of $N$ fundamental particles in the limit $g\to 0$.  According to the discussion above,  these numbers are determined from the exact Bethe equations
\bea\la{Y1m1}
Y_1(z_{*k}) = Y_{1_*}(u_{k}) = -1\,, \quad k=1,\ldots, N\,,
\eea
where $Y_1(z)$ is the Y-function of  fundamental mirror particles considered as a function on the  $z$-torus,  and  $Y_{1_*}(u)$ denotes the $Y_1$-function  analytically continued to the string $u$-plane.

In the large $J$ limit the exact Bethe equations must reduce to the BY equations, and it is indeed so because  the asymptotic $\sl(2)$ $Y_Q$-functions can be written in terms of the transfer matrices  defined in appendix \ref{appT} as follows \cite{BJ08}
\bea\la{YQasympt}
 Y_Q^o(v)=e^{-J\widetilde{\cal E}_Q(v) }{T_{Q,1}(v|\vec{u})^2\ov  \prod_{i=1}^{N} S_{\sl(2)}^{1_*Q}(u_i,v)}=e^{-J\widetilde{\cal E}_Q(v)}T_{Q,1}(v|\vec{u})^2\,  \prod_{i=1}^{N} S_{\sl(2)}^{Q1_*}(v,u_i)\,, ~~~\eea
where $v$ is the rapidity variable of the mirror $u$-plane and $\widetilde{\cal E}_Q$ is the energy of a mirror $Q$-particle. $S_{\sl(2)}^{1_*Q}$ denotes the S-matrix with the first and second arguments in the string and mirror regions, respectively.
$T_{Q,1}(v|\vec{u})$ is up to a factor the trace of the S-matrix describing the scattering on these string theory particles with a mirror Q-particle or in other words the eigenvalue of the corresponding transfer matrix. The BY equations then follow from the fact that
$\widetilde{\cal E}_{1_*}(u_k) = -ip_k$, and the following normalization of $T_{1,1}$
\bea
T_{1,1}(u_{_*k}|\vec{u}) = 1 \ \Longrightarrow\  -1=e^{iJp_k}\prod_{i=1}^{N} S_{\sl(2)}^{1_*1_*}(u_k,u_i)\,,
\eea
where $u_{_*k}=u_{k}$, and the star just indicates that one analytically continues $T_{1,1}$ to the string region. Then,  $S_{\sl(2)}^{1_*1_*}(u_k,u_i) = S_{\sl(2)}(u_k,u_i)$ is the usual $\sl(2)$ sector S-matrix used in the previous subsection.

Let us also mention that
$T_{Q,1}$ has the following large $v$ asymptotics
\bea
T_{Q,1}(v\,|\,\vec{u}) \to Q\left( 1 - \prod_{i=1}^{N} \sqrt{x_i^+\ov x_i^-} \ \right)^2\,,\quad v\to\infty\,,
\eea
and therefore it goes to 0 if the level-matching is satisfied.

Then, in the large $J$ limit all auxiliary asymptotic $\sl(2)$ Y-functions can be written in terms of the transfer matrices as follows \cite{GKV09}
\bea\nonumber
Y_-^{o}&=&-\frac{T_{2,1}}{T_{1,2}}\,,\quad
Y_+^{o}=-\frac{T_{2,3}T_{2,1}}{T_{3,2}T_{1,2}}\,,\quad
Y_{Q|vw}^{o}=\frac{T_{Q+2,1}T_{Q,1}}{T_{Q+1,2}}\,,\quad
Y_{Q|w}^{o}=\frac{T_{1,Q+2}T_{1,Q}}{T_{2,Q+1}T_{0,Q+1}}\,.
\eea
The transfer matrices $T_{a,s}$ can be computed in terms of $T_{a,1}$ by using the Bazhanov-Reshetikhin  formula \cite{BR}, see appendix \ref{appT} for all the necessary explicit formulae.

An important property of the Y-functions for $vw$- and $w$-strings is that they approach their vacuum values as $v\to\infty$
\bea\nonumber
Y_{M|vw}(v)\to M(M+2)\,,\quad Y_{M|w}(v)\to M(M+2)\,,\quad v\to\infty\,,\ \ -{M\ov g}<{\rm Im}(v)<{M\ov g}\,.
\eea

Now we are ready to analyze the dependence of asymptotic $Y$-functions on $g$.
Recall that they depend on the rapidities $u_k$ which are solutions of the BY equations.

\smallskip

We begin with the small $g$ limit where the effective length goes to infinity, and one can in fact trust all the asymptotic formulae. It is convenient to rescale $v$ and $u_k$ variables as $v\to v/g$, $u_k\to u_k/g$ because the rescaled variables are finite in this limit. Let $\kappa\equiv u_1=-u_2$ be the rescaled rapidity of a fundamental particle.
According to the previous subsection, in the small $g$ limit they are given by $u_{J,n}^o$, eq.\eqref{uJmo}.

\smallskip

The most important functions in the $\sl(2)$ case are $Y_{M|vw}$, and we find
that for $N=2$ they exhibit the following small $g$
behavior in the strip $-M<{\rm Im}(v)<M$
{\small \bea
Y_{M|vw}(v)=M(M+2)\frac{\big[M^2-1+v^2-\kappa^2\big]\big[(M+2)^2-1+v^2-\kappa^2\big]}
{\big[(M+1)^2+(v-\kappa)^2\big]\big[(M+1)^2+(v+\kappa)^2\big]}+{\cal
O}(g^2)\, .  \eea }
The leading term has the correct large $u$-asymptotics and
four apparent  zeros at
$$
v=\pm\sqrt{\kappa^2-M^2+1}\, , ~~~~ v=\pm\sqrt{\kappa^2-(M+2)^2+1}\, .
$$
One can see that $Y_{1|vw}$-function always has at least two real  zeros at $v=\pm\ka$.
Other  zeros of $Y_{M|vw}$-functions can be either real or purely imaginary depending on the values of $M$ and $\ka$. It appears that the form of simplified TBA equations depends on the imaginary part of these  zeros, and we will see in next sections that if a pair of  zeros
$v=\pm r$ fall in the strip $|{\rm Im}(r)|<1$ then the equations should be modified.

\smallskip

Thus, we are lead to consider the following three possibilities
\begin{enumerate}
\item If $M^2-2< \kappa^2<(M+2)^2-2$ then $Y_{M|vw}$ has two  zeros
at $v=\pm\sqrt{\kappa^2-M^2+1}$ that are in the strip $|{\rm Im}(v)|<1$.  In terms of the integers $J$ and $m$ characterizing two-particle states one gets the condition
\bea\la{cond1}
\sqrt{M^2-2}<\cot{\pi n\ov J+1}<\sqrt{(M+2)^2-2}\,.
\eea
\item If $
\kappa^2<M^2-2 \ \Longleftrightarrow\ \cot{\pi n\ov J+1}<\sqrt{M^2-2} $
then
$Y_{M|vw}$ does not have any   zeros in the strip $|{\rm Im}(v)|<1$.
\item If $
\kappa^2>(M+2)^2-2 \ \Longleftrightarrow\ \cot{\pi n\ov J+1}>\sqrt{(M+2)^2-2}
$
then
$Y_{M|vw}$ has four   zeros  in the strip $|{\rm Im}(v)|<1$.
\end{enumerate}
Some of these  zeros can be real, and in fact the canonical TBA equations take different forms depending on whether the roots are real or imaginary.

\smallskip

Classification of two-particle states at $g\sim 0$ is presented
in Table 1. The type of a state is determined by how many zeroes of
$Y_{M|vw}$-functions occur in the physical strip and it depends on
$J$ and $n$.

\smallskip

Consider a two-particle state with $\kappa = u_{J,n}^o$ for some
$(J,n)$. Table 1 shows that there exists a number $m \ge 1$, equal
to the maximal value of $M$ the condition \eqref{cond1} is
satisfied. Then both $Y_{m|vw}$ and $Y_{m-1|vw}$ have two  zeros,
all $Y_{k|vw}$ with $k\le m-2$ have four zeros, and all
$Y_{k|vw}$-functions with $k\ge m+1$ have no zeros
 in the strip $|{\rm Im}(v)|<1$. For example,
among the states with $(J,n=1)$ at small coupling, the states of
type I are found if and only if $J\le 4$. The type II is found for
$5\le J\le 7$, type III for $8\le J\le 11$, and type IV for $12
\le J\le 14$. In particular, $Y_{1|vw}$ for the state $(8,1)$ has
two real  zeros and two imaginary  zeros in the strip $|{\rm
Im}(v)|<1$, and $Y_{1|vw}$ for the state $(J\ge9,1)$ has four real
zeros.

\smallskip

As for the Konishi state with $J=2$ and $n=1$ only
$Y_{1|vw}$-function has two  zeros and all the other
$Y_{M|vw}$-functions have no  zeros  at small coupling.
Let us also mention that at $g=0$ the
$Y_{2|vw}$-function of the state $(5,1)$ (and in general of any
state $(6k-1,k)$) has a double zero at $v=0$. This double zero
however is an artifact of the  perturbative expansion, and in
reality $Y_{2|vw}$ has two imaginary  zeros for small values of
$g$ equal to $\approx \pm ig\sqrt3$.
For the state with $J=6$ and $n=1$ both $Y_{1|vw}$ and $Y_{2|vw}$ have
two real zeros.

\subsection{Critical values of $g$}

\subsubsection*{Evolution of zeros}

Now we would like to understand what happens with
$Y_{M|vw}$-functions  when one starts increasing $g$. To this end
one should use numerical solutions of the BY equations discussed
at the beginning of this section. We also switch back to the
original $u$ variables because they are more convenient for
general values of $g$, and refer to the strip $|$Im$(u)|<1/g$ as
the physical one.

We find that for finite $g$ any $Y_{k|vw}$-function has four
zeros which are either real or purely imaginary. We could not find
any other complex  zeros. The four  zeros of $Y_{k|vw}$ are split
into two pairs, and the two  zeros in a pair have opposite signs,
and are either real or complex conjugate to each other.
We denote the four zeroes of $Y_{k|vw}$ by $r_j^{(k)}$ and $\hat{r}_j^{(k)}$, where $j=1,2$ and $k=1\,,2\,, \ldots\,$.
If the four zeros are real then
$r_j^{(k)}$ are the zeros of $Y_{k|vw}$ which have a larger absolute value than $\hat{r}_j^{(k)}$.
If only two  zeros are real then we denote them as $r_j^{(k)}$ and
the imaginary zeros as $\hat{r}_j^{(k)}$.  Finally, if the four
zeros are imaginary then
$r_j^{(k)}$ are the ones closer to the real line than the second pair $\hat{r}_j^{(k)}$.

The locations of the  zeros depend on $g$, and we should distinguish two cases.
We observe first that if two zeros are real at $g\sim 0$  then they are of order $1/g$, and obviously outside the interval $[-2,2]$. With $g$ increasing they starts moving toward the origin, and at some value of $g$ they reach their closest position to the origin which is inside the interval $[-2,2]$. Then, for larger $g$ they remain inside the interval but begin to move to its boundaries and reach them at $g=\infty$.  In the second case, one considers  a pair of imaginary zeros at $g\sim 0$. With $g$ increasing they start moving toward the real line, and at some value of $g$ they get to the origin
and become a double zero. Then, for larger $g$ they split and begin to move to the boundaries of the interval $[-2,2]$, and reach them at $g=\infty$. The only exception from this behavior we find is the $g$-dependence of the two  zeros of $Y_{2|vw}$-function for the states $(6k-1,k)$ that are equal to $\pm i\sqrt3$ at small $g$. These  zeros become real at very small value of $g$. Then they start moving to $\pm2$, cross the boundaries of the interval $[-2,2]$, and reach their maximum. After that they behave as  zeros of all the other $Y_{k|vw}$-functions.
Thus, at very large values of $g$ all  zeros of any $Y_{k|vw}$-function are real, inside  the interval $[-2,2]$, and very close to $\pm2$.

 The pairs of the  zeros of different $Y_{M|vw}$-functions are not independent, and satisfy the following relations
\bea
\hat{r}_j^{(k-1)} = r_j^{(k+1)}\,,\quad k =2\,,\ldots\,, \infty\,.
\eea
Therefore, the  zeros of $Y_{M|vw}$-functions can be written as follows
\bea\la{zer}
\{ r_j^{(1)}\,,r_j^{(3)} \} \,; \{ r_j^{(2)}\,,r_j^{(4)} \}
\,;\ldots\,; \{ r_j^{(k-1)}\,,r_j^{(k+1)} \}\,;
\{ r_j^{(k)} \,, r_j^{(k+2)} \} \,; \{ r_j^{(k+1)}\,,r_j^{(k+3)} \} \,;\ldots \,,~~~~
\eea
so that $Y_{k|vw}$ has the  zeros $\{ r_j^{(k)} \,, r_j^{(k+2)} \}$.

These  zeros have a natural ordering. If we assume for definiteness that the  zeros with $j=1$ have negative
real or imaginary parts, then they are ordered as
\bea
r_1^{(1)} \prec r_1^{(2)} \prec r_1^{(3)} \prec \cdots  \prec r_1^{(k)} \prec r_1^{(k+1)} \prec \cdots\,,
\eea
where $r_1^{(k)} \prec r_1^{(k+1)}$ if either Re$(r_1^{(k)}) < \ $Re$(r_1^{(k+1)})$ or
Im$(r_1^{(k)}) > \ $Im$(r_1^{(k+1)})$.
It is important that
the  zeros never change the ordering they have at $g\sim 0$.
In particular, $Y_{1|vw}$ always has two real  zeros $r_j^{(1)}= u_j$ which are Bethe roots.  They are the largest (in magnitude)  zeros among all $Y_{k|vw}$-functions, and are the closest ones to $\pm 2$ at large $g$.

\begin{center}
{\small
 {\renewcommand{\arraystretch}{1.5}
\renewcommand{\tabcolsep}{0.2cm}
\begin{tabular}{|c|l|l|}
\hline Initial condition $\rightarrow$ & $Y_{1|vw}$, $Y_{2|vw}$ & 2+2 \\
  & $Y_{1|vw}$, $Y_{2|vw}$, $Y_{3|vw}$ & 4+2+2 \\
$g$ ~$\downarrow$ & $Y_{1|vw}$,$Y_{2|vw}$, $Y_{3|vw}$,
$Y_{4|vw}$  & 4+4+2+2 \\
& $Y_{1|vw}$, $Y_{2|vw}$, $Y_{3|vw}$,
$Y_{4|vw}$,  $Y_{5|vw}$  & 4+4+4+2+2 \\
 \vdots & \vdots & \vdots \\
  & $Y_{1|vw}$,
$Y_{2|vw}$,\quad\ldots & 4+4+\quad\ldots
\\ \hline
\end{tabular}}
}

\vspace{0.5cm}
\parbox{13cm}{\small Table 2. Evolution of a two-particle states in the $\sl(2)$-sector
with respect to $g$. At $g\sim 0$ a state has a certain number of
$Y_{M|vw}$-functions with zeroes in the physical strip. Increasing
the coupling, the critical points get crossed which leads to
accumulation of zeroes of $Y_{M|vw}$'s in the physical strip. This
phenomenon can be called ``Y-function democracy".  }
\end{center}

In addition we find that the functions below  have either  zeros or equal to $-1$ at locations related to $r_j^{(k)}$
\bea\la{zerorel}
Y_{k|vw}\big(r_j^{(k+1)}\pm{i\ov g}\big) = -1\,,\quad Y_{k+1}\big(r_j^{(k+1)}\big) = 0\,,\quad k=1\,,\ldots\,, \infty\,,\quad Y_{\pm}\big(r_j^{(2)}\big) = 0\,.~~~~~~
\eea
As will be discussed in the next section the equations $Y_{k|vw}\big(r_j^{(k+1)}\pm{i\ov g}\big) = -1$ lead to integral equations which play the same role as the exact Bethe equations $Y_1(u_j)=-1$ and allow one to find the exact location of the roots $r_j^{(k+1)}$.

Let us finally mention that nothing special happens to $Y_{M|w}$-functions.

\subsubsection*{Critical and subcritical values}

Let $m$ again be the maximum value of $M$ the condition \eqref{cond1} is satisfied.
According to the discussion above for any two-particle state there is a critical value of $g$ such that the function
$Y_{m+1|vw}$ which had no  zeros in the physical strip for small values of $g$,
acquires two   zeros at $u=\pm  i/g$.
At the same time  $Y_{m-1|vw}$  also acquires   zeros at $u=\pm i/g$.
At a slightly larger value of $g$  the two  zeros that were at $u=\pm i/g$ collide at the origin, and $Y_{m+1|vw}$ and $Y_{m-1|vw}$ acquire  double  zeros at $u=0$. Then, the double  zeros split, and both $Y_{m|vw}$ and $Y_{m+1|vw}$ have two real  zeros, and $Y_{m-1|vw}$ has four.
 Increasing $g$ more, one reaches the second critical value of $g$ such
 that the functions $Y_{m+2|vw}$ and $Y_{m|vw}$  acquire   zeros at $u=\pm
 i/g$, see Table 2.

\smallskip

 This pattern repeats itself, and there are infinitely many  critical values of $g$ which we  denote as $g_{J,n}^{r,m}$ and define as the smallest value of $g$ such that for a symmetric configuration of Bethe roots the function $Y_{m+r|vw}$ acquires two  zeros at $u=\pm i/g$. The subscript $J,n$ denotes a state in the $\sl(2)$ sector, and they determine $m$.

The critical values of $g$ can be also determined from the requirement that at $g=g_{J,n}^{r,m}$ the function $1+Y_{m+r-1|vw}$ has a double zero at $u=0$: $1+Y_{m+r-1|vw}(0,g_{J,n}^{r,m})=0$. This condition is particularly useful because the value of the Y-functions at $u=0$ can be found from the TBA equations, see next section for detail.

 The second set of subcritical values of $g$  can be defined as the smallest value of $g$ such
  that the function $Y_{m+r|vw}$ acquires a double zero at $u=0$. They are denoted as $\bar{g}_{J,n}^{r,m}$, and they are always greater than the corresponding critical values: $g_{J,n}^{r,m} <\bar{g}_{J,n}^{r,m}$.

\smallskip

The locations of the critical values depend on the state under consideration, and can be determined  approximately by using the asymptotic $Y$-functions discussed in the previous subsection. The values obtained this way are only approximate because for large enough values of $g$ one should take into account the deviations of the Y-functions from their large $J$ expressions.

\smallskip

We will see in next sections that the critical values $g_{J,n}^{r,m}$ play a crucial role in formulating excited states simplified TBA equations which take different form in each of the intervals
$g_{J,n}^{r,m}<g<g_{J,n}^{r+1,m}$, $r=0,1,\ldots $ where $g_{J,n}^{0,m}=0$.
The second set of $\bar{g}_{J,n}^{r,m}$  is not important for the simplified equations. The canonical TBA equations however require both sets because they take different form in each of the intervals
$g_{J,n}^{r,m}<g<\bar{g}_{J,n}^{r,m}\,$; $\ \bar{g}_{J,n}^{r,m}<g<g_{J,n}^{r+1,m}$, $\ r=0,1,\ldots$.

\smallskip

Strictly speaking the integration contour in TBA equations also
depends on $g$ and the state under consideration. Nevertheless it
appears that in simplified TBA equations the  contour can be chosen to be
the same for all values of $g$, and even for all two-particle
states from the $\sl(2)$ sector if one allows its dynamical
deformation. That means that with increasing $g$ the contour should be
deformed in such a way that it would not hit any singularity.
This also shows that one should not expect any kind of
non-analyticity  in the  energy of a state at a critical value of
$g$. What may happen is that the critical values are the
inflection points of the energy.  %as a function of $g$.

\subsubsection*{Critical values of $g$ for the Konishi state}

In this subsection we discuss in detail the critical values for Konishi state.
To analyze the dependence of Y-functions on $g$ one should first solve the BY equations with $J=2$, $n=1$, and then plug the $u_j$'s obtained into the expressions for $Y$-functions from appendix \ref{appT}.

\begin{figure}[t]
\begin{center}
\includegraphics[width=.45\textwidth]{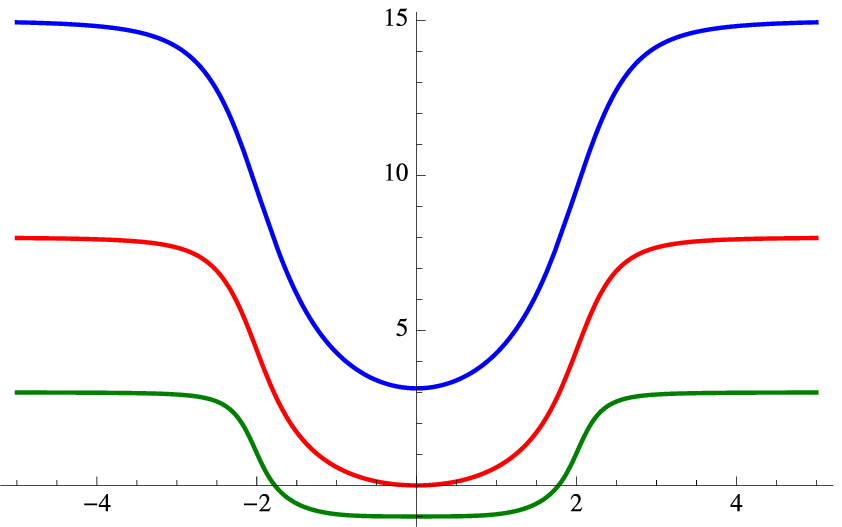}\qquad\includegraphics[width=.45\textwidth]{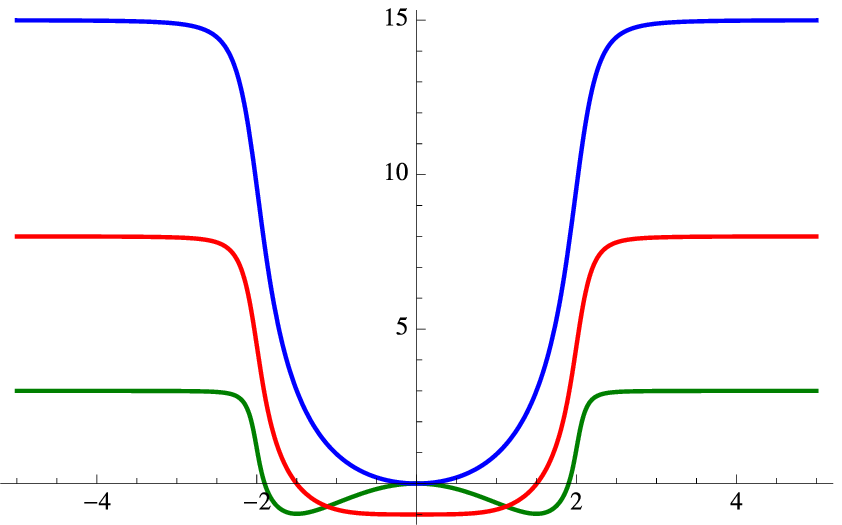}
\end{center}
\caption{On the left and right pictures  $Y_{1|vw}$, $Y_{2|vw}$ and $Y_{3|vw}$ are plotted for the Konishi state at  $\bar g_{cr}^{(1)}\approx 4.5$ and  $\bar g_{cr}^{(2)}\approx 11.5$, respectively. $Y_{2|vw}$ touches the $u$-axis at $g=\bar g_{cr}^{(1)}$, and has two real  zeros for  $\bar g_{cr}^{(1)}<g<\bar g_{cr}^{(2)}$. }
\end{figure}

Solving the equations
\bea
Y_{r+1|vw}(\pm{i\ov g},g)=0\,,\quad r=1,2,\ldots \,,
\eea
we find that there are 7 critical values of $g$ for $g<100$
\bea
g_{2,1}^{(r,1)}  = \{4.429, 11.512, 21.632,
  34.857, 51.204, 70.680,
  93.290\}\,.
\eea
Note that the distance between the critical values increases with $g$.
The first critical value is distinguished because only $Y_{2|vw}(\pm i/g,g)$ vanishes there.
For all the other critical values the function $Y_{r-1|vw}(\pm i/g,g)$ also is equal to zero
\bea
Y_{r+1|vw}(\pm {i\ov g},g_{2,1}^{(r,1)})=0\ \Longrightarrow\  Y_{r-1|vw}(\pm {i\ov g},g_{2,1}^{(r,1)})=0\,,\quad  {\rm for}\ \ r=2,3,\ldots \,.~~~~~
\eea

Then, solving the equations
\bea
Y_{r+1|vw}(0,g)=0\,,\quad r=1,2,\ldots \,,
\eea
one finds the following 7 subcritical values of $g$ for $g<100$
\bea
\bar{g}_{2,1}^{(r,1)}  = \{4.495, 11.536, 21.644,
  34.864, 51.209, 70.684,
  93.292\}\,.
\eea
Note that the distance between a critical value and a corresponding subcritical one decreases with $g$. Again,
at the first subcritical value  only $Y_{2|vw}(0,g)$ vanishes.
For all the other subcritical values the function $Y_{r-1|vw}(0,g)$ also acquires an extra double zero
\bea
Y_{r+1|vw}(0,\bar{g}_{2,1}^{(r,1)})=0\ \Longrightarrow\  Y_{r-1|vw}(0,\bar{g}_{2,1}^{(r,1)})=0\,,\quad  {\rm for}\ \ r=2,3,\ldots \,.
\eea
Once $g$ crosses a subcritical value $\bar{g}_{2,1}^{(r,1)} $ the corresponding double  zeros at $u=0$ split, and each of the functions $Y_{r-1|vw}(0,g)$ and $Y_{r+1|vw}(0,g)$ acquires two symmetrically located  zeros.
As a result at infinite $g$ all the $Y_{M|vw}$-functions have four real  zeros. Moreover, one can also see that if $g$ is between two subcritical values then all these  zeros for all the functions are inside the interval $[-2,2]$ and  approach $\pm2$ as $g\to\infty$.
\begin{figure}[t]
\begin{center}
\includegraphics[width=.45\textwidth]{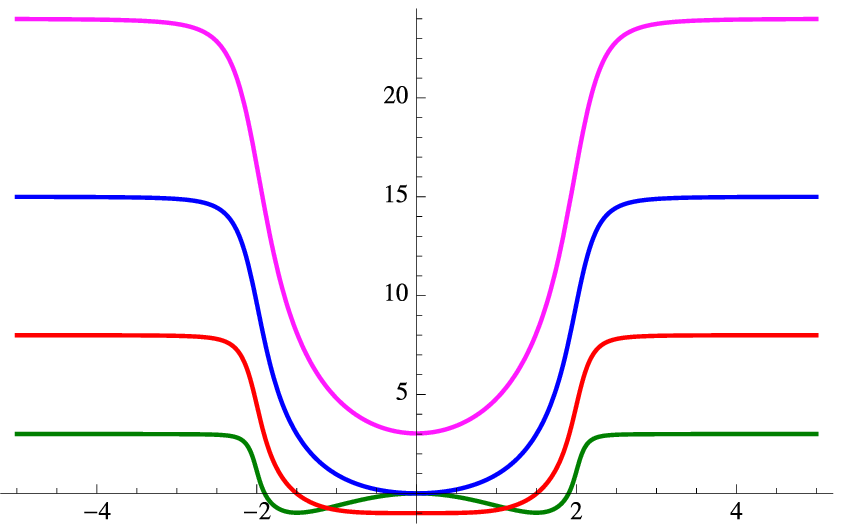}\qquad\includegraphics[width=.45\textwidth]{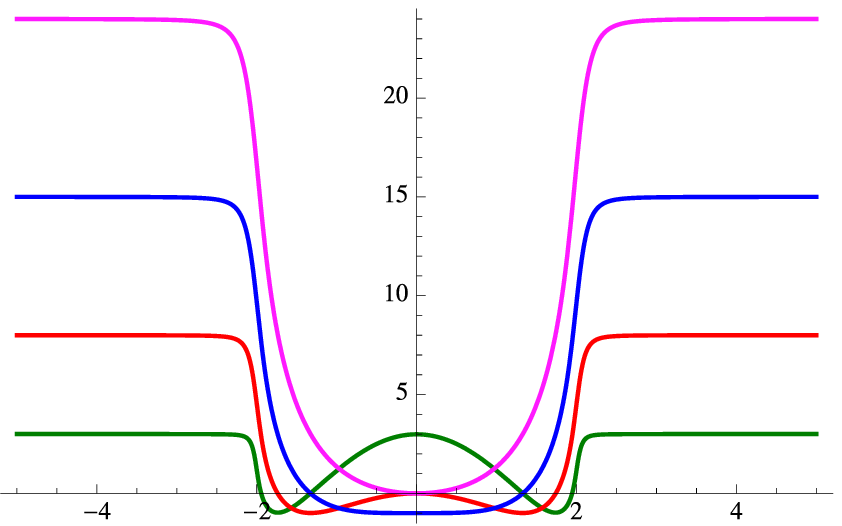}
\end{center}
\caption{On the left  and right pictures $Y_{1|vw}$, $Y_{2|vw}$, $Y_{3|vw}$  and $Y_{4|vw}$ are plotted  for the Konishi state at  $\bar g_{cr}^{(2)}\approx 11.5$ and $\bar g_{cr}^{(3)}\approx 21.6$, respectively.
$Y_{1|vw}$ and $Y_{3|vw}$ touch the $u$-axis at $g=\bar g_{cr}^{(2)}$, and $Y_{2|vw}$ and $Y_{4|vw}$
touch it at $g=\bar g_{cr}^{(3)}$. $Y_{1|vw}$
 has four real  zeros for  $g>\bar g_{cr}^{(2)}$.  }
\end{figure}
\begin{figure}[t]
\begin{center}
\includegraphics[width=.45\textwidth]{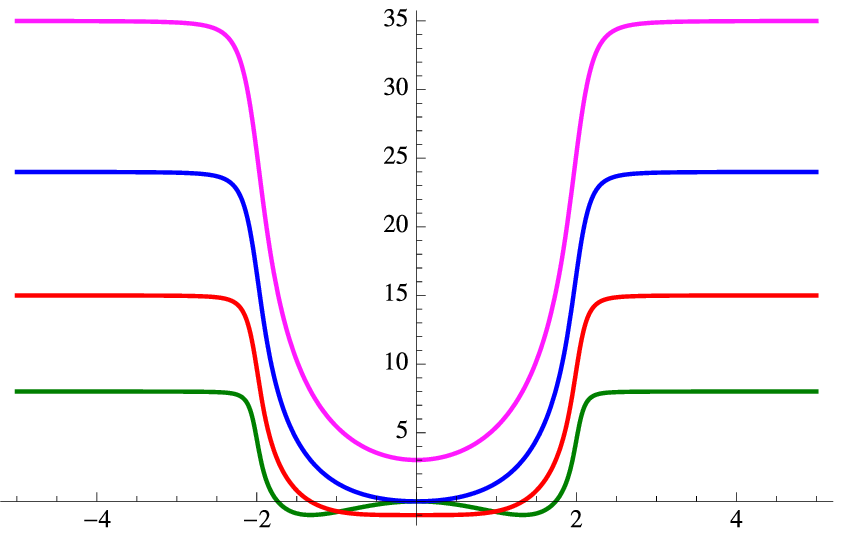}\qquad\includegraphics[width=.45\textwidth]{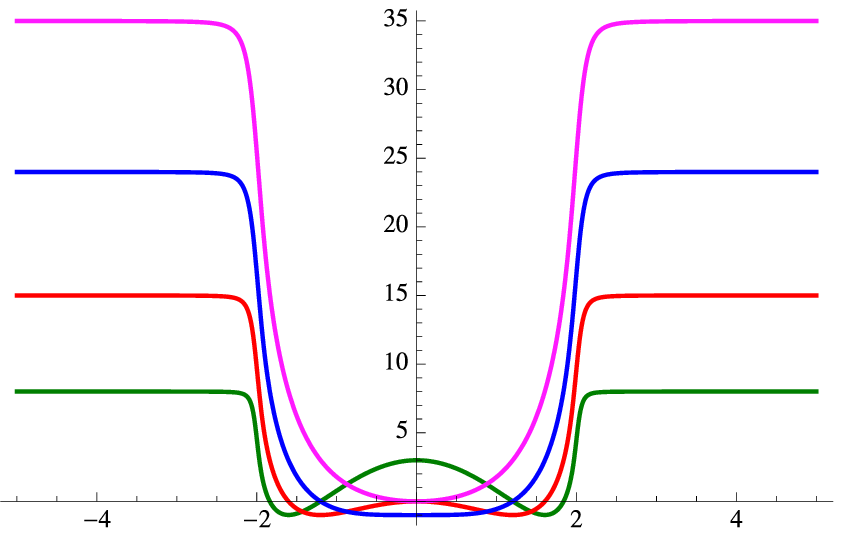}
\end{center}
\caption{On the left picture and right pictures $Y_{2|vw}$, $Y_{3|vw}$, $Y_{4|vw}$  and $Y_{5|vw}$ are plotted  for the Konishi state at $\bar g_{cr}^{(3)}\approx 21.6$ and  $\bar g_{cr}^{(4)}\approx 34.9$, respectively.
$Y_{2|vw}$ and $Y_{4|vw}$ touch the $u$-axis at $g=\bar g_{cr}^{(3)}$, and $Y_{3|vw}$ and $Y_{5|vw}$
touch it at $g=\bar g_{cr}^{(4)}$. $Y_{2|vw}$
 has four real  zeros for  $g>\bar g_{cr}^{(3)}$.  }
\end{figure}
In Figures 3-5, we show several plots of $Y_{M|vw}$-functions for the Konishi state.

In what follows a Konishi-like state  refers to any two-particle state for which only $Y_{1|vw}$-function has two real  zeros and all the other $Y_{M|vw}$-functions have no  zeros in the physical strip  at small coupling.

\subsubsection*{Critical values of $g$ for some states}

Here we analyze the $g$-dependence of Y-functions for several other states.

We begin with the state with $J=5$ and $n=1$. This is the state with the lowest value of $J$ such that both $Y_{1|vw}$ and $Y_{2|vw}$ have two  zeros in the physical strip at small $g$.
The critical and subcritical values are determined by the equations
\bea
Y_{r+2|vw}(\pm{i\ov g},g)=0\,,\quad Y_{r+2|vw}(0,\bar g)=0\,,\quad r=1,2,\ldots \,,
\eea
and we find the following values for $g<100$
\bea
{g}_{5,1}^{r,2} &=& \{6.707, 15.458, 27.233, 42.107,
 60.101, 81.222\}\\\nonumber
\bar{g}_{5,1}^{r,2} &=& \{6.764, 15.479, 27.244, 42.114,
 60.105, 81.225\}\,.
\eea

Next, we consider the state with $J=8$ and $n=1$. This is the state with the lowest value of $J$ such that $Y_{1|vw}$ has four  zeros, $Y_{2|vw}$ has two real  zeros and $Y_{3|vw}$  has two imaginary  zeros in the physical strip at small $g$. Therefore, the critical and subcritical values are determined by the equations
\bea
Y_{r+3|vw}(\pm{i\ov g},g)=0\,,\quad Y_{r+2|vw}(0,\bar{g})=0\,,\quad r=1,2,\ldots \,.
\eea
We find the following 6 critical and  7 subcritical values of $g$ for $g<100$
\bea
g_{8,1}^{r,3} &=& \{~~ -~,9.157, 19.561, 32.985, 49.505,
69.143, 91.909\}\\\nonumber
\bar g_{8,1}^{r,3} &=& \{0.116, 9.207, 19.580, 32.995, 49.511,
69.148, 91.912\}\,.
\eea
The reason why $g_{8,1}^{1,3}$ is so small is that the two imaginary roots of $Y_{3|vw}$ that are in the physical strip at $g=0$ reach the real line very quickly.

One might think that the first critical value increases with $J$. It is not so as one can see on the example of  the state with $J=9$ and $n=1$. This is again the state such that $Y_{1|vw}$ has four  zeros, and $Y_{2|vw}$ and $Y_{3|vw}$  have two  zeros at small $g$, and it has the following 6 critical values of $g$ for $g<100$
\bea
g_{9,1}^{r,3} &=& \{6.970, 16.982, 29.935, 45.968,
65.114, 87.384\}\\\nonumber
\bar g_{9,1}^{r,3} &=& \{7.052, 17.006, 29.947, 45.976,
65.119, 87.388\}\,.
\eea

%%%%%%%%%%%%%%%%%%%%%%%%%%%%%%%%%
\section{TBA equations for Konishi-like states} \la{TBAKon}

As was discussed above to formulate excited state TBA equations one
 should choose an integration contour, take it back to the real line of the mirror plane, and then check that the resulting TBA equations are solved by the large $L$ expressions for Y-functions. We begin our analysis  with the simplest case of a  Konishi-like state which appears however to be quite general and allows one to understand the structure of the TBA equations for any two-particle $\sl(2)$ sector state. To simplify the notations we denote the critical values of the state under consideration as $g_{cr}^{(r)}$.

\subsection{Excited states TBA equations:  $g<g_{cr}^{(1)}$}

Let us stress that the equations below are valid only for  physical states satisfying the level-matching condition. Since some terms in the equations below have the same form for any $N$ we keep an explicit dependence on $N$ in some of the formulae.

\subsubsection*{ Integration contour}

The integration contour for all Y-functions but $Y_\pm$ is chosen in such a way that it lies a little bit above the interval Re$(z)\in (-{\om_1\ov 2},{\om_1\ov 2})$, Im$(z)={\om_2\ov 2i}$ in the middle of the mirror theory region, and penetrates to the string theory region in the small vicinity of $z =\om_2/2$ (the centre point of the mirror region).
\begin{figure}[t]
\begin{center}
\includegraphics*[width=0.45
\textwidth]{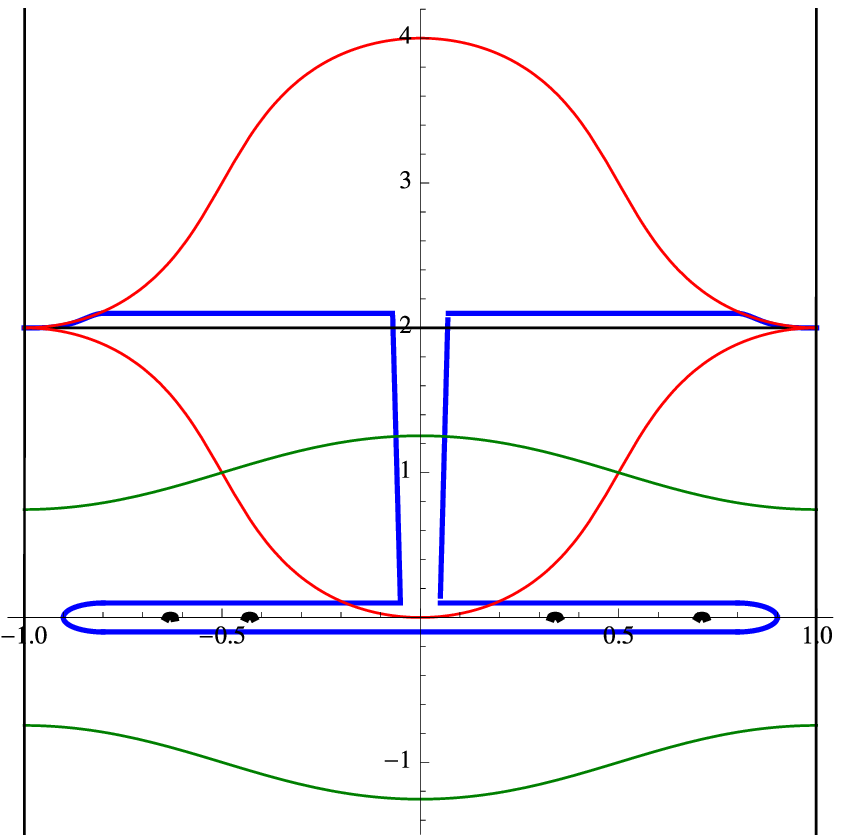}\quad \includegraphics*[width=0.4\textwidth]{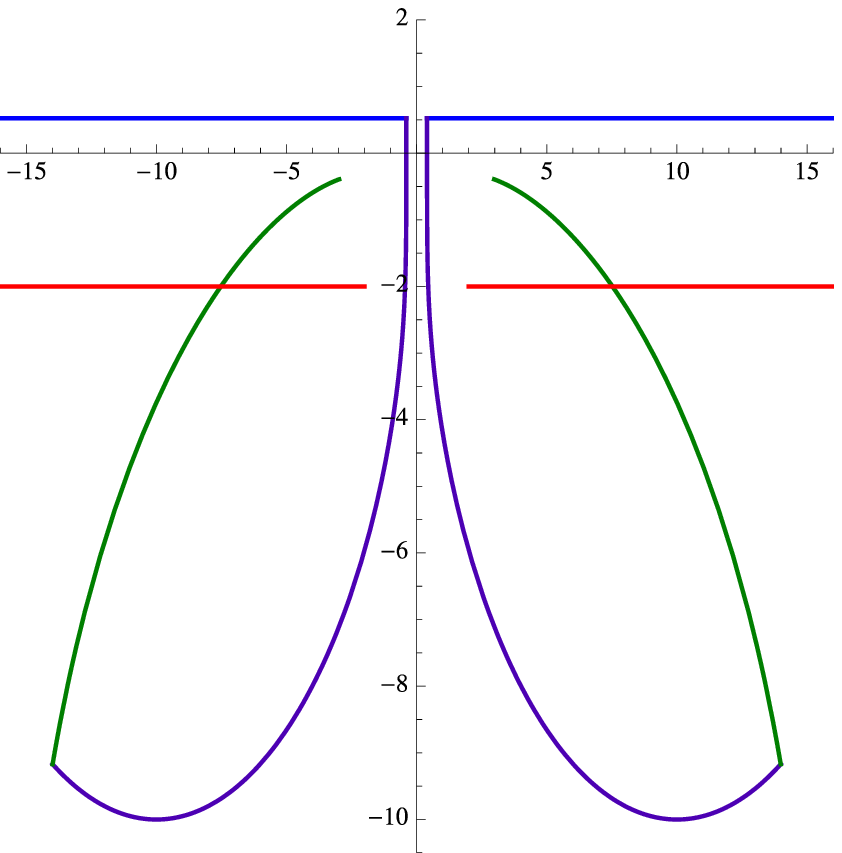}
\end{center}
\caption{On the left picture the integration contour on the $z$-torus is shown. On the right picture a part of the integration contour on the mirror u-plane going from $\pm\infty$ a little bit above the real line  to the origin, and down to the string region is plotted. The green curves correspond to the horizontal lines on the z-torus just above the real line of the string region. The red semi-lines are the cuts of the mirror $u$-plane, and on the $z$-torus they are mapped to the part of the boundary of the mirror region that lies in the string region.}
\end{figure}
Then,  it goes along the both sides of the vertical line to the string region real line, and encloses
all the points $z_{*k}$ so that they lie between the mirror theory line and the integration contour, see Figure 6.
In the  mirror $u$-plane  the contour lies above the real line, then it goes down at $u=0$, reaches a minimum value and turns back to the real line. In the equations involving the functions $Y_Q$, $Y_{Q|vw}$ and $Y_{Q|w}$ it crosses  the cuts  of the mirror $u$-plane with Im$(u)=-{Q\ov g}$, and enters another sheet, see Figure 2.
It is worth mentioning that the contour does not cross any additional cuts $Y$-functions have on the $z$-torus.

Then, one uses the TBA equations for the ground state energy, and
taking the integration contour back to the mirror region  interval  Im$(z)={\om_2\ov 2i}$, picks up $N$ extra contributions of the form $-\log S(z_{*},z)$ from any term $\log (1+Y_1)\star K$, where
$S(w,z)$ is the S-matrix corresponding to the kernel $K$: $K(w,z)={1\ov 2\pi i}{d\ov dw}\log S(w,z)$. In addition, one also gets contributions of the form $-\log S(w,z)$ from the imaginary  zeros of $1+Y_{M|vw}$ located below the real line of the mirror $u$-plane, see \eqref{zerorel}.

Finally, the integration contour for $Y_\pm$-functions should be deformed so that the points
$u_k^-=u_k-{i\ov g}$ of the mirror $u$-plane lie between the interval $[-2,2]$ of the  mirror theory line and the contour.  Then,  the terms of the form $\log (1-Y_+)\hstar K$ would produce extra contributions of the form $+\log S(u_{k}^-,z)$ because $Y_+(u_k^-)=\infty$.
In fact, this   is important only if one uses the simplified TBA equations because in the canonical TBA equations $Y_\pm$-functions appear only in the combination $1-{1\ov Y_\pm}$. Note also that $Y_\pm$-functions analytically continued to the whole mirror $u$-plane have a cut $(-\infty,-2]\cup [2,\infty)$, and they should satisfy the following important equality which, as was shown in \cite{AF09b},
is necessary for the fulfillment of the Y-system
\bea\la{ypym}
Y_+(u\pm i0) = Y_-(u\mp i0)\quad {\rm for}\ \   u\in (-\infty,-2]\cup [2,\infty)\,.
\eea
This equality shows that one can glue the two $u$-planes along the cuts, and then $Y_\pm$-functions can be thought of as two branches of one analytic function defined on the resulting surface (with extra cuts in fact). We will see that the equality \eqref{ypym} indeed follows from the TBA equations.

\subsubsection*{ Simplified TBA equations}

Using this procedure and the simplified TBA equations for the ground state derived in \cite{AF09b,AF09d}, one
gets the following set of integral equations for Konishi-like states and $g<g_{cr}^{(1)} $

\bigskip
 \noindent
$\bullet$\ $M|w$-strings: $\ M\ge 1\ $, $Y_{0|w}=0$
\bea\la{Yforws}
\log Y_{M|w}=  \log(1 +  Y_{M-1|w})(1 +  Y_{M+1|w})\star s
+\delta_{M1}\, \log{1-{1\ov Y_-}\ov 1-{1\ov Y_+} }\hstar s\,.~~~~~
\eea
These equations coincide with the ground state ones.

\bigskip
 \noindent
$\bullet$\ $M|vw$-strings: $\ M\ge 1\ $, $Y_{0|vw}=0$
\bea\la{Yforvw3}
\hspace{-0.3cm}\log Y_{M|vw}(v)&=&-\delta_{M1} \sum_{j=1}^N \log S(u_j^--v)- \log(1 +  Y_{M+1})\star s~~~~~\\\nonumber
&+& \log(1 +  Y_{M-1|vw} )(1 +  Y_{M+1|vw})\star s+\delta_{M1}  \log{1-Y_-\ov 1-Y_+}\hstar s\,,~~~~~
\eea
where $u_j^-\equiv u_j-{i\ov g}$, and the kernel $s$ and the corresponding S-matrix $S$ are defined in appendix \ref{app:rapidity}.
For $M=1$ the first term is due to our choice of the integration contour for $Y_\pm$-functions, and the pole of $Y_+$ at $u=u_j^-$.

It is worth pointing out that there is  no extra contribution in
\eqref{Yforvw3} from any term of the form $\log(1 +  Y_{k|vw}
)\star s$. In general such a term leads to a contribution equal to
$-\log S(r_1 - v)S(r_2 - v)$ where $r_1$ and $r_2$ are the two
 zeros of $1 +  Y_{k|vw}$ with negative imaginary parts. To
explain this, we notice that if the  zeros $r_j^{(k+1)}$ of
$Y_{k+1|vw}$ lie outside the physical strip, then
 $r_j$ are  $r_1=r_1^{(k+1)}-{i\ov g}$  and $r_2=r_1^{(k+1)}+{i\ov g}$,
 where Im$(r_1^{(k+1)})<-{1\ov g}$. Since $S(r-{i\ov g})S(r+{i\ov g}) = 1$ the term $\log(1 +  Y_{k|vw} )\star s$ does not
 contribute, see Figure 7.

On the other hand, if the  zeros $r_j^{(k+1)}$ of $Y_{k+1|vw}$ lie
inside the physical strip then  $r_j$ are  related to
$r_j^{(k+1)}$ as $r_j=r_j^{(k+1)}-{i\ov g}$, and the term $\log(1
+  Y_{k|vw} )\star s$ leads to the extra contribution equal to
$-\sum_j\log S(r_j^{(k+1)-} - v)$ where $r_j^{(k+1)-}\equiv
r_j^{(k+1)}-{i\ov g}$, see Figure 7.

\begin{figure}[H]
\begin{center}
\includegraphics*[width=0.9\textwidth]{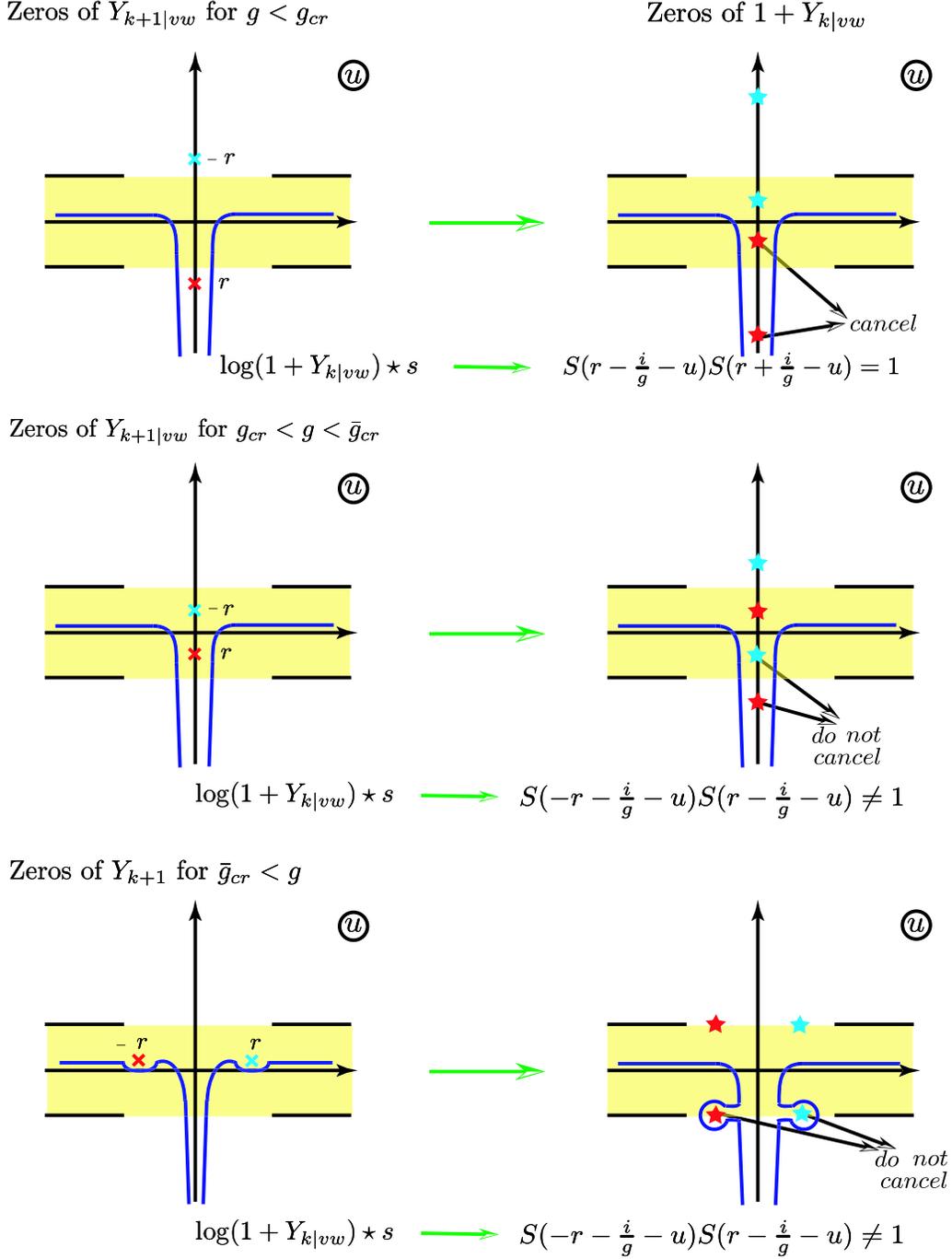}
\end{center}
\caption{In the upper pictures the positions of  zeros of
$Y_{k+1|vw}$ and of $1+Y_{k|vw}$ are shown for $g<g_{cr}$. The
contributions of $\log(1+Y_{k|vw})\star s$ to the TBA equations
cancel out for this case. The two pictures in the middle
correspond to the situation when two   zeros of $Y_{k+1|vw}$ enter
the physical region $|{\rm Im}u|<\frac{1}{g}$ (depicted in
yellow). Finally, the two pictures at the bottom are drawn for the
case when two   zeros of $Y_{k+1|vw}$ are on the real line which
corresponds to $g_{cr}<\bar{g}_{cr}<g$. When $g>g_{cr}$ the
contributions of $\log(1+Y_{k|vw})\star s$ do not cancel anymore
and lead to modification of the corresponding TBA equations. }
\end{figure}

We conclude therefore that at weak coupling and only for Konishi-like states
no term $\log(1 +  Y_{k|vw} )\star s$ gives an extra contribution to the TBA equations for $vw$-strings.

Let us also mention that  the poles of $S(u_j^--v)$ cancel the  zeros of $Y_{1|vw}(v)$ at $v=u_j$, and  \eqref{Yforvw3} is compatible with the reality condition for Y-functions.

\bigskip
 \noindent
$\bullet$\   $y$-particles \footnote{
The equation for $Y_+Y_-$ follows from eq.(4.14) and (4.26) of \cite{AF09b}, and the identity

$
K_{Qy}\hstar  K_1 = K_{xv}^{Q1} - K_{Q-1}\,.
$
Another identity
$
K_{xv}^{Q1} \star s = {1\ov 2}K_{Qy}+{1\ov 2}K_{Q} -\delta_{Q1} s - K_{Qy}^{ms}\cstar \tilde{s}\,,
$
 where $\tilde{s}(u)\equiv s(u-{i\ov g})$ is useful in deriving Y-system equations for $Y_\pm$.
 }
\bea
\la{Yfory1s}
\log {Y_+\ov Y_-}(v)&=& -\sum_{j=1}^N\, \log S_{1_*y}(u_{j},v)  +\log(1 +  Y_{Q})\star K_{Qy}\,,~~~~~~~ \\
\la{Yfory2s}
\log {Y_+ Y_-}(v) &=& -\sum_{j=1}^N\,
 \log {\big(S_{xv}^{1_*1}\big)^2\ov S_2}\star s(u_j,v)
\\\nonumber
&+&2\log{1 +  Y_{1|vw} \ov 1 +  Y_{1|w} }\star s
- \log\left(1+Y_Q \right)\star K_Q+ 2 \log(1 +  Y_{Q})\star K_{xv}^{Q1} \star s\,,~~~~
\eea
where we use the following notation
 \bea\nonumber
 &&\log  {\big(S_{xv}^{1_*1}\big)^2\ov S_2}\star s(u_{j},v) \equiv  \int_{-\infty}^\infty\, dt\, \log  {S_{xv}^{1_*1}(u_j,t)^2\ov S_2(u_j-t)}\, s(t-v)
 \,.~~~~~
\eea
Then, $S_{1_*y}(u_{j},v) \equiv S_{1y}(z_{*j},v)$ is a shorthand notation for the S-matrix with the first and second arguments in the string and mirror regions, respectively. The same convention is used for other S-matrices.  Both arguments of the kernels in these formulae are in the mirror region.

Taking into account that under the analytic continuation through the cut $|v|>2$ the S-matrix $S_{1_*y}$ and the kernel $K_{Qy}$ transforms as $S_{1_*y}\to 1/S_{1_*y}$ and $K_{Qy}\to - K_{Qy}$, one gets that the functions $Y_\pm$ are indeed analytic continuations of each other and, therefore, the equality \eqref{ypym} does hold.

It can be easily checked that the term on the first line in \eqref{Yfory2s} is real, and this makes obvious that the equations for $Y_\pm$-functions are also compatible with the reality of Y-functions.
The origin of this term can be readily understood if one uses the following identity
\bea\la{idn}
-\log {\big(S_{xv}^{1_*1}\big)^2\ov S_2}\star s(u_{j},v)= \log S_{1}(u_{j}-v)
-2 \log S_{xv}^{1_*1}\star s(u_{j},v)
\eea
which holds up to a multiple of $2\pi i$, and since $S_{xv}^{1_*1}(u,v)$ has a zero at $u=v$  the integration contour in the second term on the r.h.s. of \eqref{idn} runs above the real line.
Then, the term $\log S_{1}(u_{j}-v)$ comes from the term
$- \log\left(1+Y_1 \right)\star K_1$, and the second term come from
$2 \log(1 +  Y_{1})\star K_{xv}^{Q1} \star s$.

Eq.\eqref{Yfory2s} is very useful for checking the TBA equations in the large $J$ limit where one gets
\bea
\nonumber
\log {Y_+ Y_-} &=& -\sum_{j=1}^N\,
 \log {\big(S_{xv}^{1_*1}\big)^2\ov S_2}\star s(u_j,v)
+2\log{1 +  Y_{1|vw} \ov 1 +  Y_{1|w} }\star s
 \,.
\eea

\bigskip
 \noindent
$\bullet$\  $Q$-particles for $Q\ge 3$
\bea
\log Y_{Q}&=&\log{\left(1 +  {1\ov Y_{Q-1|vw}} \right)^2\ov (1 +  {1\ov Y_{Q-1} })(1 +  {1\ov Y_{Q+1} }) }\star s\la{YforQ2a3}
\,~~~~~~~
\eea
\bigskip
 \noindent
$\bullet$\  $Q=2$-particle
\bea\la{YforQ2a4}
\log Y_{2}&=& \sum_{j=1}^N \log S(u_{j}-v)
 -\log(1 +  {1\ov Y_{1} })(1 +  {1\ov Y_{3} }) \star  s+2\log\left(1 +  {1\ov Y_{1|vw}} \right)\star  s\,~~~~~~~
\eea
In fact by using the p.v. prescription, one gets
\bea\la{YforQ2bb}
\log Y_{2}&=& \log{\left(1 +  {1\ov Y_{1|vw}} \right)^2\ov (1 +  {1\ov Y_{1} })(1 +  {1\ov Y_{3} }) }\star_{p.v.}  s\,~~~~~~~
\eea
which makes obvious the reality of Y-functions. \bigskip

 \noindent
$\bullet$\  $Q=1$-particle
\bea
\nonumber
\log Y_{1}&=&\sum_{j=1}^N\log\check{\Sigma}_{1_*}^2\, \check{S}_{1}\,\check{\star}\, s(u_j,v)-L\, \check{\cal E}\,\check{\star}\, s
+\log\left(1-{1\ov Y_{-}} \right)^2Y_2\, \hat{\star}\, s \\\nonumber
&-&2 \log\left(1-{1 \ov Y_{-}} \right)\left(1-{1 \ov Y_{+}}\right)
Y_2\, \hat{\star}\, \check{K}\,\check{\star}\,  s +\log{Y_1}\star \check{K}_1\,\check{\star}\,  s
\\\la{YforQ1a5}
&-& \log\left(1+Y_{Q} \right)\star \big( 2\check{K}_Q^\Sigma +
\check{K}_Q  +\check{K}_{Q-2}\big)\,\check{\star}\,  s - \log(1 + Y_{2})\star s\,.~~~~~
\eea
All the kernels appearing here are defined in appendix \ref{app:rapidity}, and we also assume that $\check{K}_{0}=0$ and $\check{K}_{-1}=0$.
The reality of this equation follows from the reality of
\bea\nonumber
{S_{ss}(u-{i\ov g},v)^2 \ov \check{S}_{1}(u,v)} =  \frac{x_s(u-{i\ov g}) - x_s(v)}{x_s(u-{i\ov g}) - {1\ov x_s(v)}} \frac{x_s(u+{i\ov g}) - x_s(v)}{x_s(u+{i\ov g}) - {1\ov x_s(v)}} = S_{ss}(u-{i\ov g},v)S_{ss}(u+{i\ov g},v)\,,~~~~~~
\eea
which appears if one uses the representation \eqref{s1star} for $\check{\Sigma}_{1_*}$. Note that $\check{S}_{1}(u,v)$ is defined through the kernel $\check{K}_{1}(u,v)$ by the integral
\bea
\check{S}_{1}(u,v) = \exp\Big( 2\pi i \int_{-\infty}^u\,  du'\,  \check{K}_{1}(u',v) \Big)= {S_{ss}(u-{i\ov g},v) \ov S_{ss}(u+{i\ov g},v)}\,,
\eea
and it differs from the naive formula
\bea
S_{ms}(u-{i\ov g},v) S_{ms}(u+{i\ov g},v) =\check{S}_{1}(u,v)\, x_s(v)^2 \,,
\eea
which one could write by using the expression \eqref{ck1} for the kernel $\check{K}_{1}$.

We see that the reality of Y-functions is a trivial consequence of these equations. Moreover,
in the large $L$ limit the simplified TBA equations do not involve infinite sums at all.
As a result, they can be easily checked numerically with an arbitrary precision. We have found that for Konishi-like states  the integral equations are solved at the large $L$ limit  by
the asymptotic Y-functions given in terms of transfer matrices if the length parameter $L$ is related to the charge $J$ carried by a string state as
$$L=J+2\,.$$
We expect that for all $N$-particle states from the $\sl(2)$
sector the relation between length and charge is universal and
given by $L=J+2$.

\medskip

There is another form of the TBA equations for $Q$-particles which is
obtained by combining the simplified and canonical TBA equations.
We refer to this form as the hybrid one, and the equations can be written as follows

\bigskip

 \noindent
$\bullet$\  Hybrid equations for $Q$-particles
\begin{align}
&\log Y_Q(v) = - \sum_{j=1}^N\(  \log S_{\sl(2)}^{1_*Q}(u_j,v)
 - 2 \log S\star K^{1Q}_{vwx} (u_j^-,v) \)
\notag\\
&\quad - L\, \tH_{Q}
+ \log \left(1+Y_{Q'} \right) \star \(K_{\sl(2)}^{Q'Q} + 2 \, s \star K^{Q'-1,Q}_{vwx} \)
\label{TbaQsl2H} \\
&\quad + 2 \log \(1 + Y_{1|vw}\) \star s \hstar K_{yQ}
+ 2 \, \log \(1 + Y_{Q-1|vw}\) \star  s
\notag \\
&\quad - 2  \log{1-Y_-\ov 1-Y_+} \hstar s \star K^{1Q}_{vwx}
+  \log {1- \frac{1}{Y_-} \ov 1- \frac{1}{Y_+} } \hstar K_{Q} +  \log \big(1- \frac{1}{Y_-}\big)\big( 1- \frac{1}{Y_+} \big) \hstar K_{yQ} \,,
\notag
\end{align}
where $K^{0,Q}_{vwx}=0$, and $Y_{0|vw}=0$, and we use the notation
\bea
 \log S\star K^{1Q}_{vwx} (u_j^-,v) = \int_{-\infty}^\infty\, dt\,  \log S(u_j^- -t-i0) \star K^{1Q}_{vwx}(t+i0,v)\,.~~~
\eea
The first term on the first line of  \eqref{TbaQsl2H} comes from
$\log \left(1+Y_{Q'} \right) \star K_{\sl(2)}^{Q'Q} $, and the second one from
$- 2  \log{1-Y_-\ov 1-Y_+} \hstar s \star K^{1Q}_{vwx}$.
Eq.\eqref{TbaQsl2H}  is derived in appendix \ref{hybridQ}.

\medskip

The energy of the multiparticle state is obtained in the same way by taking the integration contour back to the real mirror momentum line, and is given by
\bea\nonumber
E_{\{n_k\}}(L)&=&  \sum_{k=1}^N\, i\tp^1(z_{*k}) -\int {\rm d}z\, \sum_{Q=1}^\infty{1\ov
2\pi}{d\tp^Q\ov dz}\log\left(1+Y_Q\right)~~~~~~\\
&=&  \sum_{k=1}^N\E_k -\int {\rm d}u\, \sum_{Q=1}^\infty{1\ov
2\pi}{d\tp^Q\ov du}\log\left(1+Y_Q\right)\,, \label{Enk}
\eea
where
\bea
\E_k = igx^-(z_{*k})-igx^+(z_{*k}) -1= igx_s^-(u_{k})-igx_s^+(u_{k}) -1\,,
\eea
is the energy of a fundamental particle in the string theory, see appendix \ref{app:rapidity} for definitions and conventions.

For practical computations the analytic continuation from the mirror region to the string one reduces to the substitution $x^{Q\pm}(u)\to x^{Q\pm}_s(u)\equiv  x_s(u\pm {i\ov g} Q)$ in all the kernels and S-matrices.
Then, as was discussed above the string theory spectrum is characterized by a set of $N$ real numbers $u_k$  (or $z_{*k}$) satisfying the exact Bethe equations \eqref{Y1m1}.
We assume for definiteness  that $u_k$ are ordered as $u_1 < \cdots < u_N$.

Finally, to derive exact Bethe equations one
should analytically continue $Y_1$ given by  either eq.\eqref{YforQ1a5} or \eqref{TbaQsl2H}. We find that it is simpler  and easier to handle the exact Bethe equations derived from the hybrid equation \eqref{TbaQsl2H} for $Y_1$. In the appendix \ref{canTBA} we also derive
exact Bethe equations  from the canonical equation for $Y_1$.

\subsubsection*{Exact Bethe equations }

Now we need to derive the integral form of the exact Bethe equations (\ref{Y1m1}). Let us note first of all that at large $L$ eq.(\ref{Y1m1}) reduces to the BY equations for the $\sl(2)$-sector by construction, and the integral form of  (\ref{Y1m1}) should be compatible with this requirement.

To derive the exact Bethe equations, we take the logarithm of
eq.(\ref{Y1m1}), and analytically continue the variable $z$ of
$Y_1(z)$ in eq.(\ref{TbaQsl2H}) to the point $z_{*k}$. On the
mirror $u$-plane it means that we go from the real $u$-line  down
below the line with Im$(u)=-{1\ov g}$  without crossing any cut,
then turn back, cross the cut with Im$(u)=-{1\ov g}$ and
$|$Re$(u)|>2$, and go back to the real $u$-line, see Figure 6. As
a result, we should make the following replacements $x(u-{i\ov g})
\to  x_s(u-{i\ov g}) = x(u-{i\ov g})$, $x(u+{i\ov g}) \to
x_s(u+{i\ov g}) = 1/x(u+{i\ov g})$ in the kernels appearing in
(\ref{TbaQsl2H}).

The analytic continuation depends on the analytic properties of the kernels and Y-functions, and its detailed consideration can be found in appendix
 \ref{app:Y1}.
As shown there, the resulting exact Bethe equations  for a string theory state from the $\sl(2)$ sector can be cast into the following integral form
\begin{align}
&\pi i(2n_k+1)=\log Y_{1_*}(u_k) =i L\, p_k- \sum_{j=1}^N\, \log S_{\sl(2)}^{1_*1_*}(u_j,u_k)\label{Tba1sl2B}\\
&\quad
 + 2 \sum_{j=1}^N\, \log {\rm Res}(S)\star K^{11_*}_{vwx} (u_j^-,u_k) -2 \sum_{j=1}^N\log\big(u_j-u_k-{2i\ov g}\big)\,
{x_j^--{1\ov x_{k}^-}\ov x_j^-- x_{k}^+}
\notag\\
&\quad
+ \log \left(1+Y_{Q} \right) \star \(K_{\sl(2)}^{Q1_*} + 2 \, s \star K^{Q-1,1_*}_{vwx} \)+ 2 \log \(1 + Y_{1|vw}\) \star \( s \hstar K_{y1_*} + \ts\)
\notag \\
&\quad - 2  \log{1-Y_-\ov 1-Y_+} \hstar s \star K^{11_*}_{vwx}
+  \log {1- \frac{1}{Y_-} \ov 1- \frac{1}{Y_+} } \hstar K_{1} +  \log \big(1- \frac{1}{Y_-}\big)\big( 1- \frac{1}{Y_+} \big) \hstar K_{y1_*} \,,
\notag
\end{align}
where we use the notations
\bea
&&\log {\rm Res}(S)\star K^{11_*}_{vwx} (u^-,v) = \int_{-\infty}^{+\infty}{\rm d}t\,\log\Big[S(u^- -t)(t-u)\Big] K_{vwx}^{11*}(t,v)\,,~~~\\
&&\ts(u)=s(u^-)\,.
\eea
The integration contours in the formulae above run a little bit above the Bethe roots $u_j$, $p_k= i \tH_{Q}(z_{*k})=-i\log{x_s(u_k+{i\ov g})\ov x_s(u_k-{i\ov g})}$ is the momentum of the $k$-th particle, and the second argument in all the kernels in \eqref{Tba1sl2B} is equal to $u_{k}$. The first argument we integrate with respect to is the original one in the mirror region.

Taking into account that the BY equations for the $\sl(2)$ sector have the form
 \bea\nonumber
\pi i(2n_k+1)=i J\, p_k - \sum_{j=1}^N \log S_{\sl(2)}^{1_*1_*}(u_{j},u_{k})\la{BYsl2}
 \,,~~~~~~
\eea
and that $Y_Q$ is exponentially small at large $J$, we conclude that if the analytic continuation has been done correctly then up to an integer multiple of $2\pi i$  the following identities between the asymptotic Y-functions should hold
\bea\nonumber
&&{\cal R}_k\equiv 2 \ssp i \, p_k + 2 \sum_{j=1}^N\, \log {\rm Res}(S)\star K^{11_*}_{vwx} (u_j^-,u_k) -2 \sum_{j=1}^N\log\big(u_j-u_k-{2i\ov g}\big)\,
{x_j^--{1\ov x_{k}^-}\ov x_j^-- x_{k}^+}
\nonumber\\
&&\quad
+ 2 \log \(1 + Y_{1|vw}\) \star \( s \hstar K_{y1_*} + \ts\)- 2  \log{1-Y_-\ov 1-Y_+} \hstar s \star K^{11_*}_{vwx}
\nonumber\\
&&\quad
+  \log {1- \frac{1}{Y_-} \ov 1- \frac{1}{Y_+} } \hstar K_{1} +  \log \big(1- \frac{1}{Y_-}\big)\big( 1- \frac{1}{Y_+} \big) \hstar K_{y1_*}  = 0\,.~\label{Rkid}
\eea
For $N=2$ and $u_1=-u_2$ one gets one equation, and by using the expressions for the Y-functions from appendix \ref{appT} one can check numerically\footnote{Since
$K^{11_*}_{vwx}(u,v)$ has a pole at $u=v$ with the residue equal to $-{1\ov 2\pi i}$
the terms of the form $2f\star K^{11_*}_{vwx} $ can be represented as $
 2f\star K^{11_*}_{vwx}  = 2f\star_{p.v.}  K^{11_*}_{vwx} + f(u_k)$ which is useful for numerics.
}
that it does hold for any real value of $u_1$ such that only $Y_{1|vw}$ has two  zeros inside the strip $|{\rm Im} \,u|<1/g$.

\subsection{Excited states TBA equations: $g_{cr}^{(1)}<g<g_{cr}^{(2)}$}

In this subsection we consider the TBA equations for values of $g$ in the first critical region $g_{cr}^{(1)}<g<g_{cr}^{(2)}$. In this region in addition to the two real  zeros of $Y_{1|vw}$ at $u_j$, two  zeros of $Y_{2|vw}$ enter the physical strip. We denote these  zeros $r_j$.\footnote{Canonical TBA equations take different forms depending  on the reality of  the  zeros, and, therefore, one has to divide the region into two subregions: $g_{cr}^{(1)}<g<\bar g_{cr}^{(1)}$ and $\bar g_{cr}^{(1)}<g<g_{cr}^{(2)}$, see the next subsection and appendix  \ref{canTBA} for detail.}

\subsubsection*{ Simplified TBA equations}

We first notice that
for $g_{cr}^{(1)}<g< g_{cr}^{(2)}$ the function $Y_{2|vw}$ has two  zeros in the physical strip. Therefore, as was discussed above, the contribution to the simplified TBA equations coming from the  zeros of  $Y_{2|vw}$ and $1+Y_{1|vw}$ does not vanish.
If $g<\bar g_{cr}^{(1)}$ no deformation of the integration contour is needed because all these  zeros are on the imaginary line of the mirror region.
As $g$ approaches $\bar g_{cr}^{(1)}$ the  two  zeros of $Y_{2|vw}$  approach $u=0$, and at
$g=\bar g_{cr}^{(1)}$ they both are at the origin. As $g>\bar g_{cr}^{(1)}$ the  zeros become real, located symmetrically, and they push the integration contour to be a little bit below them.  In addition to this at $g=\bar g_{cr}^{(1)}$ the two  zeros of $1+Y_{1|vw}$ with the negative imaginary part reach the point $u=-i/g$, and as $g>\bar g_{cr}^{(1)}$ they begin to move along the line Im$(u)=-1/g$ in opposite directions. As a result, the integration contour should be deformed in such a way that the two  zeros of $1+Y_{1|vw}$ would not cross it. Thus, the  zeros of
$1+Y_{1|vw}$ always lie between the real line and the integration contour, and the terms of the form $\log (1+Y_{1|vw})\star K$ produce the usual contribution once one takes the contour back to the real line.
%The integration contour for $g$ in the first critical region is shown on Figure ???.
Let us also mention that the points $r_j^-$ of the mirror $u$-plane are mapped to the upper boundary of the string region on the $z$-torus.

Using this integration contour, one
gets the following set of  simplified TBA equations for Konishi-like states and $g_{cr}^{(1)}<g<g_{cr}^{(2)}$

\bigskip
 \noindent
$\bullet$\ $M|w$-strings: their equations coincide with the ground state ones \eqref{Yforws}.
%$\ M\ge 1\ $, $Y_{0|w}=0$
%\bea\la{Yforwsc1}
%\log Y_{M|w}=  \log(1 +  Y_{M-1|w})(1 +  Y_{M+1|w})\star s
%+\delta_{M1}\, \log{1-{1\ov Y_-}\ov 1-{1\ov Y_+} }\hstar s\,.~~~~~
%\eea

\bigskip
 \noindent
$\bullet$\ $M|vw$-strings: $\ M\ge 1\ $, $Y_{0|vw}=0$
\bea\la{Yforvw3c1}
&&\hspace{-0.3cm}\log Y_{M|vw}(v)=-\delta_{M1} \sum_{j=1}^N \log S(u_j^--v)-\delta_{M2} \sum_{j=1}^2 \log S(r_j^--v)~~~~~\\\nonumber
&&+ \log(1 +  Y_{M-1|vw} )(1 +  Y_{M+1|vw})\star s+\delta_{M1}  \log{1-Y_-\ov 1-Y_+}\hstar s- \log(1 +  Y_{M+1})\star s\,.~~~~~
\eea
The first term is due to the pole of $Y_+$ at $u=u_j^-$, and  the second term is due to the  zeros of $1 +  Y_{1|vw}$ at $u=r_j^-$.

\bigskip
 \noindent
$\bullet$\   $y$-particles
\bea\la{Yfory1sc1}
\log {Y_+\ov Y_-}(v)&=& -\sum_{j=1}^N \log S_{1_*y}(u_{j},v)  +\log(1 +  Y_{Q})\star K_{Qy}\,,~~~~~~~
 \\
\la{Yfory2sc2}
\log {Y_+ Y_-}(v) &=& -\sum_{j=1}^N\,
 \log {\big(S_{xv}^{1_*1}\big)^2\ov S_2}\star s(u_j,v)-2\sum_{j=1}^2 \log S(r_j^--v)
\\\nonumber
&+&2\log{1 +  Y_{1|vw} \ov 1 +  Y_{1|w} }\star s
- \log\left(1+Y_Q \right)\star K_Q+ 2 \log(1 +  Y_{Q})\star K_{xv}^{Q1} \star s\,,~~~~
\eea
where the second term on the second line  is due to the  zeros of $1 +  Y_{1|vw}$ at $u=r_j^-$.

\bigskip
 \noindent
$\bullet$\  $Q$-particles for $Q\ge 3$
\bea
\log Y_{Q}&=&\log{\left(1 +  {1\ov Y_{Q-1|vw}} \right)^2\ov (1 +  {1\ov Y_{Q-1} })(1 +  {1\ov Y_{Q+1} }) }\star_{p.v.} s\la{YforQ2a3c1b}\,.
\,~~~~~~~
\eea
In fact the p.v. prescription, see appendix \ref{app:reality}, is not really needed here because for $Q=3$  the double zero of $Y_2$ cancels the  zeros of $Y_{2|vw}$, and for $Q\ge4$ everything is regular. Thus, the formula works no matter if the roots $r_j$ are real or imaginary.

\bigskip
 \noindent
$\bullet$\  $Q=2$-particle
%\bea\la{YforQ2a4c1}
%\log Y_{2}&=& \sum_{j=1}^N \log S(u_{j}-v) - 2\sum_{j=1}^2 \log S(r_{j}^--v)
%+\log{\left(1 +  {1\ov Y_{1|vw}} \right)^2\ov (1 +  {1\ov Y_{1} })(1 +  {1\ov Y_{3} }) }\star  s\,,~~~~~~~
%\eea
%By using the p.v. prescription, one gets
\bea\la{YforQ2a4c1}
\log Y_{2}&=& - 2\sum_{j=1}^2 \log S(r_{j}^--v)+\log{\left(1 +  {1\ov Y_{1|vw}} \right)^2\ov (1 +  {1\ov Y_{1} })(1 +  {1\ov Y_{3} }) }\star_{p.v.}  s\,~~~~~~~
\eea
This makes obvious the reality of Y-functions because the double zero of $Y_2$ at $v=r_j$ is cancelled by the pole of $S(r_{j}^--v)$.

\subsubsection*{ Hybrid equations}

One can easily see  that the simplified equation for $Q=1$-particles is the same as eq.\eqref{YforQ1a5} in the weak coupling region $g<g_{cr}^{(1)}$. Thus, we will only discuss the hybrid equations for $Q$-particles.
Strictly speaking,   their form is sensitive to whether the zeros of $Y_{2|vw}$ are complex or real, and the first critical region is divided into two subregions: $g_{cr}^{(1)}<g<\bar g_{cr}^{(1)}$ and $\bar g_{cr}^{(1)}<g<g_{cr}^{(2)}$.
On the other hand, to derive the exact Bethe equations we only need the hybrid $Q=1$ equation which, as we will see, takes the same form in both subregions, and, moreover, for any $g>g_{cr}^{(1)}$.

For $g_{cr}^{(1)}<g<\bar g_{cr}^{(1)}$ the function $Y_{2|vw}$ has two complex conjugate  zeros in the physical strip which, therefore, lie on the opposite sides of the real axis of the mirror $u$-plane. Thus, the zero $r_1$ with the negative imaginary part lies between the integration contour and the real line of the mirror region.
Taking the integration contour back to the real line produces an extra contribution from this zero. Since $1+Y_{1|vw}(r_j^-)=0$, and $Y_\pm=0$, the terms $2 \log \(1 + Y_{1|vw}\) \star s \hstar K_{y1}$ and $\log \big(1- \frac{1}{Y_-}\big)\big( 1- \frac{1}{Y_+} \big) \hstar K_{y1}$ in the hybrid equation \eqref{TbaQsl2H} lead to the appearance of $-2 \log S\hstar K_{y1}(r_j^-,v)$ and $2\log S_{yQ} (r_1,v)$,\footnote{Note that the S-matrix $S_{yQ}$ is normalized as $S_{yQ} (\pm 2,v)=1$.}
respectively. The integration contour for the $\hstar$-convolution in the term $-2 \log S\hstar K_{y1}(r_j^-,v)$ is the one for $Y_\pm$-functions and it should be taken to the mirror region too. Since the S-matrix $S(r_1^--t)$ has a pole at $t=r_1$, this produces the extra term $-2\log S_{yQ} (r_1,v)$ that exactly cancels the previous contribution from $\log \big(1- \frac{1}{Y_-}\big)\big( 1- \frac{1}{Y_+} \big) \hstar K_{y1}$. Thus, the only additional term that appears in the hybrid TBA equation \eqref{TbaQsl2H} for $g_{cr}^{(1)}<g<\bar g_{cr}^{(1)}$ is
$-2 \sum_j \log S\hstar K_{y1}(r_j^-,v)$.

As $g>\bar g_{cr}^{(1)}$ the two  zeros of  $Y_\pm$ (and $Y_{2|vw}$) become real and located a little bit above  the integration contour.
Therefore, the only extra contribution to the TBA equation comes from the term
$2 \log \(1 + Y_{1|vw}\) \star s \hstar K_{y1}$ and it is again is $-2 \sum_j \log S\hstar K_{y1}(r_j^-,v)$.

Thus, we see that in the first critical region $g_{cr}^{(1)}<g< g_{cr}^{(2)}$ the hybrid $Q=1$ equation takes the following form independent of
whether $g<\bar g_{cr}^{(1)}$ or $g> \bar g_{cr}^{(1)}$

%\medskip

 \noindent
$\bullet$\  Hybrid $Q=1$ equation for  $g_{cr}^{(1)}<g<g_{cr}^{(2)}$
\begin{align}
&\log Y_1(v) = - \sum_{j=1}^N\(  \log S_{\sl(2)}^{1_*1}(u_j,v)
 - 2 \log S\star K^{11}_{vwx} (u_j^-,v) \)  - 2\sum_{j=1}^2  \log S\hstar K_{y1}(r_j^-,v)
\notag\\
&\quad - L\, \tH_{1}
+ \log \left(1+Y_{Q} \right) \star \(K_{\sl(2)}^{Q1} + 2 \, s \star K^{Q-1,1}_{vwx} \)
+ 2 \log \(1 + Y_{1|vw}\) \star s \hstar K_{y1}
\notag \\
&\quad - 2  \log{1-Y_-\ov 1-Y_+} \hstar s \star K^{11}_{vwx}
+  \log {1- \frac{1}{Y_-} \ov 1- \frac{1}{Y_+} } \hstar K_{1} +  \log \big(1- \frac{1}{Y_-}\big)\big( 1- \frac{1}{Y_+} \big) \hstar K_{y1} \,.
\label{TbaQsl2H2}
\end{align}
Note also that for $g>\bar g_{cr}^{(1)}$ one can also use
 the p.v. prescription in the terms $\log S\hstar K_{y1}(r_j^-,v)
$  and $\log \big(1- \frac{1}{Y_-}\big)\big( 1- \frac{1}{Y_+} \big) \hstar K_{y1}$ because the extra terms $\pm \log S_{y1}$ cancel each other.

\subsubsection*{ Exact Bethe equations: $g_{cr}^{(1)}<g< g_{cr}^{(2)}$}

The exact Bethe equations are obtained by analytically continuing $\log Y_1$ in \eqref{TbaQsl2H2} following the same route as for the small $g$ case, and they take the following form
\begin{align}
&\pi i(2n_k+1)=\log Y_{1_*}(u_k) =i L\, p_k- \sum_{j=1}^N\, \log S_{\sl(2)}^{1_*1_*}(u_j,u_k)\label{Tba1sl2B3}\\
&\quad
 + 2 \sum_{j=1}^N\, \log {\rm Res}(S)\star K^{11_*}_{vwx} (u_j^-,u_k) -2 \sum_{j=1}^N\log\big(u_j-u_k-{2i\ov g}\big)\,
{x_j^--{1\ov x_{k}^-}\ov x_j^-- x_{k}^+}
\notag\\
&\qquad\qquad\quad  - 2\sum_{j=1}^2 \( \log S\hstar K_{y1_*}(r_j^-,u_k)- \log S(r_j-u_k)\) \notag\\
&\quad
+ \log \left(1+Y_{Q} \right) \star \(K_{\sl(2)}^{Q1_*} + 2 \, s \star K^{Q-1,1_*}_{vwx} \)+ 2 \log \(1 + Y_{1|vw}\) \star \( s \hstar K_{y1_*} + \ts\)
\notag \\
&\quad - 2  \log{1-Y_-\ov 1-Y_+} \hstar s \star K^{11_*}_{vwx}
+  \log {1- \frac{1}{Y_-} \ov 1- \frac{1}{Y_+} } \hstar K_{1} +  \log \big(1- \frac{1}{Y_-}\big)\big( 1- \frac{1}{Y_+} \big) \hstar K_{y1_*} \,,
\notag
\end{align}
We recall that in the formulae above the integration  contours run
a little bit above the Bethe roots $u_j$, and below the dynamical
roots $r_j$. The exact Bethe equation also have the same form no
matter whether $g<\bar g_{cr}^{(1)}$ or $g> \bar g_{cr}^{(1)}$.

We conclude again that the consistency with the BY equations requires
fulfillment of identities similar to \eqref{Rkid}
that we have checked numerically.

\subsubsection*{ Integral equations for the roots $r_j$}

In the first critical region the exact Bethe equations should be
supplemented by integral equations which determine the exact
location of the roots $r_j$. Let us recall that at
$r_j^\pm=r_j\pm{i\ov g}$ the function $Y_{1|vw}$ satisfies the
relations
\bea \la{y1vwm1}
Y_{1|vw}(r_j^\pm) = -1\,.
\eea
These
equations  can be used to find the roots $r_j$. They can be
brought to an integral form by analytically continuing the
simplified TBA equation \eqref{Yforvw3c1} for  $Y_{1|vw}$ to the
points $r_j^\pm$. The analytic continuation is straightforward,
and one gets for roots with non-positive imaginary parts
\bea \la{Yforvwc1}
\prod_{k=1}^2 S(u_k-r_j)\exp\( \log(1 +
Y_{2|vw})\star \ts+\log{1-Y_-\ov 1-Y_+}\hstar \ts- \log(1 +
Y_{2})\star \ts\)=-1\,,~~~~~
\eea
where we used the exponential
form of the TBA equation, and took into account that
$Y_{2|vw}(r_j)=Y_{2}(r_j)=Y_{\pm}(r_j)=0$. One can easily check
that if $r_j$ is imaginary, that is $g< \bar g_{cr}^{(1)}$, then
the product of S-matrices in \eqref{Yforvwc1} is negative, while
the exponent is real. As a result, taking the logarithm of
\eqref{Yforvwc1}  does not produce any new mode number. If $r_j$
is real, that is $g> \bar g_{cr}^{(1)}$, then both factors in
\eqref{Yforvwc1} are on the unit circle. Their phases however are
small (as they should be due to the continuity of the roots as
functions of $g$), and one does not get any mode number either.

 Let us also mention that eq.\eqref{Yforvwc1}
 can be also used to find the critical values of $g$.
If one sets $r_j=-i/g$,  one gets an integral equation for $g_{cr}^{(1)}$,
and the case of $r_j=0$ gives the one for $\bar g_{cr}^{(1)}$.
The equation \eqref{Yforvwc1} is better for numerical computation than \eqref{y1vwm1}, because all the Y-functions are evaluated on the mirror real axis.

%%%%%%%%%%%%%%%%%%%%%%%%%%%%%%%%%%%%%%%%%%
 \subsection{Excited states TBA equations: $g_{cr}^{(m)}<g<g_{cr}^{(m+1)}$}

In this subsection we propose the TBA equations for Konishi-like states for the general case where $g_{cr}^{(m)}<g<g_{cr}^{(m+1)}$ and $m\ge 2$.
Then, the functions $Y_{1|vw}\,,\ldots\, ,Y_{m-1|vw}$ have four  zeros, and $Y_{m|vw}$ and $Y_{m+1|vw}$  have two  zeros  in the physical strip. The only two roots that can be imaginary if $g<\bar g_{cr}^{(m)}$  are $r_j^{(m+1)}$.  As was discussed in section 3, these  zeros can be written in the form \eqref{zer}
\bea
\{ u_j\,,r_j^{(3)} \} \,,\ \{ r_j^{(2)}\,,r_j^{(4)} \}
\,,\ldots\,, \{ r_j^{(m-1)}\,,r_j^{(m+1)} \}\,,
\{ r_j^{(m)} \} \,,
\{ r_j^{(m+1)} \} \,,
\eea
where we indicate only those  zeros which are in the physical strip and important for formulating the TBA equations. We also recall that
the functions below  have  zeros  at locations related to $r_j^{(k)}$
\bea
Y_{\pm}\big(r_j^{(2)}\big) = 0\,,\ \ 1+Y_{k|vw}\big(r_j^{(k+1)}\pm{i\ov g}\big) = 0\,,\ \ Y_{k+1}\big(r_j^{(k+1)}\big) = 0\,,\quad k=1\,,\ldots\,, m\,.~~~~~~
\eea

\subsubsection*{ Simplified TBA equations: $g_{cr}^{(m)}<g<g_{cr}^{(m+1)}$}

The necessary modification of the integration contour is obvious, and  simplified equations take the following form

\bigskip
 \noindent
$\bullet$\ $M|w$-strings:  their equations coincide with the ground state ones \eqref{Yforws}.

%$\ M\ge 1\ $, $Y_{0|w}=0$
%\bea\la{Yforwscm}
%\log Y_{M|w}=  \log(1 +  Y_{M-1|w})(1 +  Y_{M+1|w})\star s
%+\delta_{M1}\, \log{1-{1\ov Y_-}\ov 1-{1\ov Y_+} }\hstar s\,~~~~~
%\eea
\bigskip
 \noindent
$\bullet$\ $M|vw$-strings: $\ M\ge 1\ $, $Y_{0|vw}=0$
\bea\la{Yforvw3cm}
&&\hspace{-0.3cm}\log Y_{M|vw}(v)=-\sum_{j=1}^2\left[ \log S(r_j^{(M),-}-v)+\log S(r_j^{(M+2),-}-v)\right]~~~~~\\\nonumber
&&+ \log(1 +  Y_{M-1|vw} )(1 +  Y_{M+1|vw})\star s+\delta_{M1}  \log{1-Y_-\ov 1-Y_+}\hstar s- \log(1 +  Y_{M+1})\star s\,,~~~~~
\eea
where we identify $r_j^{(1)}\equiv u_j$, and assume that $\log S(r_j^{(k),-}-v)$ is absent in the sum on the first line if
$k\ge m+2$. For $M=1$
the term $\log S(u_j-v)$ is due to the pole of $Y_+$ at $u=u_j^-$, and  the second term is due to the zero of $1 +  Y_{2|vw}$ at $u=r_j^{(3),-}$. For $M\ge 2$ both terms on the first line are due to
the  zeros of $1 +  Y_{k|vw}$ at $u=r_j^{(k+1),-}$ for $k=1,\ldots, m$.

\bigskip
 \noindent
$\bullet$\   $y$-particles:  their equations coincide with eqs.\eqref{Yfory1sc1}and \eqref{Yfory2sc2} for the first critical region
$g_{cr}^{(1)}<g<g_{cr}^{(2)}$.

%\bea\la{Yfory1scm}
%\log {Y_+\ov Y_-}(v)&=& -\sum_{j=1}^N \log S_{1_*y}(u_{j},v)  +\log(1 +  Y_{Q})\star K_{Qy}\,,~~~~~~~
% \\
%\nonumber
%\\\la{Yfory2scm}
%\log {Y_+ Y_-}(v) &=& -\sum_{j=1}^N\,
% \log {\big(S_{xv}^{1_*1}\big)^2\ov S_2}\star s(u_j,v)-2\sum_{j=1}^2 \log S(r_j^{(2)-}-v)
%\\\nonumber
%&+&2\log{1 +  Y_{1|vw} \ov 1 +  Y_{1|w} }\star s
%- \log\left(1+Y_Q \right)\star K_Q+ 2 \log(1 +  Y_{Q})\star K_{xv}^{Q1} \star s
%\,.~~~~
%\eea
%It is worth noting that these equations are the same as they are for the first critical region $g_{cr}^{(1)}<g<g_{cr}^{(2)}$.

\bigskip
 \noindent
$\bullet$\  $Q$-particles for $Q\ge 3$
\bea
\log Y_{Q}&=&-2\sum_{j=1}^2\, \log S(r_j^{(Q),-}-v) + \log{\left(1 +  {1\ov Y_{Q-1|vw}} \right)^2\ov (1 +  {1\ov Y_{Q-1} })(1 +  {1\ov Y_{Q+1} }) }\star_{p.v.} s\la{YforQ2cm}
\,.~~~~~~~
\eea
The p.v. prescription is again not really needed because  the four  zeros $\{ r_j^{(Q-1)}\,,r_j^{(Q+1)} \}$ of $Y_{Q-1|vw}$ are cancelled by the double  zeros of $Y_{Q-1}$ and $Y_{Q+1}$, and the formula works no matter if the roots $r_j^{(m+1)}$ are real or imaginary.

The equations below coincide with the corresponding eqs.\eqref{YforQ2a4c1},\eqref{YforQ1a5},\eqref{TbaQsl2H2} and \eqref{Tba1sl2B3} for the first critical region
$g_{cr}^{(1)}<g<g_{cr}^{(2)}$ with the replacement $r_j \mapsto r_j^{(2)}$.

\medskip
 \noindent
$\bullet$\  Equations for $Q=2$ and $Q=1$-particles.

\smallskip

 \noindent
$\bullet$\  Hybrid $Q=1$ equation.

\smallskip

 \noindent
$\bullet$\  Exact Bethe equations.

\smallskip

 According to these results, only some of the simplified equations for $Q$-particles and $vw$-strings change their form when crossing the $m$-th critical point $(m>1)$.
However, since all Y-functions are coupled with
each other, the existence of higher critical points still affects
the  equations above but in a less direct way.

\subsubsection*{ Integral equations for the roots $r_j^{(k)}$: $g_{cr}^{(m)}<g<g_{cr}^{(m+1)}$}

In this region the exact Bethe equations should be supplemented by integral equations which determine the exact location of the roots $r_j^{(k)}$, $k=2,\ldots, m+1$. The integral equations are obtained by
analytically continuing the simplified TBA equations \eqref{Yforvw3cm} for  $Y_{k|vw}$ to
the points $r_j^{(k+1)\pm}$, and using the conditions
\bea\la{yvwmk}
Y_{k|vw}(r_j^{(k+1)\pm}) = -1\,.
\eea
The analytic continuation is again straightforward, and one gets for roots with nonpositive imaginary parts the following equations
\bea\la{Yforvwc1m}
r_j^{(2)}:&&\quad \prod_{n=1}^2 S(u_n-r_j^{(2)})S(r_n^{(3)}-r_j^{(2)})\times\\\nonumber
&&\qquad\quad \times\exp\(\,
\log(1 +  Y_{2|vw})\star \ts+\log{1-Y_-\ov 1-Y_+}\hstar \ts- \log(1 +  Y_{2})\star \ts\, \)=-1,\\
r_j^{(k+1)}:&&\quad \prod_{n=1}^2 S(r_n^{(k)}-r_j^{(k+1)})S(r_n^{(k+2)}-r_j^{(k+1)})\times \la{Yforvwckm}
\\\nonumber
&&\qquad\quad \times
\exp\(\,
\log(1 +  Y_{k-1|vw})(1 +  Y_{k+1|vw})\star \ts- \log(1 +  Y_{k+1})\star \ts\, \)=-1,
\eea
The continuity condition of the roots as functions of $g$ again guarantees that one does not get any new mode numbers.

\subsection{Further evidence for critical values of $g$}

As was mentioned in the previous section the exact location of the critical values can be found only by solving the TBA and exact Bethe equations. One may wonder if the critical values can be absent at all due to the contribution of $Y_Q$ functions.
Indeed, $Y_Q$-functions at $g\approx g_{cr}$ are not small as one can see from Figure 8.
\begin{figure}[t]
\begin{center}
\includegraphics*[width=0.48\textwidth]{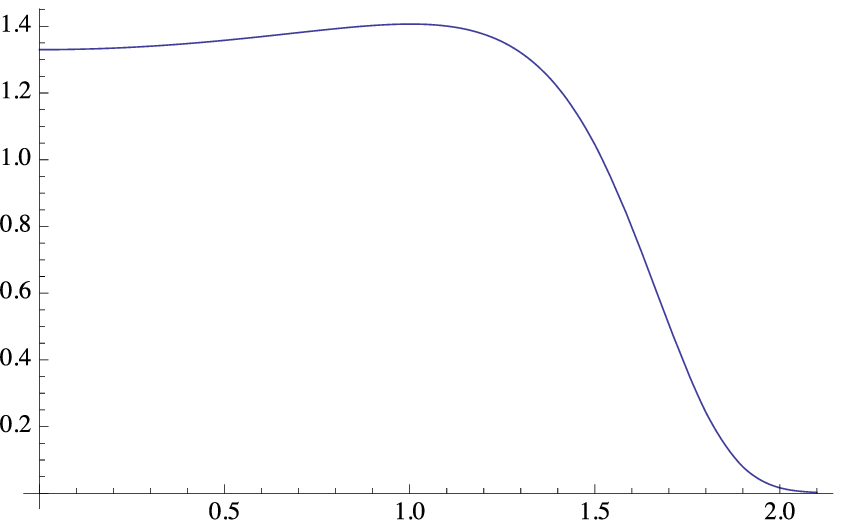}\quad
\includegraphics*[width=0.48\textwidth]{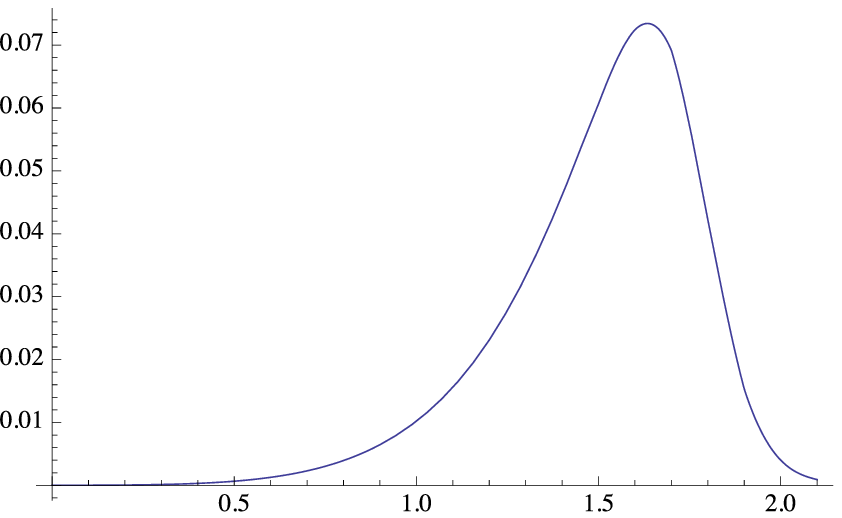}
\end{center}
\caption{In the left  and right pictures the profiles of the asymptotic
$Y_{1}$ and $Y_{2}$ functions for the Konishi state at $g=g_{cr}^{(1)}=4.429$ are depicted, respectively. }
\end{figure}
However, their contribution to the quantities of interest is still
small. To illustrate this,  in Figure 9 we present the plots of
asymptotic $Y_\pm$, $Y_{1|vw}$ and $Y_\pm$, $Y_{1|vw}$ (only for
$u>0$ because all Y-functions are even) obtained after the first
iteration of the simplified TBA equations by taking into account
the contribution of the first eight $Y_Q$-functions for
$g=4.38671$, $\lambda = 759.691$.
\begin{figure}[t]
\begin{center}
\hspace{-0.40cm}\includegraphics*[width=0.32\textwidth]{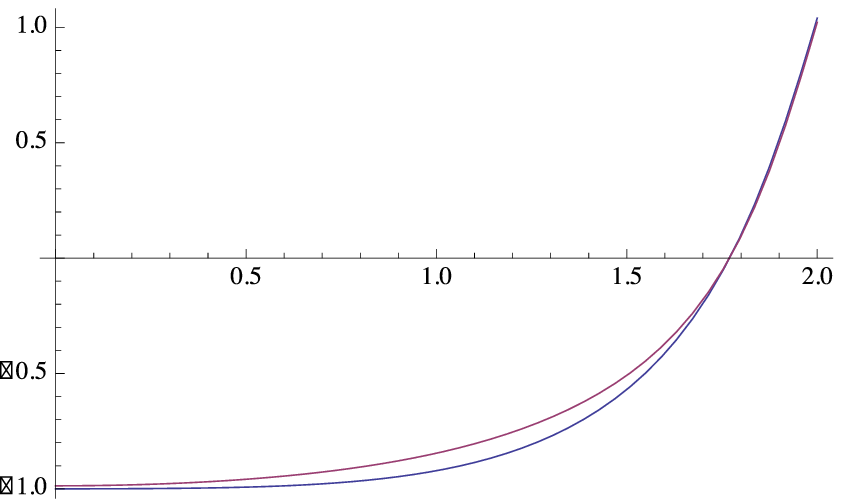}\quad
\includegraphics*[width=0.32\textwidth]{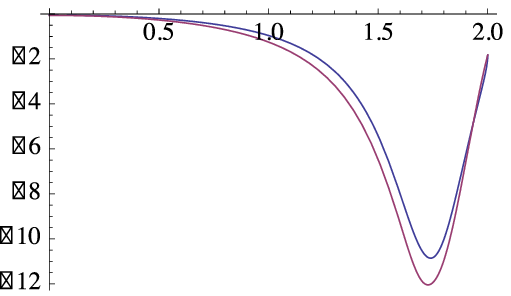}
\
\includegraphics*[width=0.32\textwidth]{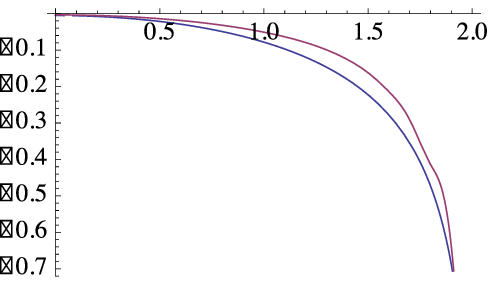}
\end{center}
\caption{In the left picture the profiles of the asymptotic
$Y_{1|vw}$ (blue) and $Y_{1|vw}$  after the first iteration
(purple) are depicted. In the middle and right pictures similar
profiles for $Y_+$ and $Y_-$ are presented. }
\end{figure}
The blue curves are plots of the asymptotic Y-functions. As we
discussed in this section, at the first critical value
$Y_{1|vw}(0)=-1$, and at the first subcritical value
$Y_{\pm}(0)=0$.  One can see from the plots that the influence of
$Y_Q$-functions at $u \approx 0$ is extremely small\footnote{Note,
that for $g$ far enough from the critical value, e.g. $g = 2$, the
contribution of $Y_Q$-functions around $u=0$ results in a more
visible change.}. This is of course expected because the finite
contribution of $Y_Q$-functions cannot balance the infinities
originating from the zeroes of Y-functions. We also see from the
plot of $Y_{1|vw}$ that there is a tendency for the actual
critical value to be higher than the asymptotic one.

To show that  the zeroes of Y-functions unavoidably enter the
physical strip when the coupling increases, we also provide a plot of
asymptotic $Y_Q$-functions for large value of $\lambda=10^8$, see
Figure 10. One can see that the functions are very small for
almost all values of $u$ except those close to $\pm2$. Thus, at
strong coupling all Y-functions are well-approximated around $u=0$
by their asymptotic value. On the other hand, if the zeroes are
outside the physical strip then they are purely imaginary,  and
their positions can be determined by analytically continuing the
corresponding TBA equations, see e.g. eq.\eqref{Yforvwc1}.
Assuming the roots $r_j$ are imaginary, at large $g$ the main
contribution of the TBA kernel $\tilde s$ appearing in \eqref{Yforvwc1}
originates from the region around $u=0$. In this region, as was
explained above, it is legitimate to use the asymptotic solution
which implies that the roots $r_j$ are real and close to $\pm2$.
This leads to a contradiction with the assumption that the roots
$r_j$ are imaginary at sufficiently large $g$. Thus, one concludes that for large $g$ the roots are real and close to $\pm2$.

\begin{figure}[t]
\begin{center}
\includegraphics*[width=0.99\textwidth]{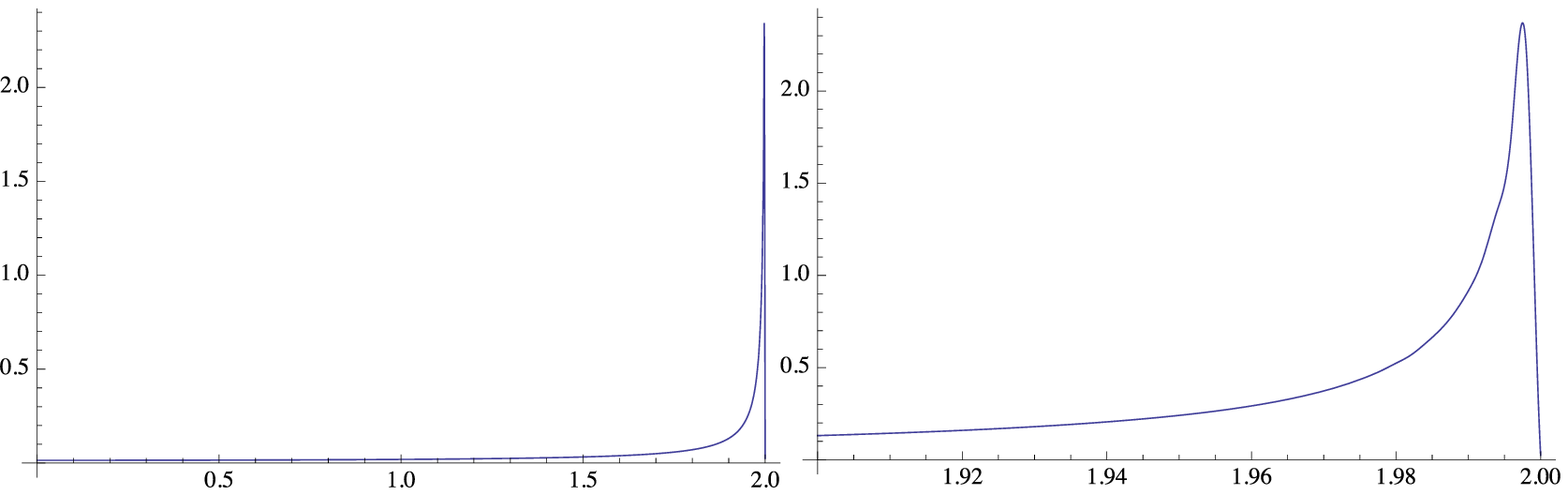}\\[2mm]
\hspace{-0.40cm}\includegraphics*[width=0.99\textwidth]{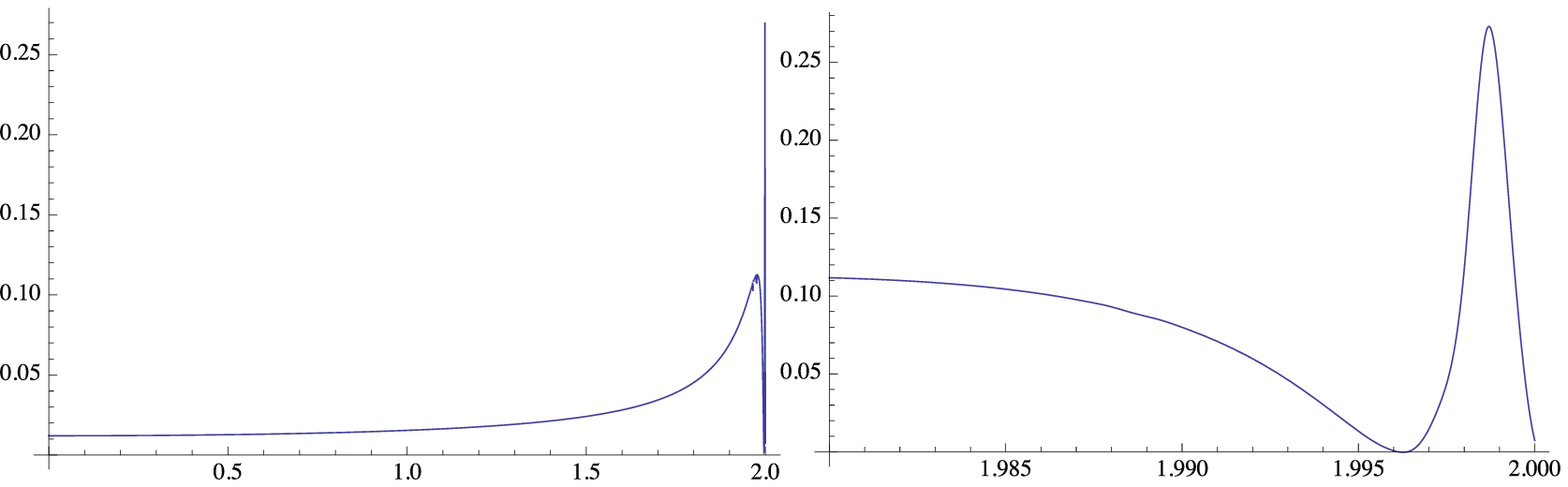}
%\\[2mm]
%\hspace{-0.40cm}\includegraphics*[width=0.99\textwidth]{YoQ3}
%\\[2mm]
%\hspace{-0.40cm}\includegraphics*[width=0.99\textwidth]{YoQ4}
\end{center}
\caption{Profiles of the asymptotic $Y_{Q}$,
$Q=1,2$  are depicted for $\lambda=10^8$. }
\end{figure}

%%%%%%%%%%%%%%%%%%%%%%%%%%%%%%%%%
\section{TBA equations for arbitrary two-particle $\sl(2)$ states}

Our consideration of the TBA equations for
  Konishi-like states can be easily used to formulate TBA equations for an arbitrary two-particle state from the $\sl(2)$ sector. One starts again with analyzing the structure of  zeros of $Y_{M|vw}$-functions at weak coupling.  Then, as was discussed in section 3, for every state $(J,n)$ there is a number $m$ such that the first $m-1$ $Y_{k|vw}$-functions have four   zeros,
   both  $Y_{m|vw}$ and $Y_{m+1|vw}$ have two  zeros, and  all $Y_{k|vw}$-functions with $k\ge m+2$ have no  zeros
 in the physical strip. This is exactly the structure of  zeros  of $Y_{M|vw}$-functions for
 a Konishi-like state with the coupling constant being in the $m$-th critical region: $g_{cr}^{(m)}<g<  g_{cr}^{(m+1)}$.

\smallskip

Thus, at weak coupling the simplified TBA equations  for this
$(J,n)$ state have exactly the same form as the TBA equations for
a Konishi-like state in the $m$-th critical region. As to the
canonical TBA and exact Bethe equations, one should take into
account
 that real  zeros $r_j^{(k)}$ of $Y_{k|vw}$-functions  are located outside
  the interval $[-2,2]$ for small $g$ for most of the $\sl(2)$ states. In particular, if the  zeros $r_j^{(2)}$
 of $Y_{2|vw}$ are outside the interval then the corresponding  zeros $r_j^{(2)-}$
 of $1+Y_{1|vw}$ are located on the cuts of  $S_{vwx}^{1Q}(r_j^{(2)-},v)$, and
 since the integration contour runs below these  zeros, one should add $+i0$ to them in all the expressions. In addition, the integration contour for $Y_\pm$-functions runs over the interval $[-2,2]$ (and a little bit above it), and if $|r_j^{(2)}|>2$ then there is no singularity of the integrand and the p.v. prescription is not necessary. Strictly speaking, this means that in addition to the critical and subcritical values of $g$ we discussed above, one should also consider such values of $g$ that
the  zeros $r_j^{(2)}$ of $Y_{2|vw}$ are equal to $\pm2$, and
distinguish the  cases with the  zeros being inside and outside
the interval $[-2,2]$. It is straightforward to do it, and we
refrain from presenting explicit formulae for these cases here.

\smallskip

Increasing $g$, one reaches the first critical  value
$g_{J,n}^{1,m+1}$ of the $(J,n)$ state, and the TBA equations
would have to be modified. It is in fact clear that in the region
$g_{J,n}^{r,m+1}<g <g_{J,n}^{r+1,m+1}$ the TBA equations for the
$(J,n)$ state coincide with the ones for a Konishi-like state in
the $m+r$-th critical region.

\section{Remarks on the Y-system}
For reader's convenience, in this section we summarize  the
most essential properties of the Y-system implied by the TBA
equations under study.

\smallskip

The kernel $s$ has the uniquely defined inverse $s^{-1}$ which
acts as the following operator
$$
(f\star s^{-1})(u)=\lim_{\epsilon\to
0^+}\big[f(u+\frac{i}{g}-i\epsilon)+f(u-\frac{i}{g}+i\epsilon)\big]\,
.
$$
Applying $s^{-1}$ to the set of the simplified TBA equations, one
can easily derive the Y-system equations \cite{GKV09}  for
$-2<u<2$. Since the Y-functions appear to be non-analytic, for
$|u|>2$ (here and in what follows we mean
$u \in (-\infty,-2) \cup (2, \infty)$ by $|u|>2$)
one gets instead new equations which, as we will explain below, 
encode the jump discontinuities of the Y-functions across
the cuts \cite{AF09b}. To exemplify this statement, it is enough
to consider $Y_{1|w}$ and $Y_1$-functions.

\smallskip

We start with eq.\eqref{Yforws}  for $Y_{1|w}$. Applying $s^{-1}$ to this equation, one finds
\begin{align}
Y_{1|w}^{(\a)}\big(u+\frac{i}{g}-i0\big)Y_{1|w}^{(\a)}\big(u-\frac{i}{g}+i0\big)&=
\Big(1+Y_{2|w}^{(\a)}\Big)\frac{1-\frac{1}{Y_-^{(\a)}}}
{1-\frac{1}{Y_+^{(\a)}}}(u)\, , ~~~|u|<2
\label{TBAforY} \\
Y_{1|w}^{(\a)}\big(u+\frac{i}{g}-i0\big)Y_{1|w}^{(\a)}\big(u-\frac{i}{g}+i0\big)&=
1+Y_{2|w}^{(\a)}(u)\, , ~~~~~~~~~~~~~~~~~~|u|>2\, .
\label{TBAforY outside}
\end{align}
We stress that these equations are unambiguously {\it derived}
from the TBA equation and the $\epsilon$-prescription on the left
hand side is fixed by that of $s^{-1}$. We recall that in the TBA
equations the functions $Y_{\pm}$ have their support on $[-2,2]$.

\smallskip

The TBA equation  \eqref{Yforws} for $Y_{1|w}$  shows that this
function has branch points located at $u=\pm 2\pm \frac{i}{g}$.
Since we are dealing with the mirror $u$-plane, it is natural to
choose the cuts to run from $\pm \infty$ to $\pm 2\pm \frac{i}{g}$
parallel to the real axis.  In the same way the TBA equations show
that various Y-functions have branch points located at $\pm 2\pm
\frac{i}{g}Q$, $Q=0,1,2,\ldots,\infty$, and, therefore, all the
cuts of all the Y-functions can be chosen to be outside  the strip
$|{\rm Re}\,u|<2$, and running parallel to the real axis. Then,
Y-functions are analytic in the strip $|{\rm Re}\,u|<2$, in
eq.(\ref{TBAforY}) the $\epsilon$-prescription can be dropped, and
$u$ can be considered as a complex variable taking values in the
strip \bea\label{TBAforY1}
Y_{1|w}^{(\a)}\big(u+\frac{i}{g}\big)Y_{1|w}^{(\a)}\big(u-\frac{i}{g}\big)=
\Big(1+Y_{2|w}^{(\a)}\Big)\frac{1-\frac{1}{Y_-^{(\a)}}}
{1-\frac{1}{Y_+^{(\a)}}}(u)\, , ~~~|{\rm Re}\, u|<2\, . \eea Thus,
we see that with this (and only this) choice of the cuts the
Y-system takes its standard form.

In fact,  analytically continuing Y-functions outside the
strip $|{\rm Re}\, u|<2$, one concludes that eq.\eqref{TBAforY1} is valid for all complex values of $u$ but those which belong to the cuts.
Approaching then the real axis for $|u|>2$, say, from above, one
arrives at the following prescription \cite{FS} \footnote{
Naively one could try to define a new ``inverse'' operator
$s^{-1}_{\small G}$\,,
% which is not inverse to $s$, by

$
(f\star s^{-1}_{\small G})(u)=\lim_{\epsilon\to
0^+}\big[f(u+\frac{i}{g}+i\epsilon)+f(u-\frac{i}{g}+i\epsilon)\big]\,,
$
and claim that applying it to the simplified TBA equation, one gets eq.\eqref{TBAforY1b} for all values of $u$. The problem with this is that the operator $s^{-1}_{\small G}$ is not inverse to $s$. In fact it annihilates $s$: $s\star s^{-1}_{\small G}=0$.
}
\bea\label{TBAforY1b}
Y_{1|w}^{(\a)}\big(u+\frac{i}{g}+i0\big)Y_{1|w}^{(\a)}\big(u-\frac{i}{g}+i0\big)=
\Big(1+Y_{2|w}^{(\a)}\Big)\frac{1-\frac{1}{Y_-^{(\a)}}}
{1-\frac{1}{Y_+^{(\a)}}}(u+i0)\, , ~~ |u|>2\, .
\eea
Such an {\it analytically continued} equation can be used to determine the jump discontinuities of the Y-functions across the cut. From the compatibility of the equation \eqref{TBAforY1b} with eq.\eqref{TBAforY outside}, we find that
\bea
\frac{Y_{1|w}^{(\a)}\big(u+\frac{i}{g}+i0\big)}{Y_{1|w}^{(\a)}\big(u+\frac{i}{g}-i0\big)}=
\frac{1-\frac{1}{Y_-^{(\a)}}}
{1-\frac{1}{Y_+^{(\a)}}}(u+i0),~~~~~~~|u|>2\,,
\eea
and analogously
\bea
\frac{Y_{1|w}^{(\a)}\big(u+\frac{i}{g}-i0\big)}{Y_{1|w}^{(\a)}\big(u+\frac{i}{g}+i0\big)}=
\frac{1-\frac{1}{Y_-^{(\a)}}}
{1-\frac{1}{Y_+^{(\a)}}}(u-i0),~~~~~~~|u|>2\, .
\eea
We also find that these discontinuity equations are consistent with the relation (\ref{ypym}) which follows from the TBA equations for $Y_{\pm}$.

\smallskip

Thus, the TBA
equations tell us that, because of the absence of analyticity, the
 Y-system equations must be supplemented by proper jump
discontinuity conditions. This was one of the important
observations made in \cite{AF09b}.

\smallskip

Finally, we recall that for $Y_1$ one finds
\bea\begin{aligned}
Y_{1}\big(u+\frac{i}{g}-i0\big)Y_{1}\big(u-\frac{i}{g}+i0\big)&=
\frac{\Big(1-\frac{1}{Y_-^{(1)}}\Big)\Big(1-\frac{1}{Y_-^{(2)}}\Big)}{1+\frac{1}{Y_2}}\,
, ~~~|u|<2 \\
Y_{1}\big(u+\frac{i}{g}-i0\big)Y_{1}\big(u-\frac{i}{g}+i0\big)&=\frac{e^{-\Delta}}{1+\frac{1}{Y_2}}\,
, ~~~~~~~~~~~~~~~~~~~~~~|u|>2\,,
\end{aligned}\eea where the
explicit form of the quantity $\Delta$ is given in \cite{AF09b,AF09d} for the ground state case, and  in the excited state case it can be extracted from eq.\eqref{YforQ1a5} or obtained from the ground state one by using the contour deformation trick.
Once again, the second equation here determines the jump
discontinuity of $Y_1$ across the cut. However, a new feature
here is that the jump discontinuity, that is $\Delta$, does depend
on a state under consideration!

\smallskip

To summarize, the Y-system exhibits the following properties
dictated by the underlying TBA equations
\begin{itemize}
\item The Y-system is not analytic on the $u$-plane and for this
reason it must be supplemented by jump discontinuity conditions;

\item In general, jump discontinuities depend on a
state of interest and they can be only fixed by the TBA equations;

\item Different Y-functions have different cut structure. In total
there are infinitely many cuts on the mirror theory $u$-plane with
the branch points located at $\pm 2\pm \frac{i}{g}Q$,
$Q=0,1,2,\ldots,\infty$. As a result, the Y-system lives on a
Riemann surface of infinite genus.

\end{itemize}

Needless to say, these intricate analyticity properties discovered
in \cite{AF09b,FS} render the AdS/CFT Y-system rather
different from its known relativistic cousins and, for this
reason, make it much harder to solve.

%%%%%%%%%%%%%%%%%%%%%%%%%%%%%%%%%%%%%%%

\section{Conclusions}

In this paper we have analyzed the TBA equations  for the $\AdS$
mirror model. We provided an evidence that for any excited string
state and, therefore, for any ${\cal N}=4$ SYM operator there
could be infinitely many critical values of 't Hooft's  coupling
constant at which the TBA equations have to be modified. It is a
demanding but also a rather challenging problem to locate the
exact position of the critical points and it is similar in spirit
to determination of the exact positions of Bethe roots. At the
same time, we also want to give a word of caution. Our approach is
based on the optimistic assumption that the analytic structure of
the exact TBA solution emulates the one of the large $L$
asymptotic solution. The possibility that the exact solution might
develop new singularities in comparison to the asymptotic one
would lead to even more complicated scenario than we described
here.

\smallskip

One could also speculate about the physical origin of critical
points. It is known that the $\sl(2)$ sector is closed to any
order of perturbation theory. One possibility would be that the
first critical value of $g$ of an $\sl(2)$ state is the one where
this state begins to mix with states from other sectors of the
theory. It would be interesting to understand if this is indeed
the case.

\smallskip

The TBA equations and the contour deformation trick  we have
formulated allow one to discuss many interesting problems, and we
list some of them below.

\smallskip

Concerning the issue of critical points, it would be very
interesting to analyze the TBA equations,
 and compute numerically the scaling dimension of the Konishi operator
  in the vicinity of and beyond  a critical point.
The simplified TBA equations seem to be much better suited for
such an analysis than the canonical ones.

\smallskip

Then, one should solve analytically the TBA equations for
two-particle states  at large $g$.  The large $g$ expansion should
contain no $\log g$ which follow from the BY equations
\cite{Bec,RS09}. It should also fix the coefficient of the
subleading $1/g$ term in the energy expansion. There are currently
two different predictions for the coefficient \cite{AF05,RT09k}
obtained by using some string theory methods and relaying on
certain assumptions. Thus, it is important to perform a rigorous
string theory computation of this coefficient.

\smallskip

Recently, the TBA approach has been applied to obtain the 5-loop
anomalous dimension of the Konishi operator by a combination of analytical and numerical means
\cite{Arutyunov:2010gb} and the corresponding result was found to
be in a perfect agreement with the one based on the generalization
of L\"uscher's formulae \cite{BJ09}. This constitutes an important
test of the TBA equations we propose in this paper (the hybrid
equations). It would be nice to support this numerical agreement
by an analytic proof.

\smallskip

The TBA equations we have proposed are not valid for the
two-particle $(J,{J+1\ov 2})$ state. In the  semi-classical string
limit $g\to\infty$ and $J/g$ fixed it should correspond to the
folded string rotating in S$^2$ \cite{GKP02}. It would be
interesting to analyze these states along the lines of our paper,
write TBA equations, and solve them.

\smallskip

We have discussed only two-particle states.  It is certainly
important to generalize our analysis to arbitrary $N$-particle
$\sl(2)$ states. For $J=2$ the lowest energy  $N$-particle state
is dual to the twist-two operator that plays an important role in
field theory, see e.g. \cite{Kotikov:2007cy}.

\smallskip

It would  be also of interest to
 consider  the one-particle case at large $g$ and finite $J/g$. Its energy should match the string theory result \cite{AFZmag}.

\smallskip

Let us finally mention that one should also consider other sectors
and exhibit the $\psu(2,2|4)$ invariance of the spectrum.

%%%%%%%%%%%%%%%%%%%%%%%%%%%%%%%%%%%%%%%
\section*{Acknowledgements}
The work of G.~A. was supported in part by the RFBR grant
08-01-00281-a, by the grant NSh-672.2006.1, by NWO grant 047017015
and by the INTAS contract 03-51-6346. The work of S.F. was
supported in part by the Science Foundation Ireland under Grants
No. 07/RFP/PHYF104 and 09/RFP/PHY2142. The work of R.S. was
supported  by the Science Foundation Ireland under Grants
No. 07/RFP/PHYF104.

\section{Appendices}

\subsection{Kinematical variables, kernels and S-matrices}\label{app:rapidity}

All kernels and S-matrices we are using are expressed in terms of the function
$x(u)$
 \bea\la{basicx}
x(u)=\frac{1}{2}(u-i\sqrt{4-u^2}), ~~~~{\rm Im}\, x(u)<0\, , \eea
which maps the $u$-plane with the cuts $[-\infty, -2]\cup [2,\infty]$ onto the physical region of the mirror theory,
and the function $x_s(u)$
\bea\la{stringx}
x_s(u)={u\ov 2}\Big(1 + \sqrt{1-{4\ov u^2}}\Big)\,,\quad |x_s(u)|\ge 1\,,
\eea
which maps the $u$-plane with the cut $[-2,2]$ onto the physical region of the string theory.

The momentum $\tilde{p}^Q$ and the energy $\tilde{\cal{E}}_Q$ of a
mirror $Q$-particle are expressed in terms of $x(u)$ as follows
\bea \tp_Q=g x\big(u-\frac{i}{g}Q\big)-g
x\big(u+\frac{i}{g}Q\big)+i Q\, ,
~~~~~\tilde{\cal{E}}_Q=\log\frac{x\big(u-\frac{i}{g}Q\big)}{x\big(u+\frac{i}{g}Q\big)}\,
. \eea

The kernels act from the right, and the three types of star operations used in this paper are defined as follows
\bea\nonumber
&&f\star K(v) \equiv \int_{-\infty}^\infty\, du\, f(u) \, K(u,v)\,,\quad f\hstar K(v) \equiv \int_{-2}^2\, du\, f(u) \, K(u,v)\,,\\
&&f\cstar K(v) \equiv \left(\int_{-\infty}^{-2} +\int_{2}^\infty\right)\, du\, f(u) \, K(u,v)\,.
\eea
The TBA equations discussed in this paper  involve convolutions with a number of kernels
which we specify below, see also \cite{AF09b} for more details. First, the following universal kernels appear in the TBA equations
\begin{alignat}{2}
s (u) & = \frac{1}{2 \pi i} \, \frac{d}{du} \log S(u)= {g \ov 4\cosh {\pi g u \ov 2}}\,,\quad S(u)=-\tanh[ \frac{\pi}{4}(u g - i)]\,,
\nonumber \\
K_Q (u) &= \frac{1}{2\pi i} \, \frac{d}{du} \, \log S_Q(u) = \frac{1}{\pi} \, \frac{g\, Q}{Q^2 + g^2 u^2}\,,\quad S_Q(u)= \frac{u - \frac{iQ}{g}}{u + \frac{i Q}{g}} \,, \nonumber\\
K_{MN}(u) &= \frac{1}{2\pi i} \, \frac{d}{du} \, \log S_{MN}(u)=K_{M+N}(u)+K_{N-M}(u)+2\sum_{j=1}^{M-1}K_{N-M+2j}(u)\,,\nonumber\\
S_{MN}(u) &=S_{M+N}(u)S_{N-M}(u)\prod_{j=1}^{M-1}S_{N-M+2j}(u)^2 \label{KMMp}=S_{NM}(u)\,.
\end{alignat}
Then, the kernels $K_\pm^{Qy}$ are related to the scattering matrices $S_\pm^{Qy}$ of $Q$- and $y_\pm$-particles in the usual way
\bea\nonumber
K^{Qy}_-(u,v)&=&1\ov 2\pi i}{d\ov du}\log S^{Qy}_-(u,v)\,,\quad  S^{Qy}_-(u,v) = \frac{x(u-i{Q\ov g})-x(v)}{x(u+i{Q\ov g})-x(v)} \sqrt{{\frac{x(u+i{Q\ov g})}{x(u-i{Q\ov g})}} \,,\\\nonumber
K^{Qy}_+(u,v)&=&1\ov 2\pi i}{d\ov du}\log S^{Qy}_+(u,v)\,,\quad  S^{Qy}_+(u,v) =  \frac{x(u-i{Q\ov g})-{1\ov x(v)}}{x(u+i{Q\ov g})-{1\ov x(v)}} \sqrt{{\frac{x(u+i{Q\ov g})}{x(u-i{Q\ov g})}}\,.
%\\\la{KpmQy}
\eea
These kernels can be expressed in terms of the kernel $K_Q$, and the  kernel
 \bea
 K(u,v) = \frac{1}{2 \pi i} \, \frac{d}{du} \, \log S (u,v) = \frac{1}{2 \pi i} \, \frac{ \sqrt{4-v^2}}{\sqrt{4-u^2}}\, {1\ov u-v} \,,\ \  S(u,v)=\frac{x(u) - x(v)}{x(u) - 1/x(v)}\,,~~~
\label{Kuv}
 \eea
 as follows
 \bea
 K^{Qy}_\mp(u,v)&=&{1\ov 2}\Big( K_Q(u-v) \pm  K_{Qy}(u,v)\Big)\,,
 \eea
 where  $K_{Qy}$ is given by
 \bea\la{sqy}
K_{Qy}(u,v)&=&{1\ov 2\pi i}{d\ov du}\log S_{Qy}(u,v)=K(u-\frac{i}{g}Q,v)-K(u+\frac{i}{g}Q,v)\,,\\\nonumber
 S_{Qy}(u,v) &=&{S_-^{Qy}(u,v) \ov S_+^{Qy}(u,v) }={x(u-{i\ov g}Q) - x(v)\ov x(u-{i\ov g}Q) - {1\ov x(v)}}\,{x(u+{i\ov g}Q) - {1\ov x(v)}\ov x(u+{i\ov g}Q) - x(v)}= {S(u-\frac{i}{g}Q,v)\ov S(u+\frac{i}{g}Q,v)}\, .~~~~~
\la{kqy}
\eea
The S-matrices $S^{Qy}_\pm$ and  $S_{Qy}$ (and kernels $K^{Qy}_\pm$ and  $K_{Qy}$) can be easily continued to the string region by using the substitution $x(u\pm i{Q\ov g})\to x_s(u\pm i{Q\ov g})$, and the resulting S-matrices are denoted as $S^{Q_*y}_\pm$ and  $S_{Q_*y}$. Notice that $S_{Q_*y}(-\infty,v) =1$. One can also replace $x(v)$ by $x_s(v)$ in the formulae above. Then, one gets the S-matrices and kernels which are denoted as $K_{Qy}^{ms}$ ($ms$ for mirror-string) and so on.

The following kernels and S-matrices are similar to $K_\pm^{Qy}$
\bea\nonumber
K^{yQ}_-(u,v)&=&{1\ov 2\pi i}{d\ov du}\log S^{yQ}_-(u,v)\,,\quad  S^{yQ}_-(u,v) = \frac{x(u)-x(v+i{Q\ov g})}{x(u)-x(v-i{Q\ov g})}\sqrt{\frac{x(v-i{Q\ov g})}{x(v+i{Q\ov g})}}\,,\\\nonumber
K^{yQ}_+(u,v)&=&{1\ov 2\pi i}{d\ov du}\log S^{yQ}_+(u,v)\,,\quad  S^{yQ}_+(u,v) = \frac{{1\ov x(u)}-x(v-i{Q\ov g})}{{1\ov x(u)}-x(v+i{Q\ov g})}\sqrt{
\frac{x(v+i{Q\ov g})}{x(v-i{Q\ov g})}
}\,.\eea
They satisfy the following relations
\bea\nonumber &&K^{yQ}_\pm(u,v) ={1\ov 2}\Big(K_{yQ}(u,v)\mp K_Q(u-v)\Big)\, ,\\\nonumber
 K_{yQ}(u,v)&=&{1\ov 2\pi i}{d\ov du}\log S_{yQ}(u,v)= \la{KyQuv}
K(u,v+{i\ov g}Q)-K(u,v-{i\ov g}Q)\\\nonumber
S_{yQ}(u,v) &=&{S_-^{yQ}(u,v)S_+^{yQ}(u,v) }=\frac{x(u)-x(v+i{Q\ov g})}{x(u)-x(v-i{Q\ov g})}\frac{{1\ov x(u)}-x(v-i{Q\ov g})}{{1\ov x(u)}-x(v+i{Q\ov g})} \\
&=&{S(u,v+\frac{i}{g}Q)\ov S(u,v-\frac{i}{g}Q)} {x(v-i{Q\ov g})\ov x(v+i{Q\ov g})}
\la{kyq}
\, .~~~~~~
\eea
It is worth mentioning that $S_{yQ}(\pm2,v) =1$.

Next, we define the following S-matrices and kernels
\bea\nonumber S_{xv}^{QM}(u,v) &=&
\frac{x(u-i{Q \ov g })-x(v+i{M \ov g})}{x(u+i{Q \ov g })-x(v+i{M
\ov g})}\, \frac{x(u-i{Q \ov g })-x(v-i{M \ov g})}{x(u+i{Q \ov g
})-x(v-i{M \ov g})}\, \frac{x(u+i{Q \ov g })}{x(u-i{Q \ov g
})}~~~~\\\la{Sxv} &\times
&\prod_{j=1}^{M-1}\frac{u-v-\frac{i}{g}(Q-M+2j)}{u-v+\frac{i}{g}(Q-M+2j)}\,,\\\nonumber
K_{xv}^{QM}(u,v) &=&{1\ov 2\pi i}{d\ov du}\log S_{xv}^{QM}(u,v)\,,
\eea
and
\bea\nonumber S_{vwx}^{QM}(u,v) &=&
\frac{x(u-i{Q \ov g })-x(v+i{M \ov g})}{x(u-i{Q \ov g })-x(v-i{M\ov g})}\,
 \frac{x(u+i{Q \ov g })-x(v+i{M \ov g})}{x(u+i{Q \ov g})-x(v-i{M \ov g})}\,
 \frac{x(v-i{M \ov g })}{x(v+i{M \ov g})}~~~~
\\\la{Svwx} &\times
&\prod_{j=1}^{M-1}\frac{u-v-\frac{i}{g}(M-Q+2j)}{u-v+\frac{i}{g}(M-Q+2j)}\,,\\\nonumber
K_{vwx}^{QM}(u,v) &=&{1\ov 2\pi i}{d\ov du}\log S_{vwx}^{QM}(u,v)\,.
\eea

Then, the $\sl(2)$ S-matrix $S_{\sl(2)}^{QM}$ in the uniform
light-cone gauge \cite{AFrev} with the gauge parameter $a=0$ can
be written in the form
\bea\la{Ssl2}
S_{\sl(2)}^{QM}(u,v)= S^{QM}(u-v)^{-1} \,
\Sigma_{QM}(u,v)^{-2}\,, \eea
where  $\Sigma^{QM}$ is the improved dressing factor \cite{AF09c}.
The corresponding $\sl(2)$ and dressing kernels are defined as usual
 \bea
K_{\sl(2)}^{QM}(u,v)= \frac{1}{2\pi i}\frac{d}{du}\log
S_{\sl(2)}^{QM}(u,v) \,,\quad K_{QM}^{\Sigma}(u,v)=\frac{1}{2\pi
i}\frac{d}{du}\log \Sigma_{QM}(u,v)\,.~~~~ \eea The analytically
continued $\sl(2)$ S-matrix is given by \bea\nonumber
S_{\sl(2)}^{1_*M}(u,v)&=&{1\ov
S_{1M}(u-v)\Sigma_{1_*M}(u,v)^2}\,,~~~~ \eea where the improved
dressing factor is given by \cite{AF09c} \bea\la{sigtot3}
\begin{aligned} {1\ov i}\log\Sigma_{1_*M}(u,v)
 &=
\Phi(y_1^+,y_2^+)-\Phi(y_1^+,y_2^-)-\Phi(y_1^-,y_2^+)+\Phi(y_1^-,y_2^-)
\\
&+{1\ov
2}\left(\Psi(y_{2}^+,y_1^+)+\Psi(y_{2}^-,y_1^+)-\Psi(y_{2}^+,y_1^-)
-\Psi(y_{2}^-,y_1^-) \right)~~~~~
 \\
&+\frac{1}{2i}\log\frac{(y_1^--y_2^+)\big(y_1^-
-\frac{1}{y_2^-}\big)\big(y_1^+
-\frac{1}{y_2^-}\big)}{(y_1^+-y_2^+)\big(y_1^-
-\frac{1}{y_2^+}\big)^2} \,.~~~~~
\end{aligned}\eea
Here $y_{1}^{\pm}=x_s(u\pm{i\ov g})$ are parameters
of a fundamental particle in string theory, and $y_{2}^{\pm}=x(v\pm{i\ov g}M)$ are parameters
of an $M$-particle bound state in the mirror theory, see \cite{AF09c} for details.

Next, we introduce the following kernel and S-matrix
\bea \la{bK}\nonumber
\bar{K}(u,v)= \frac{1}{2 \pi i} \, \frac{d}{du} \, \log S_{ms}(u,v)={1\ov
2\pi} \frac{\sqrt{1-{4\ov v^2}}}{\sqrt{4-u^2}}{v\ov u-v}\,,\ \ S_{ms}(u,v)=\frac{x(u) - x_s(v)}{x(u) - {1\ov x_s(v)}}\,.
\eea
With the help of this kernel we  define\footnote{The definitions of the kernels $\check{K}$ and $\check{K}_Q$ differ by the sign from the ones used in \cite{AF09b}.}
 \bea
\check{K}(u,v)&=&\bar{K}(u,v)\big[\theta(-v-2)+\theta(v-2)\big]\,,\quad
~~~~\\
 \la{ck1} \check{K}_Q (u,v)&=& \big[\bar{K}(u+{i\ov
g}Q,v) + \bar{K}(u-{i\ov g}Q,v) \big]\big[\theta(-v-2)
+\theta(v-2)\big]\, , \eea
where $\theta(u)$ is the standard unit
step function. Obviously, both $\check{K}$ and $\check{K}_Q$
vanish for $v$ being in the interval $(-2,2)$ and are equal to (twice) the jump discontinuity of the kernels ${K}$ and ${K}_{Qy}$ across the real semi-lines $|v|>2$.

We also use
\bea \la{Kss}\nonumber
K_{ss}(u,v)= \frac{1}{2 \pi i} \, \frac{d}{du} \, \log S_{ss}(u,v)={1\ov
2\pi i} \frac{\sqrt{1-{4\ov v^2}}}{\sqrt{1-{4\ov u^2}}}{v\ov (u-v)}\,,\ \ S_{ss}(u,v)=\frac{x_s(u) - x_s(v)}{x_s(u) - {1\ov x_s(v)}}\,,
\eea
and define $\check{S}_{Q}(u,v)$ as
\bea
\check{S}_{Q}(u,v) = {S_{ss}(u-{i\ov g}Q,v) \ov S_{ss}(u+{i\ov g}Q,v)}\,,\quad  \check{K}_Q (u,v)=  \frac{1}{2 \pi i} \, \frac{d}{du} \, \log \check{S}_{Q}(u,v)\,,
\eea
to ensure that $\check{S}_{Q}(-\infty,v)=1$.

The quantity
$\check{\cal E}$ is defined as
\bea\la{cEu}
\check{\cal E}(u)=\log  \frac{x (u - i0)}{x (u + i0)} =
2\log x_s(u) \neq 0 \quad {\rm for} \ \ u \in (-\infty,-2) \cup (2, \infty) \,.
\eea
The TBA equations for $Y_Q$-particles involve the
  kernel
  \bea\la{Ksig}
 {\check K}^\Sigma_{Q} = {1\ov 2\pi i} {\pa\ov \pa u} \log{\check \Sigma}_{Q}= - K_{Qy}\hstar  \check{I}_0 + \check{I}_Q
  \eea
  where
   \bea\la{checkIQ}
 &&\check{I}_Q=\sum_{n=1}^\infty
\check{K}_{2n+Q}(u,v)=K_\Gamma^{[Q+2]}(u-v)+2\int_{-2}^2 {\rm d}t
\, K_\Gamma^{[Q+2]}(u-t)\check{K}(t,v) \,,~~~~~~~\\\la{KG0}
&&K_\Gamma^{[Q]}(u)={1\ov 2\pi i} {d\ov d u} \log
\frac{\Gamma\big[{Q\ov 2}-\frac{i}{2}g u\big]}{\Gamma\big[{Q\ov
2}+\frac{i}{2}g u\big]}={g\g\ov2\pi}+ \sum_{n=1}^\infty\Big(K_{2n+Q-2}(u)-{g\ov 2\pi n}\Big)  \, .~~~~~~~~~~\eea
The kernel \eqref{Ksig} is related to the dressing kernel
$K_{QM}^{\Sigma}$ as follows
\bea\la{KSKi1}
\check{K}_{Q}^\Sigma(u,v)= K^\Sigma_{Q1}(u,v+{i\ov g}-i0)+K^\Sigma_{Q1}(u,v-{i\ov g}+i0) - K^\Sigma_{Q2}(u,v)
\,. \eea
The analytically continued kernel $\check{K}_{1_*}^\Sigma$ is given by
\bea\la{k1starsigma}
\check{K}_{1_*}^\Sigma(u,v) =- K_{1_*y}\hstar
\check{I}_0  -K_{ss}(u-{i\ov g}\,,v)\,.~~~~~~~ \eea
Finally, integrating \eqref{k1starsigma} over the first argument, one gets
\bea\la{s1star} \log\check{\Sigma}_{1_*}(u,v) =- \log S_{1_*y}\hstar
 \check{I}_0  -\log S_{ss}(u-{i\ov g},v)\,.~~~~~~~ \eea
Note that the first term here is real, but the last one is not.

%%%%%%%%%%%%%%%%%%%%%%%%%%%%%%
\subsection{Solution of the Bethe-Yang equation for the Konishi state}\la{appBY}
\subsubsection*{Perturbative solution}

The equation \eqref{BYe} can be solved in perturbative theory, and one gets up to $g^{16}$
{\smaller
\bea
p&=& \frac{2 \pi }{3}-\frac{\sqrt{3} g^2}{4}+\frac{9 \sqrt{3}g^4}{32}-\frac{3\sqrt{3}}{8} g^6 \left(
   \zeta (3)+1\right)
   +\frac{g^8 \sqrt{3}\left(960
   \zeta (3)+960\zeta (5)+671
   \right)}{1024}
   \\\nonumber
   &+&
   g^{10}\sqrt{3} \left(-\frac{141
   \zeta (3)}{64}-\frac{309  \zeta
   (5)}{128}-\frac{315 \zeta (7)}{128}-\frac{3807
  }{2560}\right)
   \\\nonumber
   &+&
   g^{12}\sqrt{3}
   \left(\frac{2799\zeta (3)}{512}+\frac{9
   \zeta (3)^2}{64}+\frac{1527  \zeta
   (5)}{256}+\frac{3339 \zeta (7)}{512}+\frac{441
   \zeta (9)}{64}+\frac{7929
}{2048}\right)
   \\\nonumber
   &+&
   g^{14}\sqrt{3} \left(-\frac{30015
\zeta (3)}{2048}-\frac{81 \zeta
   (3)^2}{128}-\frac{7929 \zeta
   (5)}{512}-\frac{45  \zeta (3) \zeta
   (5)}{64}-\frac{17127  \zeta
   (7)}{1024} \right.
   \\\nonumber
   &&\hspace{7cm} -\left.\frac{1197 \zeta
   (9)}{64}-\frac{10395  \zeta
   (11)}{512}-\frac{303837 }{28672}\right)
   \\\nonumber
   &+&
   g^{16}\sqrt{3} \left(\frac{340785 \zeta
   (3)}{8192}+\frac{2349\zeta
   (3)^2}{1024}+\frac{350505 \zeta
   (5)}{8192}+\frac{891 \zeta (3) \zeta
   (5)}{256}+\frac{225  \zeta
   (5)^2}{256}\right.
   \\\nonumber
   &+&\left. \frac{183519  \zeta
   (7)}{4096}+\frac{945  \zeta (3) \zeta
   (7)}{512}+\frac{100863  \zeta
   (9)}{2048}+\frac{230175  \zeta
   (11)}{4096}+\frac{127413  \zeta
   (13)}{2048}+\frac{15543873
  }{524288}\right)\,.
\eea}
Approximately one gets
\bea
p&=& 2.0944-0.433013 g^2+0.487139 g^4-1.43028 g^6+4.77062
   g^8\\\nonumber
   &&\hspace{3cm}-15.7964 g^{10}+52.5014 g^{12}-176.638 g^{14}+602.45
   g^{16}\,.
\eea
The corresponding expansion of the $u$-variable is given by
{\smaller
\bea
u&=&\frac{1}{\sqrt{3} g}\Big[1+2 g^2-\frac{5 g^4}{4}+g^6
   \left(\frac{7}{4}+\frac{3
   \zeta(3)}{4}\right)-\frac{1}{128} g^8 (461+144
   \zeta(3)+240 \zeta(5))
     \\\nonumber
   &+&
   g^{10}
   \left(\frac{1133}{128}+\frac{63
   \zeta(3)}{32}+\frac{189
   \zeta(5)}{64}+\frac{315
   \zeta(7)}{64}\right)
    \\\nonumber
   &-&
   g^{12}
   \left(\frac{23835}{1024}+\frac{1167
   \zeta(3)}{256}+\frac{9
   \zeta(3)^2}{64}+\frac{729
   \zeta(5)}{128}+\frac{2079
   \zeta(7)}{256}+\frac{441
   \zeta(9)}{32}\right)
    \\\nonumber
   &+&g^{14}
   \left(\frac{64731}{1024}+\frac{3429
   \zeta(3)}{256}+\frac{9
   \zeta(3)^2}{128}+\frac{897
   \zeta(5)}{64}+\frac{45}{64} \zeta(3)
   \zeta(5)+\frac{8559
   \zeta(7)}{512}
   \right.
   \\\nonumber
   &&\hspace{7cm} \left.
   +\frac{189
   \zeta(9)}{8}+\frac{10395
   \zeta(11)}{256}\right)
    \\\nonumber
   &+&g^{16}
   \left(-\frac{1441077}{8192}-\frac{176445
   \zeta(3)}{4096}+\frac{405
   \zeta(3)^2}{512}-\frac{169371
   \zeta(5)}{4096}-\frac{477}{512} \zeta(3)
   \zeta(5)
   \right.
   \\\nonumber
   &-& \left.
   \frac{225
   \zeta(5)^2}{256}-\frac{87417
   \zeta(7)}{2048}-\frac{945}{512} \zeta(3)
   \zeta(7)-\frac{51975
   \zeta(9)}{1024}-\frac{147015
   \zeta(11)}{2048}-\frac{127413
   \zeta(13)}{1024}\right)\Big]\,.
\eea
}
Approximately one gets
\bea
u&=& \frac{0.57735}{g}+1.1547 g-0.721688 g^3+1.53087
   g^5-3.98263 g^7
   \\\nonumber
   &&\hspace{2cm}+11.1101 g^9
   -32.8297 g^{11}+101.602
   g^{13}-325.587 g^{15}\,.
\eea
The dimension of the Konishi operator is
{\smaller
\bea
\Delta&=& 2+3 g^2-3 g^4+\frac{21 g^6}{4}+g^8
   \left(-\frac{705}{64}-\frac{9
   \zeta(3)}{8}\right)+g^{10}
   \left(\frac{6627}{256}+\frac{135
   \zeta(3)}{32}+\frac{45
   \zeta(5)}{16}\right)
    \\\nonumber
   &+&
   g^{12}
   \left(-\frac{67287}{1024}-\frac{27
   \zeta(3)}{2}-\frac{1377
   \zeta(5)}{128}-\frac{945
   \zeta(7)}{128}\right)
    \\\nonumber
   &+&
   g^{14}
   \left(\frac{359655}{2048}+\frac{10899
   \zeta(3)}{256}+\frac{27
   \zeta(3)^2}{128}+\frac{18117
   \zeta(5)}{512}+\frac{7371
   \zeta(7)}{256}+\frac{1323
   \zeta(9)}{64}\right)
    \\\nonumber
   &+&
   g^{16}
   \left(-\frac{7964283}{16384}-\frac{278505
   \zeta(3)}{2048}-\frac{621
   \zeta(3)^2}{512}-\frac{58491
   \zeta(5)}{512}-\frac{135}{128} \zeta(3)
   \zeta(5)
    \right.
   \\\nonumber
   &&\hspace{6cm} \left.-\frac{198207
   \zeta(7)}{2048}-\frac{20979
   \zeta(9)}{256}-\frac{31185
   \zeta(11)}{512}\right)
    \\\nonumber
   &+&
   g^{18}
   \left(\frac{22613385}{16384}+\frac{3600585
   \zeta(3)}{8192}+\frac{1539
   \zeta(3)^2}{256}+\frac{1520127
   \zeta(5)}{4096}+\frac{7101 \zeta(3)
   \zeta(5)}{1024} \right.
   \\\nonumber
   &+& \left.
   \frac{675
   \zeta(5)^2}{512}+\frac{2605095
   \zeta(7)}{8192}+\frac{2835 \zeta(3)
   \zeta(7)}{1024}+\frac{573237
   \zeta(9)}{2048}+\frac{1002375
   \zeta(11)}{4096}+\frac{382239
   \zeta(13)}{2048}\right)\,.
\eea
}
Approximately one gets
\bea
\Delta_{\rm Konishi}&=& 2+3 g^2-3g^4+5.25 g^6-12.3679 g^8+33.8743
   g^{10}-100.537 g^{12}~~~~\\\nonumber
   &+&313.532 g^{14} -1011.73
   g^{16}+3348.11 g^{18}\,.
\eea
\begin{figure}[t]
\begin{center}
\includegraphics[width=.46\textwidth]{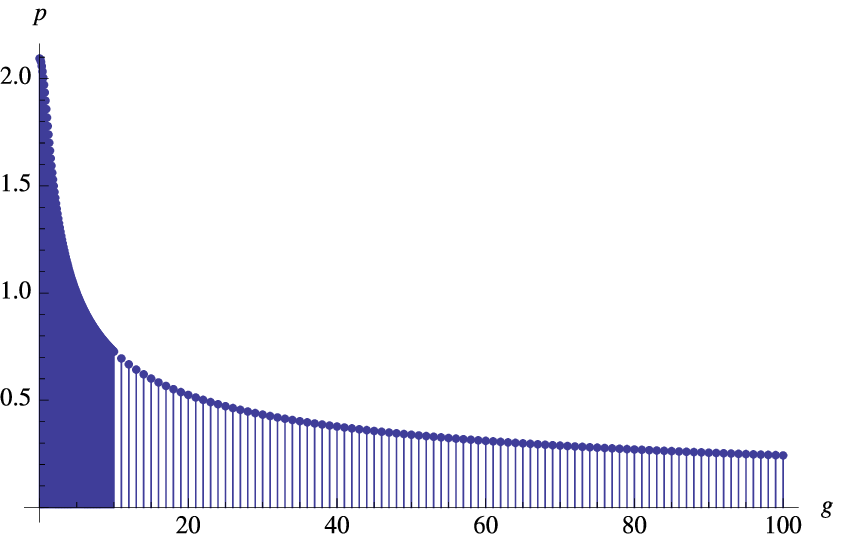}\quad \includegraphics[width=.40\textwidth]{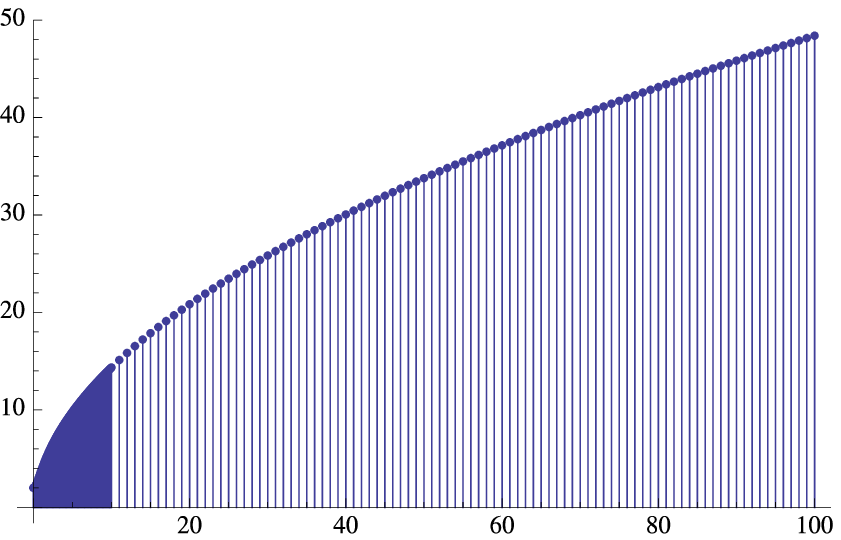}
\end{center}
\caption{Numerical solution of the BYE for the Konishi operator. The momentum was computed for ${1\ov 10}\le g\le 10$ with the step $\Delta g =  {1\ov 10}$, and for
$10\le g\le 100$ with $\Delta g =  1$. On the right picture the asymptotic dimension of the Konishi operator is plotted. It approaches $2\sqrt{2\pi g}$ as expected from \cite{AFS}.}
\end{figure}
We have also solved the BY equation numerically for small values of $g$, and its numerical solution perfectly agrees with the analytic one. The perturbative solution works very well at least up to $g= {1\ov 5}$. For $g={1\ov 5}$ the difference between the analytic solution and numerical one is $\approx 5\times  10^{-10}$.

\subsubsection*{Numerical solution for ${1\ov 10}\le g\le 100$}
The BY equation can be solved numerically up to very large values of $g$, and one gets the  plot on Figure 7.
for the momentum $p$ as a function of $g$.
For large values of $g \sim 100$ the momentum is approximated by
\bea
p_{\rm AFS} = \sqrt{2\pi\ov g} - {1\ov g}\,,
\eea
with a good precision as expected from \cite{AFS,AF05}. For $g=100$ the difference between the numerical solution and the AFS formula is equal to $-0.0016902$. If one uses the following asymptotic expression from \cite{RS09}
\bea
p_{\rm RS} = \sqrt{2\pi\ov g} - {1\ov g}+\frac{0.931115+0.199472 \log
   (g)}{g^{3/2}}\,,
   \eea
  that is one
   includes the next subleading terms then the agreement is much better and  for $g=100$ the difference between the numerical solution and the Rej-Spill formula is equal to $0.0001587$. To match the coefficients in $p_{\rm RS}$ one has to solve the BY equation for larger values of $g$.

The corresponding plots of the $u$-variable are shown on Figure 3, and were discussed in section 3.

 %%%%%%%%%%%%%%%%%%%%%%%%%%%%%%
\subsection{Transfer matrices and asymptotic Y-functions}\label{appT}
Here we remind the construction of the asymptotic Y-functions in
terms of transfer-matrices corresponding to various
representations of the centrally extended $\su(2|2)$ superalgebra.
Consider $K^I$ physical particles (excited states) of string
theory characterized by the $u_*$-plane rapidities $u_1,\ldots,
u_N$, $N\equiv K^{\rm I}$ or, equivalently, by the $p_1,\ldots,
p_N$ physical momenta. Each of these particles transforms in the
fundamental representation of $\su(2|2)$. Consider also a single
auxiliary particle with rapidity $v$ corresponding to a bound
state (atypical) representation of $\su(2|2)$ with the bound state
number $a$. Scattering the bound state representation through the
chain of $N$ physical particles gives rise to the following
monodromy matrix
$$
\mathbb{T}(v|\vec{u})=\prod_{i=1}^N \mathbb{S}_{0i}(v,u_i)\, .
$$
Here $\mathbb{S}_{0i}(v,u_i)$ is the S-matrix describing the
scattering of the auxiliary particle in the bound state
representation with a physical particle with rapidity $u_i$. The
transfer-matrix $T_a(v|\vec u)$ corresponding to this scattering
process is defined as the trace of $\mathbb{T}(v|\vec u)$ over the
auxiliary space of the $a$-particle bound state representation $T_a(v|\vec{u})={\rm tr}_0\mathbb{T}(v|\vec{u})$.
We are mostly interested in the situation where the auxiliary
particle is in the mirror theory. An eigenvalue of this transfer
matrix for an anti-symmetric bound state representation of the
mirror particle is given by the formula found in \cite{ALST}
\begin{eqnarray}\label{eqn;FullEignvalue}
&&T_a(v\,|\,\vec{u})=\prod_{i=1}^{K^{\rm{II}}}{\textstyle{\frac{y_i-x^-}
{y_i-x^+}\sqrt{\frac{x^+}{x^-}}
\, +}}\\
&&
{\textstyle{+}}\prod_{i=1}^{K^{\rm{II}}}{\textstyle{\frac{y_i-x^-}{y_i-x
^+}\sqrt{\frac{x^+}{x^-}}\left[
\frac{x^++\frac{1}{x^+}-y_i-\frac{1}{y_i}}{x^++\frac{1}{x^+}-y_i-\frac{1
}{y_i}-\frac{2i
a}{g}}\right]}}\prod_{i=1}^{K^{\rm{I}}}
{\textstyle{\left[\frac{(x^--x^-_i)(1-x^-
x^+_i)}{(x^+-x^-_i)(1-x^+
x^+_i)}\frac{x^+}{x^-}  \right]}}\nonumber\\
&&{\textstyle{+}}
\sum_{k=1}^{a-1}\prod_{i=1}^{K^{\rm{II}}}{\textstyle{\frac{y_i-x^-}{y_i-
x^+}\sqrt{\frac{x^+}{x^-}}
\left[\frac{x^++\frac{1}{x^+}-y_i-\frac{1}{y_i}}{x^++\frac{1}{x^+}-y_i-\frac{1}{y_i}-\frac{2ik}{g}}\right]}}
\left\{\prod_{i=1}^{K^{\rm{I}}}{\textstyle{\lambda_+(v,u_i,k)+}}\right.\left.\prod_{i=1}^{K^{\rm{I}}}{\textstyle{\lambda_-(v,u_i,k)}}\right\}\nonumber\\
&&\quad -\sum_{k=0}^{a-1}\prod_{i=1}^{K^{\rm{II}}}
{\textstyle{\frac{y_i-x^-}{y_i-x^+}\sqrt{\frac{x^+}{x^-}}\left[\frac{x^+
-\frac{1}{x^+}-y_i-\frac{1}{y_i}}
{x^+-\frac{1}{x^+}-y_i-\frac{1}{y_i}-\frac{2ik}{g}}\right]}}\prod_{i=1}^
{K^{\rm{I}}}{\textstyle{\frac{x^+-x^+_i}{x^+-x^-_i}\sqrt{\frac{x^-_i}{x^
+_i}}
\left[1-\frac{\frac{2ik}{g}}{v-u_i+\frac{i}{g}(a-1)
}\right]}}\times\nonumber\\
&&\quad\times
\left\{\prod_{i=1}^{K^{\rm{III}}}{\textstyle{\frac{w_i-x^+-\frac{1}{x^+}
+\frac{i(2k-1)}{g}}{w_i-x^+-\frac{1}{x^+}+\frac{i(2k+1)}{g}}+
}}\prod_{i=1}^{K^{\rm{II}}}{\textstyle{\frac{y_i+\frac{1}{y_i}-x^+-\frac
{1}{x^+}+\frac{2ik}{g}}{y_i+\frac{1}{y_i}-x^+-\frac{1}{x^+}+\frac{2i(k+1
)}{g}}}}\prod_{i=1}^{K^{\rm{III}}}{\textstyle{\frac{w_i-x^+-\frac{1}{x^+
}+\frac{i(2k+3)}{g}}{w_i-x^+-\frac{1}{x^+}+\frac{i(2k+1)}{g}}}}\right\}.
\nonumber
\end{eqnarray}
Eigenvalues are parametrized by solutions of the auxiliary Bethe
equations:
\begin{eqnarray}
\label{bennote}
\prod_{i=1}^{K^{\rm{I}}}\frac{y_k-x^-_i}{y_k-x^+_i}\sqrt{\frac{x^+_i}{x^
-_i}}&=&
\prod_{i=1}^{K^{\rm{III}}}\frac{w_i-y_k-\frac{1}{y_k}+\frac{i}{g}}{w_i-y
_k-\frac{1}{y_k}-\frac{i}{g}},\\
\prod_{i=1}^{K^{\rm{II}}}\frac{w_k-y_i-\frac{1}{y_i}+\frac{i}{g}}{w_k-y_
i-\frac{1}{y_i}-\frac{i}{g}}
&=& \prod_{i=1,i\neq
k}^{K^{\rm{III}}}\frac{w_k-w_i+\frac{2i}{g}}{w_k-w_i-\frac{2i}{g}}.\nonumber
\end{eqnarray}
In the formulae above
$$
v=x^++\frac{1}{x^+}-\frac{i}{g}a=x^-+\frac{1}{x^-}+\frac{i}{g}a\,,
$$
and $v$ takes values in the mirror theory $v$-plane, so $x^\pm =
x(v \pm {i\ov g}a)$ where $x(v)$ is the mirror theory
$x$-function. As was mentioned above, $u_j$ take values in string
theory $u$-plane, and therefore $x_j^\pm = x_s(u_j \pm {i\ov g})$
where $x_s(u)$ is the string theory $x$-function. Finally, the
quantities $\lambda_{\pm}$ are
\begin{eqnarray}\nonumber \hspace{-1cm}
\lambda_\pm(v,u_i,k)&=&\frac{1}{2}\left[1-\frac{(x^-_ix^+-1)
  (x^+-x^+_i)}{(x^-_i-x^+)
  (x^+x^+_i-1)}+\frac{2ik}{g}\frac{x^+
  (x^-_i+x^+_i)}{(x^-_i-x^+)
  (x^+x^+_i-1)}\right.\\ \label{eqn;lambda-pm}
&&~~~~~~~~~~~~\left.\pm\frac{i x^+
  (x^-_i-x^+_i)}{(x^-_i-x^+)
 (x^+x^+_i-1)}\sqrt{4-\left(v-\frac{i(2k-a)}{g}\right)^2}\right]\,
 .
\end{eqnarray}
For the $\sl(2)$-sector $K^{\rm II}=0=K^{\rm III}$ and the
expression above simplifies to
\bea\label{TS}
T_a(v\,|\,\vec{u})&=&1+\prod_{i=1}^{K^{\rm{I}}}
\frac{(x^--x^-_i)(1-x^- x^+_i)}{(x^+-x^-_i)(1-x^+
x^+_i)}\frac{x^+}{x^-}\\
&&\hspace{-0.5cm} -2\sum_{k=0}^{a-1}\prod_{i=1}^{K^{\rm{I}}}
\frac{x^+-x^+_i}{x^+-x^-_i}\sqrt{\frac{x^-_i}{x^+_i}}
\left[1-\frac{\frac{2ik}{g}}{v-u_i+\frac{i}{g}(a-1)}\right]+\sum_{m=\pm}
\sum_{k=1}^{a-1}\prod_{i=1}^{K^{\rm
I }}\lambda_m(v,u_i,k)\, . \nonumber
\eea
For a two-particle physical
state, the last
formula appears to coincides up to a gauge transformation with \cite{Beisert06b}
\bea\nonumber\hspace{-0.5cm}
T_a(v\,|\,\vec{u})&=&\frac{x^+}{x^-}\left[(1+a)\prod_{i=1}^2\frac{x^--x^-_i}{x^+-x^-_i}
+(a-1)\prod_{i=1}^2\frac{x^--x^+_i}{x^+-x^-_i}\frac{x^-_i-\frac{1}{x^+}}
{x^+_i-\frac{1}{x^+}}\right.\\
&&~~~~~~~~~-\left. a\prod_{i=1}^2\frac{x^--x^+_i}{x^+-x^-_i}
-a\prod_{i=1}^2\frac{x^--x^-_i}{x^+-x^-_i}\frac{x^-_i-\frac{1}{x^+}}{x^+_i-\frac{1}{x^+}}\right]\,
, \label{TA}\eea which is nothing else but an eigenvalue of the
transfer matrix evaluated on the fermionic vacuum and continued to
the mirror region for the auxiliary variable. In numerical
computations formula (\ref{TA}) works much faster than (\ref{TS})
and, therefore, we use it for constructing the asymptotic
Y-functions.

\smallskip

The transfer matrices
$T_{a,s}$ are used to introduce Y-functions $Y_{a,s}$ which solve Y-system equations in  the following standard  way  \cite{Kuniba:1993cn,Tsuboi, GKV09}
\bea
Y_{a,s}=\frac{T_{a,s+1}T_{a,s-1}}{T_{a+1,s}T_{a-1,s}} \eea The
transfer matrices $T_{a,s}$ solve the so-called Hirota equations
\cite{Hirota} and they can be computed via  $T_{a,1}\equiv
T_a(v\,|\,\vec{u})$ through the Bazhanov-Reshetikhin formula  \cite{BR} which we present here for the asymptotic solution
\bea
T_{a,s}=\det_{1\leq i,j\leq
s}T_{a+i-j}\Big(v+\frac{i}{g}(s+1-i-j)\mid\vec{u}\Big)\, . \eea If
here $a+i-j< 0$ then the corresponding element
$T_{a+i-j}\Big(v+\frac{i}{g}(s+1-i-j)\mid\vec{u}\Big)$ is regarded
as zero. Also, $T_0\Big(v+\frac{i}{g}(s+1-i-j)\mid
\vec{u}\Big)=1$. In other words, asymptotic $T_{a,1}$ and $T_{1,s}$ satisfy
the boundary conditions: $T_{0,s} =1=T_{a,0}$, $T_{a<0,s}=0$ and $T_{a,s<0}=0$.

\smallskip

Here for reader's convenience we recall the relationship
of the Y-functions introduced in \cite{AF09b} and those of
\cite{GKV09}. For the auxiliary Y-functions we have
\bea\begin{aligned}
&Y_-^{(1)}=-\frac{1}{Y_{1,1}}=-\frac{T_{2,1}T_{0,1}}{T_{1,2}T_{1,0}}\,
, & Y_-^{(2)}&
=-\frac{1}{Y_{1,-1}}=-\frac{T_{2,-1}T_{0,-1}}{T_{1,0}T_{1,-2}}\,
,
\\
&Y_+^{(1)}=-Y_{2,2}=-\frac{T_{2,3}T_{2,1}}{T_{3,2}T_{1,2}}\, ,
&Y_+^{(2)}&=-Y_{2,-2}=-\frac{T_{2,-1}T_{2,-3}}{T_{3,-2}T_{1,-2}}\, ,\\
&Y_{Q|vw}^{(1)}=\frac{1}{Y_{Q+1,1}}=\frac{T_{Q+2,1}T_{Q,1}}{T_{Q+1,2}T_{
Q+1,0}}\,
,
&Y_{Q|vw}^{(2)}&=\frac{1}{Y_{Q+1,-1}}=\frac{T_{Q+2,-1}T_{Q,-1}}{T_{Q+1,0
}T_{Q+1,-2}}\, ,\\
&Y_{Q|w}^{(1)}=Y_{1,Q+1}=\frac{T_{1,Q+2}T_{1,Q}}{T_{2,Q+1}T_{0,Q+1}}\,
,
&Y_{Q|w}^{(2)}&=Y_{1,-Q-1}=\frac{T_{1,-Q}T_{1,-Q-2}}{T_{2,-Q-1}T_{0,-Q-1
}}\,
.
\end{aligned}
\eea The asymptotic functions $Y_Q(v)$ in the $\sl(2)$-sector are
given by (\ref{YQasympt}). Note that in this expression the
transfer matrix naturally comes with a prefactor
 \bea S_Q(v\,|\,\vec{u})= \prod_{i=1}^{K^{\rm{I}}}
\sqrt{S_{\sl(2)}^{Q1_*}(v,u_i)} = \prod_{i=1}^{K^{\rm{I}}} {1\ov
\sqrt{S_{\sl(2)}^{1_*Q}(u_i,v)}} \, .\eea  This prefactor can be
split as follows \bea \label{split} S_Q(v\,|\,\vec{u}) =
\prod_{i=1}^{K^{\rm{I}}} {1\ov
\sqrt{S_{\sl(2)}^{1_*Q}(u_i,v)h_Q(u_i,v)}} \prod_{k=1}^{K^{\rm{I}}}
\sqrt{h_Q(u_k,v)}
 \, ,
\eea where $h_Q(u,v)$ is introduced in \eqref{siqvwx}.
This quantity satisfies the equation
$$
h_Q\Big(u,v+\frac{i}{g}\Big)h_Q\Big(u,v-\frac{i}{g}\Big)=h_{Q+1}(u,v)h_{
Q-1}(u,v)
$$
and is a pure phase when $u$ and $v$ are in the string and mirror
regions, respectively. The splitting (\ref{split}) can be used to
introduce the normalized transfer-matrix $\widetilde{T}_{Q,1}(v) $
$$
\widetilde{T}_{Q,1}(v\mid \vec{u}) \equiv\prod_{k=1}^{K^{\rm{I}}}
\sqrt{h_Q(u_k,v)}\, T_{Q}(v\,|\,\vec{u})
$$
which renders the corresponding $\widetilde{Y}_Q=e^{-J\tilde{\cal
E}_Q}\widetilde{T}_{Q,1}^2(v | \vec{u})$ real. The functions
$\widetilde{Y}_Q$ represent a simple and useful tool for checking
the TBA equations.

%%%%%%%%%%%%%%%%%%%%%%%%%%%%%%%%
\subsection{Hybrid equations for $Y_Q$-functions}\la{hybridQ}
In this appendix we derive the hybrid TBA equations for  $Y_Q$-functions. We discuss only the ground state TBA equations because the excited states equations can be obtained by using the contour deformation trick.

The canonical TBA equation for $Q$-particles reads
\begin{align}
&\log Y_Q = - L\, \tH_{Q} + \log\left(1+Y_{Q'} \right) \star K_{\sl(2)}^{Q'Q}
\label{TbaQsl2app} \\[1mm]
&\quad + 2 \log\left(1+ \frac{1}{Y_{M'|vw}} \right) \star K^{M'Q}_{vwx}
+ 2 \log \left(1- \frac{1}{Y_-} \right) \hstar K^{yQ}_-
+ 2 \log \left(1- \frac{1}{Y_+} \right) \hstar K^{yQ}_+ \,.
\notag
\end{align}
To derive the hybrid equation we need to compute the first sum on the second line of this equation. It is done by using the simplified equations for $Y_{M|vw}$
\begin{align}\la{SimYvw}
\log Y_{M|vw} &= \log(1 +  Y_{M-1|vw})(1 +  Y_{M+1|vw}) \star s \\
&\notag - \log(1 + Y_{M+1})\star s + \delta_{M1} \log{1-Y_-\ov 1-Y_+} \hstar s .
\end{align}
Let us assume that we have
 some kernels ${\sf K}_M$ which satisfy the following identities
 \begin{equation}
{\sf K}_M - s \star \({\sf K}_{M+1} + {\sf K}_{M-1}\) = \delta {\sf K}_M \quad \(M \ge 2\), \qquad
{\sf K}_1 - s \star {\sf K}_2 = \delta {\sf K}_1 \,,
\label{def:delta KM}
\end{equation}
where $ \delta {\sf K}_M$ are known kernels.
Then, applying the kernel ${\sf K}_M$ to both sides of \eqref{SimYvw}, and taking the sum over $M$ from 1 to $\infty$, we obtain the formula
\begin{align}\label{identity sum over Mvw}
\sum_{M=1}^\infty \log \(1 + {1 \over Y_{M|vw}}\) \star {\sf K}_M &=
\sum_{M=1}^\infty \log \(1 + Y_{M|vw}\) \star \delta {\sf K}_M
\\[1mm]
&+ \sum_{M=1}^\infty \log(1 + Y_{M+1}) \star s \star {\sf K}_M
-  \log{1-Y_-\ov 1-Y_+} \star {\sf K}_1 \,.
\notag
\end{align}
Now choosing the kernel ${\sf K}_M$ to be $K^{MQ}_{vwx}$ and using the formula (6.31) from \cite{AF09b}
\begin{align}
K^{MQ}_{vwx} - s \star \( K^{M+1,Q}_{vwx} + K^{M-1,Q}_{vwx} \) &= \delta_{M+1,Q} \, s \qquad \(M \ge 2\),
\notag \\
K^{1Q}_{vwx} - s \star K^{2Q}_{vwx} &= s \hstar K_{yQ} + \delta_{2,Q} \, s\,,
\end{align}
and the identity \eqref{identity sum over Mvw}, we get
\begin{align}\notag
&\sum_{M=1}^\infty \log \(1 + {1 \over Y_{M|vw}}\) \star K^{MQ}_{vwx} =
\log \(1 + Y_{1|vw}\) \star s \hstar K_{yQ}
+ \delta_{M+1,Q} \log \(1 + Y_{M|vw}\) \star  s
\label{identity sum over Mvw3} \\[1mm]
&\quad + \sum_{M=1}^\infty \log(1 + Y_{M+1}) \star s \star K^{MQ}_{vwx}
-  \log{1-Y_-\ov 1-Y_+} \hstar s \star K^{1Q}_{vwx} \,.
\end{align}
Finally, substituting this formula into \eqref{TbaQsl2app}, we get the hybrid ground state TBA equation.

%%%%%%%%%%%%%%%%%%%%%%%%%%%%%%%%
\subsection{Analytic continuation of $Y_1(v)$}\la{app:Y1}

To derive the exact Bethe equations  \eqref{Tba1sl2B} one has to
 analytically continue $Y_1(z)$ in eq.(\ref{TbaQsl2H}) to the point $z_{*k}$. On the $v$-plane it means that we go from the real $v$-line  down below the line with Im$(v)=-{1\ov g}$  without crossing any cut, then turn back, cross the cut with Im$(v)=-{1\ov g}$ and $|$Re$(v)|>2$, and go back to the real $v$-line. As a result we should make the following replacements
$x(v-{i\ov g}) \to  x_s(v-{i\ov g}) = x(v-{i\ov g})$, $x(v+{i\ov g}) \to  x_s(v+{i\ov g}) = 1/x(v-{i\ov g})$ in the kernels appearing in (\ref{Y1m1}).

The analytic continuation depends on the analytic properties of the kernels and Y-functions. Let us consider the terms in  eq.(\ref{TbaQsl2H}) order by order

\subsubsection*{Terms with $Y_\pm(v)$-functions}
For the analytic continuation of the last two terms in \eqref{TbaQsl2H}  it is convenient to use the kernels $K^{y1}_\pm$ given by
\bea\la{Ky1m}
K^{y1}_-(u,v)&=&{1\ov 2\pi i}{d\ov du}\log S^{y1}_-(u,v)\,,\quad S^{y1}_-(u,v)= \frac{x(u)-x(v+{i\ov g})}{x(u)-x(v-{i\ov g})}\sqrt{\frac{x(v-{i\ov g})}{x(v+{i\ov g})}}\,,~~~~~\\
\la{Ky1p}
K^{y1}_+(u,v)&=&{1\ov 2\pi i}{d\ov du}\log S^{y1}_+(u,v)\,,\quad S^{y1}_+(u,v)= \frac{{1\ov x(u)}-x(v-{i\ov g})}{{1\ov x(u)}-x(v+{i\ov g})}\sqrt{
\frac{x(v+{i\ov g})}{x(v-{i\ov g})}
}
\,.\eea
It is clear from these equations that  $K^{y1}_+(u,v)$ remains regular in the analytic continuation.  $K^{y1}_-(u,v)$ on the other hand has a pole at $v= u-{i\ov g}$ and behaves as
\bea
K^{y1}_-(u,v)&=& {1\ov 2\pi i}{1\ov u-v-{i\ov g}} +\ regular\ at\ v \sim u-{i\ov g}\,.
\eea
Thus, one needs to analyze the analytic continuation of a function defined by the following integral for real $v$
\bea
F(v) = {1\ov 2\pi i}\int_{-2}^2\, du\, f(u){1\ov u-v-{i\ov g}} \,.
\eea
The consideration is the same as the one for the dressing phase in \cite{AF09c}, and after $v$ crosses the interval $[-2-{i\ov g},2-{i\ov g}]$ one gets the following expression for $F(v)$
\bea\la{Fvafter}
F(v) =f(v+{i\ov g})+  {1\ov 2\pi i} \int_{-2}^2\, du\, f(u){1\ov u-v-{i\ov g}}\,,\quad {\rm Im}(v)<-{1\ov g} \,.
\eea
Then we go back to the real $v$-line but we do not cross the interval $[-2-{i\ov g},2-{i\ov g}]$, and therefore \eqref{Fvafter} remains the same. However, we should also analytically continue $f(v+{i\ov g})$ back to real values of $v$.  Thus we conclude that the analytic continuation of
\bea
\log \left(1- \frac{1}{Y_{-}} \right) \hstar K^{y1}_- + \log\left(1- \frac{1}{Y_+} \right) \hstar K^{y1}_+
\eea
is given by
\bea\la{ypmc}
\log \left(1- \frac{1}{Y_{*-}(v+{i\ov g})} \right) +\log \left(1- \frac{1}{Y_-} \right) \hstar K^{y1_*}_- + \log\left(1- \frac{1}{Y_+} \right) \hstar K^{y1_*}_+ \,,~~~~~
\eea
where $Y_{*-}(v)$ is the $Y_{-}(v)$-function analytically continued to the upper half-plane through the cut $|v|>2$.

As was discussed in section 4, $Y_{*-}$ coincides with $Y_{+}$, and, therefore,
to find $Y_{*-}(v+{i\ov g})$ one can just use the analytic continuation of the TBA equation \eqref{Tbaysl2} for  $Y_{+}(v)$ to $Y_{+}(v+{i\ov g})$. Since the S-matrix $S_-^{1_*y}(u_k,v)$ has a pole at $v=u_k+{i\ov g}$, see appendix \ref{app:rapidity}, one concludes that
${Y_{*-}(u_k+{i\ov g})}=\infty$, and, therefore, the contribution of the first term in \eqref{ypmc} vanishes in the exact Bethe equation $Y_{1_*}(u_k) = -1$.

Nothing dangerous happens with the term  $\log{1-Y_-\ov 1-Y_+} \hstar s \star K^{11}_{vwx}$ because there is no singularity in the analytic continuation process of $K^{11}_{vwx}$
until we get back to the real $v$-line. Then, we get for the analytically continued  $K^{11_*}_{vwx}$
 \bea\la{K11svwx}
&&K^{11_*}_{vwx}(u,v)={1\ov 2\pi i}{d\ov du}\log S^{11_*}_{vwx}(u,v)\,,\quad\\\nonumber
 &&S^{11_*}_{vwx}(u,v)= \frac{x(u-{i\ov g})-x_s(v+{i\ov g})}{x(u+{i\ov g})-x_s(v-{i\ov g})} \,\frac{{1\ov x(u-{i\ov g})}-x_s(v-{i\ov g})}{{1\ov x(u+{i\ov g})}-x_s(v+{i\ov g})}\,,~~~~\eea
and it shows a pole at $v=u$
 \bea\la{Km1svwx2}
&&K^{11_*}_{vwx}(u,v)=-{1\ov 2\pi i}\,{1\ov u-v} +\ regular\ at\ v \sim u\,.~~~~\eea
Since we integrate over the line which is above the real line, the pole is not dangerous if the function we integrate with the kernel is regular as it is the case for the case under consideration.

\subsubsection*{Terms with $Y_{M|vw}$-functions}

The analytic continuation of the term $\log \(1 + Y_{1|vw}\) \star s \hstar K_{y1}$ is given by
\bea
\log \(1 + Y_{1|vw}\) \star (s \hstar K_{y1} + \ts)
\eea
because $K_{y1}(u,v)$ has a pole at $v=u-{i\ov g}$.

\subsubsection*{Terms with $Y_Q$-functions}
The kernel $K_{\sl(2)}^{Q1}$ is given by
\bea\la{Kq1sl2}
K_{\sl(2)}^{Q1}(u,v)&=& - K_{Q1}(u-v) - 2 K^\Sigma_{Q1}(u,v)\\\nonumber
 &=& - K_{Q-1}(u-v)- K_{Q+1}(u-v)  - 2 K^\Sigma_{Q1}(u,v) \,,
\eea
Since $K^\Sigma_{Q1}(u,v)$ is a holomorphic function if $u$ is in the mirror region and $v$ is in the string one,  only  $K_{Q1}$ can cause any  problem with the analytic continuation.  Moreover, we see immediately that only the $Q=2$ case should be treated with care. It is easy to show however that since the integral over $u$ is taken from $-\infty$ to $\infty$ the analytic continuation does not give any extra term because $v$ crosses the line Im$(u)=-{1\ov g}$ twice. This is the difference of this case from the $Y_\pm$ one. The term with $K^{Q'-1,1}_{vwx}(v) $ is harmless too because $s$ is regular.

\subsubsection*{The term  $\log S\star K^{11}_{vwx} (u_j^-,v)$}

This term is similar to  $\log\left(1+ \frac{1}{Y_{1|vw}}  \right)\star  K^{11}_{vwx}$ considered above, and its analytic continuation is given by
\bea
 \log {\rm Res}(S)\star K^{11_*}_{vwx} (u_j^-,u_k) -\sum_{j=1}^N\log\big(u_j-u_k-{2i\ov g}\big)\,
{x_j^--{1\ov x_{k}^-}\ov x_j^-- x_{k}^+}\,.
\eea

\medskip

Summing up all the contributions, one gets the exact Bethe equations \eqref{Tba1sl2B} and  \eqref{Tba1sl2}.
We have shown that the r.h.s. of  \eqref{Tba1sl2B} is purely imaginary, and, therefore, the exact Bethe equations can be also written in the form
 \begin{align}
&\pi (2n_k+1)=L\, p_k +i\sum_{j=1}^N\, \log S_{\sl(2)}^{1_*1_*}(u_j,u_k)\label{Tba1sl2Bb}\\
&\quad
+2 \sum_{j=1}^N\, \log {\rm Res}(S)\star {\rm Im}K^{11_*}_{vwx} (u_j^-,u_k) -2 \sum_{j=1}^N{\rm Im}\log\big(u_j-u_k-{2i\ov g}\big)\,
{x_j^--{1\ov x_{k}^-}\ov x_j^-- x_{k}^+}
\notag\\
&\quad
-2 \log \left(1+Y_{Q} \right) \star \({\rm Im}K_{Q1_*}^\Sigma -  s \star {\rm Im}K^{Q-1,1_*}_{vwx} \)- 2i \log \(1 + Y_{1|vw}\) \star \(  \ts +s \hstar K_{y1_*} \)
\notag \\
&\quad - 2  \log{1-Y_-\ov 1-Y_+} \hstar s \star {\rm Im}K^{11_*}_{vwx}
-i  \log \big(1- \frac{1}{Y_-}\big)\big( 1- \frac{1}{Y_+} \big) \hstar K_{y1_*} \,.
\notag
\end{align}

\subsubsection*{Analytic continuation of the canonical TBA equation for $Y_1$}
Let us also consider the analytic continuation of the canonical TBA equation for $Y_1$.
Then, the kernel $K^{M1}_{vwx}$ is given by
\bea\la{Km1vwx}
&&K^{M1}_{vwx}(u,v)={1\ov 2\pi i}{d\ov du}\log S^{M1}_{vwx}(u,v)\,,\quad\\\nonumber
 &&S^{M1}_{vwx}(u,v)= \frac{x(u-{i\ov g}M)-x(v+{i\ov g})}{x(u+{i\ov g}M)-x(v-{i\ov g})} \,\frac{{1\ov x(u-{i\ov g}M)}-x(v-{i\ov g})}{{1\ov x(u+{i\ov g}M)}-x(v+{i\ov g})}\,,~~~~\eea
and nothing dangerous happens for $M>1$. So,  we just have to consider the $M=1$ case.
Since  $Y_{1|vw}(u_j)=0$, and we should take care of the resulting log-singularity before the analytic continuation.

Introducing $Z$-functions as in \eqref{YvwZvw},
one gets for the term in the canonical TBA equation (\ref{TbaQsl2})
 \bea\nonumber
&&2 \log\left(1+ \frac{1}{Y_{M|vw}}  \right)\star  K^{M1}_{vwx} =
 2 \log{M^2\ov M^2-1}\star  K^{M-1,1}_{vwx}  - 2 \log\prod_{j=1}^N \left( u - {u_j}\right)\star  K^{11}_{vwx}~~~~~ \\
  &&~~~~~~~~~~~~~~~~~~~~~~~~~~~~~~~~~~~~~~~~~~+2 \log Z_{M|vw}\star  K^{M1}_{vwx}\\\nonumber
  &&~~~~~~~=2\log 2 -2 \sum_{j=1}^N\log\big(u_j-v-{2i\ov g}\big)\,
{x_j^--{1\ov x^-}\ov x_j^--{1\ov  x^+}} +2\log Z_{M|vw}\star K^{M1}_{vwx}
 \,,~~~~~~~~~
\eea
 where we used $ \log{M^2\ov M^2-1}\star  K^{M-1,1}_{vwx}=\log 2$, and  the following formula
 \bea\nonumber
\int_{-\infty}^\infty\, dt\, \log( t +i0- u_j)\,  K^{11}_{vwx}(t,v) =  \log\big(u_j-v-{2i\ov g}\big)\,
{x_j^--{1\ov x^-}\ov x_j^--{1\ov  x^+}}
\,.~~~~~\la{tmuKvwx}
\eea
The analytic continuation in $v$ then gives
  \bea\nonumber
2\log\left(1+ \frac{1}{Y_{M|vw}}  \right)\star  K^{M1_*}_{vwx}(v) &=&
2\log 2 -2 \sum_{j=1}^N\log\big(u_j-v-{2i\ov g}\big)\,
{x_j^--{1\ov x^-}\ov x_j^-- x^+} \\\nonumber
&+&2 \log Z_{M|vw}\star_{p.v.}  K^{M1_*}_{vwx} + \log  Z_{1|vw}(v)
 \,.
\eea
The equation seems to coincide with the one derived in \cite{GKV09b} after one performs the parity transformation $u_j\to -u_j$, and sets $g=2$.

%%%%%%%%%%%%%%%%%%%%%%%%%%%%%%
\subsection{Canonical TBA equations} \la{canTBA}

\subsubsection*{Canonical TBA equations: $g<g_{cr}^{(1)} $}

The canonical ground state TBA equations \cite{AF09b} are derived from the string hypothesis for the mirror model \cite{AF09a} by following a textbook route, see e.g. \cite{Korepin}, and can be written in the form

\medskip \noindent
$\bullet$\ $Q$-particles
\begin{multline}
V_Q\equiv \log Y_Q + L\, \tH_{Q}  - \log\left(1+Y_{Q'} \right) \star K_{\sl(2)}^{Q'Q} - 2\log\left(1+ \frac{1}{Y_{M'|vw}} \right) \star K^{M'Q}_{vwx} \\
- 2\log \left(1- \frac{1}{Y_-} \right) \hstar K^{yQ}_- - 2\log\left(1- \frac{1}{Y_+} \right) \hstar K^{yQ}_+ = 0\,,
\label{TbaQsl2v}
\end{multline}

\noindent
$\bullet$\ $y$-particles
\begin{equation}
V_\pm\equiv \log Y_\pm + \log\left(1+ Y_Q \right) \star K^{Qy}_\pm -\log {1+ \frac{1}{Y_{M|vw}} \over 1+ \frac{1}{Y_{M|w}} } \star K_{M}=0\,.
\label{Tbaysl2v}
\end{equation}

\noindent
$\bullet$\ $M|vw$-strings
\begin{multline}
V_{M|vw}\equiv \log Y_{M|vw}+ \log\left(1+Y_{Q'} \right) \star K^{Q'M}_{xv} \\
- \log\left(1+ \frac{1}{Y_{M'|vw}} \right) \star K_{M'M}
- \log {1- \frac{1}{Y_-} \over 1- \frac{1}{Y_+} } \hstar K_M =0\,.
\label{Tbavwsl2v}
\end{multline}

\noindent
$\bullet$\ $M|w$-strings
\begin{equation}
V_{M|vw}\equiv \log Y_{M|w} -  \log \left(1+ \frac{1}{Y_{M'|w}} \right)\star K_{M'M}
- \log{1- \frac{1}{Y_-} \over 1- \frac{1}{Y_+} } \hstar K_M =0\,.
\label{Tbawsl2v}
\end{equation}

\medskip

Applying the contour deformation trick to the canonical TBA equations, one gets the following set of  integral equations for Konishi-like states and $g<g_{cr}^{(1)} $
\bea
&&\bullet \ Q{\rm -particles}:\quad\quad V_Q = - \sum_{*} \log S_{\sl(2)}^{1_*Q}(u_j,v)\,,\label{TbaQsl2}\\
&&\bullet \ y{\rm -particles}:\quad\quad\ V_\pm = \sum_{*} \log S^{1_*y}_\pm(u_j,v)\,,\la{Tbaysl2}\\
&&\bullet \ M|vw{\rm -strings}: \quad V_{M|vw}= \sum_{*} \log S^{1_*M}_{xv}(u_j,v)\,, \label{Tbavwsl2}\\
&&\bullet  \ M|w{\rm -strings}: \quad\ \ V_{M|w} =0\label{Tbawsl2}\,.
\eea
Here summation over repeated indices is assumed. The sums in the formulae run over the set of $N$ particles,
all Y-functions depend on the real $u$ variable (or $z$)  of the mirror region.
We recall that $S_{\sl(2)}^{1_*Q}$ and $S^{1_*M}_{xv}$ denote the S-matrices with the first and second arguments in the string and mirror regions, respectively, and both arguments of the kernels in these formulae are in the mirror region.

Then, the integrals are taken over the interval $[-2,2]$ for convolutions involving $Y_\pm$,  and over the horizontal line that is a little bit above the real $u$ line (or the interval Re$(z)\in (-{\om_1\ov 2},{\om_1\ov 2})$, Im$(z)={\om_2\ov 2i}$ on the $z$-torus) for all other convolutions.

The reason why one should choose the integration contour to run a little bit above the real line of the mirror $u$-plane is that
$Y_{1|vw}$ has  zeros at $u=u_k$,
and, therefore, the terms  $\log\left(1+ \frac{1}{Y_{1|vw}} \right) \star K$ with any kernel $K$ should be treated carefully. The prescription above for the integration contour  guarantees the reality of all Y-functions as we show in appendix \ref{app:reality}.
The $\log S$-term in the equation for vw-strings is in fact necessary to cancel the corresponding singularity on the l.h.s. of this equation.

\medskip

One can show that the imaginary  zeros of $Y_{k|vw}$ and $1 +  Y_{k|vw}$ do not contribute to the canonical equations for Konishi-like states at weak coupling.
In appendix \ref{appSimple} we also show that the simplified equations can be derived from the canonical ones following the same route as in \cite{AF09b,AF09d}.

\medskip

The canonical TBA equations  are rather complicated and involve infinite sums that makes the high-precision numerical tests very time-consuming.
We have checked numerically that for Konishi-like states  they are solved at the large $L$ limit  by
the asymptotic Y-functions if $L=J+2$.

\medskip

Let us stress again that the TBA equations above are valid only up to the first critical value of $g$ where the function $Y_{1|vw}(u)$ has real  zeros  only for $u=u_k$, and  $Y_{M|vw}$-functions with $M\ge 1$ do not have other  zeros in the physical strip.

\subsubsection*{Exact Bethe equations: $g<g_{cr}^{(1)} $ }

To derive the exact Bethe equations we take the logarithm of eq.(\ref{Y1m1}), and analytically continue the variable $z$
of $Y_1(z)$ in eq.(\ref{TbaQsl2}) to the point $z_{*k}$.
The analytic continuation is similar to the one in section \ref{TBAKon}, and its detailed consideration can be found in appendix
 \ref{app:Y1}.
As shown there, the resulting exact Bethe equations  for a string theory state from the $\sl(2)$ sector can be cast into the following integral form
\bea\nonumber
&&\pi i(2n_k+1)=\log Y_{1_*}(u_k) =i L\, p_k - \sum_{j=1}^N \log S_{\sl(2)}^{1_*1_*}(u_{j},u_{k})+ \log\left(1+Y_Q \right)\star K_{\sl(2)}^{Q1_*}~~~~~~\\\nonumber
 &&~~~~~~~~~~~~~+2\log 2 -2 \sum_{j=1}^N\log\big(u_j-u_k-{2i\ov g}\big)\,
{x_j^--{1\ov x_{k}^-}\ov x_j^-- x_{k}^+}+2 \log Z_{M|vw}\star K^{M1_*}_{vwx} \\\la{Tba1sl2}
 &&~~~~~~~~~~~~~+ 2 \log \Big(1- \frac{1}{Y_-}\Big) \hstar K^{y1_*}_- + 2 \log\Big(1- \frac{1}{Y_+}\Big) \hstar K^{y1_*}_+\,.~~~~~~
\eea
Here the integration contours run a little bit above the Bethe roots $u_j$, $p_k= i \tH_{Q}(z_{*k})=-i\log{x_s(u_k+{i\ov g})\ov x_s(u_k-{i\ov g})}$ is the momentum of the $k$-th particle, and the second argument in all the kernels in this equation is equal to $u_{k}$ but the first argument we integrate with respect to is the original one in the mirror region.
The Z-functions are defined in the same way as in \cite{GKV09b}
\bea\la{YvwZvw}
1 + {1\ov Y_{1|vw}} \equiv Z_{1|vw}\, {4\ov 3}\,  {1\ov \prod_{j=1}^N \left( u - {u_j}\right)}\,,\quad  1+ {1\ov Y_{M|vw}} \equiv Z_{M|vw}\, {(M+1)^2\ov M(M+2)}\,,~~~~~
\eea
they are positive for real $u$, and $Z_{M|vw}$, $M=2,3,\ldots$ asymptote to 1 at $u\to\infty$.

Taking into account that the BY equations for the $\sl(2)$ sector have the form
 \bea\nonumber
\pi i(2n_k+1)=i J\, p_k - \sum_{j=1}^N \log S_{\sl(2)}^{1_*1_*}(u_{j},u_{k})\la{BYsl2b}
 \,,~~~~~~
\eea
and that $Y_Q$ is exponentially small at large $J$, we conclude that if the analytic continuation has been done correctly then up to an integer multiple of $2\pi i$  the following identities between the asymptotic Y-functions should hold
\bea\nonumber
&&{\cal R}_k\equiv 2 \ssp i\, p_k +2\log 2 -2 \sum_{j=1}^N\log\big(u_j-u_k-{2i\ov g}\big)\,
{x_j^--{1\ov x_{k}^-}\ov x_j^--x_{k}^+}+2 \log Z_{M|vw}\star K^{M1_*}_{vwx}~~~\\\la{Rksl2}
 &&~~~~~~~+ 2 \log \Big(1- \frac{1}{Y_-}\Big) \hstar K^{y1_*}_- + 2 \log\Big(1- \frac{1}{Y_+}\Big) \hstar K^{y1_*}_+ = 0\,.~
\eea
For $N=2$ and $u_1=-u_2$ one gets one equation, and by using the expressions for the Y-functions from appendix \ref{appT} one can check numerically
%\footnote{Since
%$K^{11_*}_{vwx}(u,v)$ has a pole at $u=v$ with the residue equal to $-{1\ov 2\pi i}$ the term $2\log Z_{1|vw}\star K^{11_*}_{vwx} $ can be represented as $
%2\log Z_{1|vw}\star K^{11_*}_{vwx}  = 2\log Z_{1|vw}\star_{p.v.}  K^{11_*}_{vwx} + \log  Z_{1|vw}(u_k)$ which is useful for numerics.}
that it does hold for any real value of $u_1$ such that only $Y_{1|vw}$ has two  zeros for real $u$.

We believe that  the canonical TBA equations (\ref{TbaQsl2}-\ref{Tbawsl2}) and eqs.\eqref{Tba1sl2} are equivalent to the ones proposed in \cite{GKV09b} for states having the vanishing total momentum  but a detailed comparison is hard to perform due to different notations and conventions. We see however that they are definitely different for physical states which satisfy the  level-matching condition
but do not have the vanishing total momentum.

\subsubsection*{ Canonical TBA equations: $g_{cr}^{(1)}<g<\bar g_{cr}^{(1)}$}

Here is the set of  canonical TBA equations for Konishi-like states and $g_{cr}^{(1)}<g<\bar g_{cr}^{(1)}$
\bea
&&\bullet \ Q{\rm -particles}:\quad V_Q = - \sum_{*} \log S_{\sl(2)}^{1_*Q}(u_j,v)-2 \sum_{j=1}^2 \log S_{vwx}^{1Q}(r_j^-,v)~~~~~~~~ \\
&&\hspace{6cm}+2 \log S_{vwx}^{2Q}(r_1,v)+ 2 \log S_{yQ} (r_1,v)\,,
\label{TbaQsl2c1}\nonumber\\
&&\bullet \  y{\rm -particles}: \quad\ V_\pm = \sum_{j=1}^N \log S^{1_*y}_\pm(u_j,v)- \sum_{j=1}^2 \log S_1(r_j^--v)+\log S_2(r_1-v)\,,~~~~~\nonumber\\
&&\bullet \  M|vw{\rm -strings}:  V_{M|vw}= \sum_{j=1}^N\log S^{1_*M}_{xv}(u_j,v) - \sum_{j=1}^2 \log S_{1M}(r_j^-,v)  +\log S_{2M}(r_1,v) \,, \nonumber\\
&&\bullet  \  M|w{\rm -strings}: \ \ V_{M|w} =0\nonumber\,,
\eea
where $2 \log S_{yQ} (r_1,v)$ appears due to the imaginary zero $r_1$ of $Y_\pm$, and we take into account that the S-matrix $S_{yQ}$ is normalized as $S_{yQ} (\pm 2,v)=1$.

\subsubsection*{ Exact Bethe equations: $g_{cr}^{(1)}<g< \bar g_{cr}^{(1)}$}

The exact Bethe equations are obtained by analytically continuing $\log Y_1$ in \eqref{TbaQsl2c1} following the same route as for the small $g$ case, and they take the following form
\bea\nonumber
\pi i(2n_k+1)&=&\log Y_{1_*}(u_k) =i L\, p_k - \sum_{j=1}^N \log S_{\sl(2)}^{1_*1_*}(u_{j},u_{k})+ \log\left(1+Y_Q \right)\star K_{\sl(2)}^{Q1_*}~~~~~~~~~\\\nonumber
 &-&2 \sum_{j=1}^2 \log S_{vwx}^{11_*}(r_j^-,u_k) +2 \log S_{vwx}^{21_*}(r_1,u_k) + 2 \log S_{y1_*} (r_1,u_k)\\\nonumber
 &-& 2 \sum_{j=1}^N\log\big(u_j-u_k-{2i\ov g}\big)\,
{x_j^--{1\ov x_{k}^-}\ov x_j^-- x_{k}^+}
 +2\log 2+2 \log Z_{M|vw}\star  K^{M1_*}_{vwx}
 \\\la{Eba1sl2bc1}
 &+& 2 \log \Big(1- \frac{1}{Y_-}\Big) \hstar K^{y1_*}_-
 + 2 \log\Big(1- \frac{1}{Y_+}\Big) \hstar K^{y1_*}_+\,.~~~~~~
\eea
We conclude again that the consistency with the BY equations requires
the fulfillment of the identities between the asymptotic Y-functions similar to
\eqref{Rksl2}
that we have checked numerically for the Konishi-like states.

\subsubsection*{ Canonical TBA equations: $\bar g_{cr}^{(1)}<g< g_{cr}^{(2)}$}

The canonical TBA equations for Konishi-like states and $\bar g_{cr}^{(1)}<g<g_{cr}^{(2)}$ take the form
\bea
&&\bullet \ Q{\rm -particles}:\quad V_Q = - \sum_{j=1}^N \log S_{\sl(2)}^{1_*Q}(u_j,v)-2 \sum_{j=1}^2 \log S_{vwx}^{1Q}(r_j^-,v)\,,\label{TbaQsl2c1b}\\
&&\bullet \  y{\rm -particles}: \quad\ V_\pm = \sum_{j=1}^N \log S^{1_*y}_\pm(u_j,v)- \sum_{j=1}^2 \log S_1(r_j^--v)\,,~~~~~\nonumber\\
&&\bullet \  M|vw{\rm -strings}:  V_{M|vw}=\sum_{j=1}^N\log S^{1_*M}_{xv}(u_j,v) - \sum_{j=1}^2 \log S_{1M}(r_j^-,v)   \,, \nonumber\\
&&\bullet  \  M|w{\rm -strings}: \ \ V_{M|w} =0\nonumber\,,
\eea
The equations  for $Y_Q$-particles differ from \eqref{TbaQsl2c1} by the absence of the term
$2\log S_{vwx}^{2Q}(r_1,v)$. The p.v. prescription however would give instead additional terms due to the real  zeros $r_j$ of $Y_{2|vw}$ and $Y_\pm$.

\subsubsection*{ Exact Bethe equations: $\bar g_{cr}^{(1)}<g< g_{cr}^{(2)}$}

Analytically continuing $\log Y_1$ in \eqref{TbaQsl2c1b}, one gets  the exact Bethe equations
\bea\nonumber
&&\pi i(2n_k+1)=\log Y_{1_*}(u_k) =i L\, p_k - \sum_{j=1}^N \log S_{\sl(2)}^{1_*1_*}(u_{j},u_{k})+ \log\left(1+Y_Q \right)\star K_{\sl(2)}^{Q1_*}~~~~~~~~~\\\nonumber
 &&~~~-2 \sum_{j=1}^2 \log S_{vwx}^{11_*}(r_j^-,u_k) +2\log 2 -2 \sum_{j=1}^N\log\big(u_j-u_k-{2i\ov g}\big)\,
{x_j^--{1\ov x_{k}^-}\ov x_j^-- x_{k}^+} \\\la{Eba1sl2c1b}
 &&~~~+2 \log Z_{M|vw}\star  K^{M1_*}_{vwx}+ 2 \log \Big(1- \frac{1}{Y_-}\Big) \hstar K^{y1_*}_- + 2 \log\Big(1- \frac{1}{Y_+}\Big) \hstar K^{y1_*}_+\,.~~~~~~
\eea
We recall that the integration contours should run a little bit above the Bethe roots $u_j$, and below the dynamical roots $r_j$, and the consistency with the BY equations leads to identities of the form \eqref{Rksl2}
that we have checked numerically.

%%%%%%%%%%%%%%%%%%%%%%%%%%%%%%%%%%%%%%%%

\subsubsection*{Canonical TBA equations: $g_{cr}^{(m)}<g<\bar g_{cr}^{(m)}$}

The necessary modification of the integration contour follows the one for the first critical region, and the integration contour runs a little bit above the Bethe roots $u_j$, below the  zeros $r_j^{(k)}$, $k=1,\ldots,m$, and between the  zeros $r_j^{(m+1)}$, Im$(r_1^{(m+1)})<0$.  All the roots $r_j^{(k)} - {i\ov g}$  are between the integration contour and the real line of the mirror region. The contour for $Y_\pm$ functions lies above the  zeros $r_j^{(2)}$.

Here are the canonical TBA equations for Konishi-like states and $g_{cr}^{(m)}<g<\bar g_{cr}^{(m)}$
\bea\nonumber
&&\bullet \ Q{\rm -particles}:\quad V_Q = - \sum_{j=1}^N \log S_{\sl(2)}^{1_*Q}(u_j,v)
-2 \sum_{k=1}^{m-1}\sum_{j=1}^2 \log S_{vwx}^{kQ}(r_j^{(k+1)-},v)\\
&&\hspace{4cm}+
2 \log \frac{S_{vwx}^{m-1,Q}(r_1^{(m+1)},v) \, S_{vwx}^{m+1,Q}(r_1^{(m+1)},v)}
{S_{vwx}^{m,Q}(r_1^{(m+1)-},v)}\,,\label{TbaQsl2c1n}\\\nonumber
&&\bullet \  y{\rm -particles}: \quad\ V_\pm = \sum_{j=1}^N \log S^{1_*y}_\pm(u_j,v)- \sum_{k=1}^{m-1}\sum_{j=1}^2 \log S_k(r_j^{(k+1)-}-v)\\
&&\hspace{4cm}+\log {S_{m-1}(r_1^{(m+1)}-v)S_{m+1}(r_1^{(m+1)}-v)\ov S_{m}(r_1^{(m+1)-}-v)}\,,~~~~~\nonumber\\\nonumber
&&\bullet \  M|vw{\rm -strings}:  V_{M|vw}=\sum_{j=1}^N\log S^{1_*M}_{xv}(u_j,v) - \sum_{k=1}^{m-1}\sum_{j=1}^2 \log S_{kM}(r_j^{(k+1),-},v)\\
&&\hspace{5cm}+ \log {S_{m-1,M}(r_1^{(m+1)},v)S_{m+1,M}(r_1^{(m+1)},v)\ov S_{m,M}(r_1^{(m+1)-},v)} \,, \nonumber\\
&&\bullet  \  M|w{\rm -strings}: \ \ V_{M|w} =0\nonumber\,,
\eea
%%%%%%%%%%%%%%%%%%%%%%%%%%%%%%%%%%

\subsubsection*{ Exact Bethe equations: $g_{cr}^{(m)}<g< \bar g_{cr}^{(m)}$}

The exact Bethe equations take the following form
\begin{multline}
\pi i(2n_k+1)=\log Y_{1_*}(u_k) =i L\, p_k - \sum_{j=1}^N \log S_{\sl(2)}^{1_*1_*}(u_{j},u_{k})+ \log\left(1+Y_Q \right)\star K_{\sl(2)}^{Q1_*}\\
-2 \sum_{k=1}^m\sum_{j=1}^2 \log S_{vwx}^{k1_*}(r_j^{(k+1)-},u_k) +
2 \log \frac{S_{vwx}^{m-1,1_*}(r_1^{(m+1)},u_k) \, S_{vwx}^{m+1,1_*}(r_1^{(m+1)},u_k)}
{S_{vwx}^{m1_*}(r_1^{(m+1)-},u_k)}\\ + 2\log 2 -2 \sum_{j=1}^N\log\big(u_j-u_k-{2i\ov g}\big)\,
{x_j^--{1\ov x_{k}^-}\ov x_j^-- x_{k}^+}
 +2 \log Z_{M|vw}\star  K^{M1_*}_{vwx}\\+ 2 \log \Big(1- \frac{1}{Y_-}\Big) \hstar K^{y1_*}_- + 2 \log\Big(1- \frac{1}{Y_+}\Big) \hstar K^{y1_*}_+\,,
\la{Tba1sl2cm}
\end{multline}
where $\tilde{x}_j^-\equiv x(r_j^{(3)}-{i\ov g})$,
and the integration contours  run just above the Bethe roots, and below the  zeros $r_j^{(k)}$, $k=1,\ldots,m$.
The functions $Z_{M|vw}$ are defined as in \eqref{YvwZvw}.

The consistency with the BY equations again implies
that the sum of the terms on the last three lines of \eqref{Tba1sl2cm} is equal to $-2i p_k$
that we have checked numerically for the Konishi-like states.

\subsubsection*{Canonical TBA equations: $\bar g_{cr}^{(m)}<g< g_{cr}^{(m+1)}$}

In this case all the  zeros $r_j^{(k)}$, $k=1,\ldots,m+1$ are real, the integration contour runs below all of them but the Bethe roots $u_j$, and  the canonical TBA equations for Konishi-like states and $g_{cr}^{(m)}<g<\bar g_{cr}^{(m)}$ take the following form
\bea
&&\bullet \ Q{\rm -particles}:\quad V_Q = - \sum_{j=1}^N \log S_{\sl(2)}^{1_*Q}(u_j,v)
-2 \sum_{k=1}^m\sum_{j=1}^2 \log S_{vwx}^{kQ}(r_j^{(k+1)-},v) \,,~~~~~~~~\label{TbaQsl2c1nb}\\\nonumber
&&\bullet \  y{\rm -particles}: \quad\ V_\pm = \sum_{j=1}^N \log S^{1_*y}_\pm(u_j,v)- \sum_{k=1}^m\sum_{j=1}^2 \log S_k(r_j^{(k+1)-}-v) \,,~~~~~\nonumber\\\nonumber
&&\bullet \  M|vw{\rm -strings}:  V_{M|vw}=\sum_{j=1}^N\log S^{1_*M}_{xv}(u_j,v) - \sum_{k=1}^m\sum_{j=1}^2 \log S_{kM}(r_j^{(k+1)-},v)  \,, \nonumber\\
&&\bullet  \  M|w{\rm -strings}: \ \ V_{M|w} =0\nonumber\,.
\eea

\subsubsection*{ Exact Bethe equations: $\bar g_{cr}^{(m)}<g<  g_{cr}^{(m+1)}$}

The exact Bethe equations take the following form

\bea\nonumber
&&\pi i(2n_k+1)=\log Y_{1_*}(u_k) =i L\, p_k - \sum_{j=1}^N \log S_{\sl(2)}^{1_*1_*}(u_{j},u_{k})+ \log\left(1+Y_Q \right)\star K_{\sl(2)}^{Q1_*}~~~~~~~~~\\\nonumber
 &&~~~-2 \sum_{k=1}^m\sum_{j=1}^2 \log S_{vwx}^{k1_*}(r_j^{(k+1),-},u_k) +2\log 2 -2 \sum_{j=1}^N\log\big(u_j-u_k-{2i\ov g}\big)\,
{x_j^--{1\ov x_{k}^-}\ov x_j^-- x_{k}^+} \\\la{Tba1sl2cmb}
 &&~~~+2 \log Z_{M|vw}\star  K^{M1_*}_{vwx}+ 2 \log \Big(1- \frac{1}{Y_-}\Big) \hstar K^{y1_*}_- + 2 \log\Big(1- \frac{1}{Y_+}\Big) \hstar K^{y1_*}_+ \,.~~~~~~
\eea
The consistency with the BY equations again implies the fulfillment of
identities of the form \eqref{Rksl2}
that we have checked numerically for the Konishi-like states.

%%%%%%%%%%%%%%%%%%%%%%%%%%%%%%%%
\subsection{Reality of Y-functions}\la{app:reality}

In this appendix we show that the reality condition for Y-functions is compatible with the canonical TBA equations for Konishi-like states. We consider only the weak coupling region. The generalization to other regions and general $\sl(2)$ states is straightforward.

To start with we introduce the principle value prescription for the integrals involving $\log f(u)$ where $f(u)$ is real for real $u$, and has first-order  zeros (or poles) in the interval $[a,b]$ at $u_k$, $k=1,\ldots, N$
\bea\la{pvp}
\log f \star_{p.v} K \equiv \lim_{\e\to 0}
\sum_{k=0}^{N}\int_{u_k+\e}^{u_{k+1}-\e}\, du\, \log\left| f(u)\right|  K(u,v)\,,~~~~
\eea
where $u_0=a$, $u_{N+1}=b$. In the cases of interest $a=-\infty\,,\ b=\infty$ or $a=-2\,,\ b=2$.

Assuming for definiteness that $f(u)$ has $N$  zeros, and $f(\infty)>0$, one can write
\bea
f(u) = \tilde{f}(u)\prod_{k=1}^N (u-u_k)\,,
\eea
where $\tilde{f}(u)>0$ for any $u\in{\bf R}$.

Then the convolution terms of the form $\log f \star K$ with the integration contour running a little bit above the real line can be written as follows
\bea\la{logfk}
&&\log f \star K = \int_{a}^{b}\, du\, \log f(u+i0)   K(u+i 0,v)\\\nonumber
&&~~~=
 \int_{a}^{b}\, du\, \log \tilde{f}(u) K(u,v) + \sum_{k=1}^N \int_{a}^{b}\, du\, \log (u-u_k+i0)   K(u+i 0,v) \\\nonumber
&&~~~= \log f \star_{p.v} K \ +\ \pi i\, \sum_{k=1}^N \int_{a}^{u_k}\, du\,  K(u,v)
= \log f \star_{p.v} K \ +\ {1\ov 2} \sum_{k=1}^N \log {S(u_k,v)\ov S(a,v)}\,,
\eea
where
$\log S(u,v) \equiv 2\pi i  \int^{u}\, du'\,  K(u',v)$, and it can differ from the S-matrix defining the kernel $K$ by a function of $v$.
It is convenient to choose the normalization $S(a,v)=1$, and
most of our S-matrices satisfy this condition.

Let us now use \eqref{logfk} to show the reality of Y-functions. We start with
\eqref{Tbaysl2} that we write as follows
\bea\la{Yfory1}
&&\log {Y_+ \ov Y_-}(v)= - \sum_{j=1}^N \log S_{1_*y}(u_{j},v)  +\log(1 +  Y_{Q})\star K_{Qy} \,,\\
\la{Yfory2}
&&\log {Y_+  Y_-}(v) = \sum_{j=1}^N \log S_{1}(u_{j}-v)  - \log\left(1+Y_Q \right)\star K_Q+2\log {1+{1\ov Y_{M|vw}}
 \ov1+{1\ov Y_{M|w}}}\star K_M\nonumber\,,~~~~
\eea
where
\bea\la{sqsy}
S_{Q_*y}(u,v) = {x_s(u-{i\ov g}Q) - x(v)\ov x_s(u-{i\ov g}Q) - {1\ov x(v)}}\,{x_s(u+{i\ov g}Q) - {1\ov x(v)}\ov x_s(u+{i\ov g}Q) - x(v)}\,,
\eea
and we used that
\bea
S_{Qy}(u,v) = {S_-^{Qy}(u,v) \ov S_+^{Qy}(u,v) }\,,\quad S_{Q}(u-v) = {S_-^{Qy}(u,v) S_+^{Qy}(u,v) } = {u-v-{i\ov g}Q\ov u-v+{i\ov g}Q}\,,~~~~~
\eea
and that $S_{Q}(u-v)$ analytically continued to the string-mirror region is equal to itself.
Taking into account that
\bea
2\log {1+{1\ov Y_{M|vw}}
 \ov1+{1\ov Y_{M|w}}}\star K_M =2\log {1+{1\ov Y_{M|vw}}
 \ov1+{1\ov Y_{M|w}}}\star_{p.v} K_M -\sum_{k=1}^N \log S_1(u_k-v)\,,
\eea
we see that the equations for $Y_\pm$ can be written as
\bea\la{Yfory1b}
&&\log {Y_+ \ov Y_-}(v)= - \sum_{j=1}^N \log S_{1_*y}(u_{j},v)  +\log(1 +  Y_{Q})\star K_{Qy} \,,\\
\la{Yfory2b}
&&\log {Y_+  Y_-}(v) =  - \log\left(1+Y_Q \right)\star K_Q+2\log {1+{1\ov Y_{M|vw}}
 \ov1+{1\ov Y_{M|w}}}\star_{p.v.} K_M \,.~~~~
\eea
Assuming now that $Y_Q$, $Y_{M|vw}$ and $Y_{M|w}$ are real, the reality of $Y_\pm$ just follows from the reality and positivity of  $S_{Q_*y}(u,v)$ that can be easily checked by using \eqref{sqsy}.

Next we consider $Y_{M|vw}$. By using the p.v. prescription we write \eqref{Tbavwsl2} as follows
\begin{multline}
\log Y_{M|vw}(v) = {1\ov 2}\sum_{j=1}^N \log {S^{1_*M}_{xv}(u_j,v)^2\ov S^{1M}(u_j-v)} - \log\left(1+Y_{Q'} \right) \star K^{Q'M}_{xv} \\
+ \log\left(1+ \frac{1}{Y_{M'|vw}} \right) \star_{p.v.} K_{M'M}
+ \log {1- \frac{e^{ih_\a}}{Y_-} \over 1- \frac{e^{ih_\a}}{Y_+} } \star K_M \,,
\label{Tbavwsl2b}
\end{multline}
where
\bea
S^{1_*M}_{xv}(u,v)&=& {x_s(u-{i\ov g})-x(v+{i\ov g}M)\ov x_s(u+{i\ov g})- x(v-{i\ov g}M) }\,  {x_s(u+{i\ov g})-{1\ov x(v+{i\ov g}M)}\ov  x_s(u-{i\ov g})-{1\ov x(v-{i\ov g}M)}}\,,\\
S^{1M}(u)&=&  \frac{u -{i\ov g}(1+M)}{u +{i\ov g}(1+M)}  \frac{u -{i\ov g}(M-1)}{u +{i\ov g}(M-1)}\,.
\eea
Thus, we get
\bea
 {S^{1_*M}_{xv}(u,v)^2\ov S^{1M}(u-v)} &=&
 {x_s(u-{i\ov g})-x(v-{i\ov g}M)\ov x_s(u+{i\ov g})- x(v+{i\ov g}M) }\,  {x_s(u+{i\ov g})-{1\ov x(v+{i\ov g}M)}\ov  x_s(u-{i\ov g})-{1\ov x(v-{i\ov g}M)}}
 \\\nonumber
 &&~~~~~~~~\times  {x_s(u-{i\ov g})-x(v+{i\ov g}M)\ov x_s(u+{i\ov g})- x(v-{i\ov g}M) }\,  {x_s(u+{i\ov g})-{1\ov x(v-{i\ov g}M)}\ov  x_s(u-{i\ov g})-{1\ov x(v+{i\ov g}M)}}\,,
\eea
which is obviously real and positive, and at $M=1$ it has a double zero at $v=u$ as it should be.

Finally we consider $Y_Q$-functions. We write \eqref{TbaQsl2} as follows
\begin{multline}
 \log Y_Q(v) =- L\, \tH_{Q} - \sum_{j=1}^N \log S_{\sl(2)}^{1_*Q}(u_j,v)S^{1Q}_{vwx}(u_j,v)  + \log\left(1+Y_{Q'} \right) \star K_{\sl(2)}^{Q'Q}~~~ \\
~~~~~~~~~~~~~+ \log\left(1+ \frac{1}{Y_{M'|vw}} \right) \star_{p.v.} K^{M'Q}_{vwx} + \log \left(1- \frac{e^{ih_\a}}{Y_-} \right) \star K^{yQ}_- + \log\left(1- \frac{e^{ih_\a}}{Y_+} \right) \star K^{yQ}_+\,.
\label{TbaQsl2b}
\end{multline}
Here
\bea\nonumber
S^{1Q}_{vwx}(u,v) &=&  {x(u-{i\ov g})-x(v+{i\ov g}Q)\ov x(u-{i\ov g})- x(v-{i\ov g}Q) }\,  {x(u+{i\ov g})-x(v+{i\ov g}Q)\ov  x(u+{i\ov g})-x(v-{i\ov g}Q)}\, {x(v-{i\ov g}Q)\ov x(v+{i\ov g}Q)}\\\label{siqvwx}
&=&{x_s(u-{i\ov g})-x(v+{i\ov g}Q)\ov x_s(u-{i\ov g})- x(v-{i\ov g}Q) }\,  {1 - {1\ov x_s(u+{i\ov g})x(v+{i\ov g}Q)}\ov  1-{1\ov x_s(u+{i\ov g})x(v-{i\ov g}Q)}}=h_Q(u,v)\,,~~~~
\eea
where $h_Q(u,v)$ is the function that appears in the crossing relations, see \cite{AF09c}.
Note that $S^{1Q}_{vwx}$ is unitary: $\left(S^{1Q}_{vwx}\right)^* = 1/S^{1Q}_{vwx} = S^{1Q}_{vwx}/h_Q^2$.

To prove the reality of $S_{\sl(2)}^{1_*Q}(u,v)S^{1Q}_{vwx}(u,v)$ we use the crossing relation.\footnote{It follows closely the proof of the unitarity of the mirror S-matrix \cite{AF07}.} To this end it is convenient to go to the $z$-torus. Then, the crossing relation for $S_{\sl(2)}^{1Q}$ can be written in the form \cite{AF06}
\bea
S_{\sl(2)}^{1Q}(z_{1}-{\om_2\ov 2},z_2)S_{\sl(2)}^{1Q}(z_{1}+{\om_2\ov 2},z_2) = {1\ov h_Q(u,v)^2}\,,
\eea
and we have
\bea\nonumber
\left(S_{\sl(2)}^{1_*Q}(u,v)\right)^* &\equiv& \left(S_{\sl(2)}^{1Q}(z_{*1},z_2)\right)^* =
 \left(S_{\sl(2)}^{1Q}(z_{1}-{\om_2\ov 2},z_2)\right)^* = {1\ov S_{\sl(2)}^{1Q}(z_{1}+{\om_2\ov 2},z_2) }\\
 &=& S_{\sl(2)}^{1_*Q}(u,v) h_Q(u,v)^2\,,
\eea
where we have chosen $z_k$ to be real, and used the generalized unitarity condition for the mirror model. Taking into account  \eqref{siqvwx}, one concludes that  $S_{\sl(2)}^{1_*Q}S^{1Q}_{vwx}$ is real. It is also possible to show that
the product is positive by representing $S_{\sl(2)}^{1_*Q}S^{1Q}_{vwx}$ as a product $s s^*$, where $s \sim \sigma$.

%%%%%%%%%%%%%%%%%%%%%%%%%%%%%%%%%%%%%%%%%
\subsection{From canonical to simplified TBA equations}\la{appSimple}

In this appendix we discuss the derivation of the simplified TBA equations from the canonical ones for Konishi-like states in the weak coupling region.
Since it basically repeats the one done in \cite{AF09b,AF09d} we will be sketchy.

\subsubsection*{Simplifying the TBA equations for $vw$-strings}

To derive the simplified equation \eqref{Yforvw3} for $vw$-strings from the canonical one, one applies the inverse kernel $(K+1)^{-1}_{MN}$ to   \eqref{Tbavwsl2}
 and uses the following identity
\bea\la{SxvKI}
\log S^{1_*Q}_{xv}\star(K+1)_{QM}^{-1} &=& \delta_{M1}\Big(\log S(u-v+{i\ov g}-i0) + \log S_{1_*y}\star s\Big)\,. ~~~~~
\eea
Note also that one should understand $\log S^{1_*Q}_{xv}$ as
\bea
\log S^{1_*Q}_{xv}(u,v) = \log(v-u) + \log {S^{1_*Q}_{xv}(u,v) \ov v-u}\,.
\eea
Then, even though the last formula is valid up to a multiple of $i \pi$, it agrees exactly with Mathematica choice of branches.

\subsubsection*{Simplifying the TBA equations for $Q$-particles}
We want to apply $(K+1)^{-1}_{MN}$ to the equation \eqref{TbaQsl2}.
This requires computing
\bea\nonumber
\log S_{\sl(2)}^{1_*M}\star (K+1)^{-1}_{MQ} &=&-\log S_{1M}\star (K+1)^{-1}_{MQ}-2\log \Sigma_{1_*M}\star (K+1)^{-1}_{MQ} \\
&=&-\delta_{Q2}\log S(u-v)-2\delta_{Q1}\log \check{\Sigma}_{1_*}\cstar s\,.~~~~
\eea
As was shown in \cite{AF09c}, the
improved dressing factor is a holomorphic function of the first
argument in the string region, and the second one in the intersection of the region
$\{|y^{+}_{2}|<1,|y^{-}_{2}|>1\}$ with the mirror region ${\rm
Im}\,y_2^{\pm}<0$, which includes the real momentum line of the
mirror theory. This immediately implies that
\bea
\log\Sigma_{1_*M}\star (K+1)^{-1}_{MQ}=0~~~~{\rm for}~~~Q\neq 1.
\eea
To find  $\log\Sigma_{1_*M}\star (K+1)^{-1}_{M1}$, and, therefore,
$\check{\Sigma}_{1_*}$ we start with the kernel
$K^\Sigma_{1_*M}(u,v)$, and
 use the following identity derived in \cite{AF09d}
\bea\la{KSKi} K^\Sigma_{Q'M}\star (K + 1)^{-1}_{MQ}
=\delta_{1Q}\check{K}_{Q'}^\Sigma\star s\,, \eea where the
kernel $\check{K}_{Q}^\Sigma(u,v)$ does not vanish for
$|v|<2$, and
is given by \eqref{Ksig}.
Setting $Q'=1=Q$ in \eqref{KSKi} and analytically continuing the first argument to the string region, one gets
\bea\la{KSKism} K^\Sigma_{1_*M}\star (K + 1)^{-1}_{MQ}
=\delta_{1Q}\check{K}_{1_*}^\Sigma\star s\,, \eea
where the analytically continued kernel $\check{K}_{1_*}^\Sigma$ is given by
\eqref{k1starsigma}.
To obtain eq.\eqref{k1starsigma} from  \eqref{Ksig} one uses the following formula
\bea
\check{I}_0(u+{i\ov g}\,,v) &=&\check{I}_1(u\,,v) - {1\ov 2\pi i}{1\ov u-v-{i\ov g}}
-  {1\ov \pi i}\int_{-2}^2\, dt\, {1\ov u-t-{i\ov g}}\check{K}(t,v)~~~~ \nonumber\\
&=&\check{I}_1(u\,,v) - K_{ss}(u-{i\ov g}\,,v)\,.~~~~~~~
\eea
Finally, integrating \eqref{k1starsigma} over the first argument, one gets
\eqref{s1star}.

The remaining part of the derivation of the simplified equations for Q-particles repeats \cite{AF09d}.

%%%%%%%%%%%%%%%%%%%%%%%%%%%%%%%%%%%%%%%

\end{document}